%% file: article.tex
\documentclass[a4paper,12pt]{article}
\pdfoutput=1
\usepackage{amssymb}
\usepackage{amsmath}
\usepackage{graphicx}
\usepackage{epsfig}
\usepackage{slashed}
\usepackage{braket}
\usepackage{siunitx}
\usepackage{lineno}
\usepackage{setspace}
\usepackage{mathtools}
\usepackage[utf8]{inputenc}
\usepackage[top=15mm,bottom=20mm,left=15mm,right=15mm]{geometry}
\usepackage[colorlinks=true,hypertexnames=false]{hyperref}
\usepackage{color}
\usepackage{amstext}
\usepackage{multirow,bigdelim}
\usepackage{array}
\newcolumntype{C}{>{$}c<{$}}
\newcolumntype{L}{>{$}l<{$}}
\newcolumntype{R}{>{$}r<{$}}
\usepackage{arydshln}
\usepackage{xcolor,colortbl}

\makeatletter
\g@addto@macro\bfseries{\boldmath}
\makeatother

\def\Tr{\,\mathrm{Tr}}
\def\mev{\mathrm{Me\kern-0.1em V}}
\def\gev{\mathrm{Ge\kern-0.1em V}}
\def\tev{\mathrm{Te\kern-0.1em V}}
\def\fm{\mathrm{fm}}
\def\refcite#1{Ref.~\cite{#1}}
\def\refscite#1{Refs.~\cite{#1}}
\def\reff#1{\ref{#1}}

\def\eq#1{Eq.~(\reff{#1})}

\def\fig#1{Fig.~\reff{#1}}

\def\epemtohad{e^+e^-\to\text{hadrons}}
\def\epemtopippim{e^+e^-\to\pi^+\pi^-}
\def\amulohvp{a_{\mu}^\text{LO-HVP}}

\def\chidof{\chi^2/\mathrm{dof}}

\usepackage[sorting=none,style=nature,backend=bibtex]{biblatex}
\AtEveryBibitem{\clearfield{doi}}
\AtEveryBibitem{\clearfield{url}}

\input{figures/values}

\begin{document}

\clearpage
\setcounter{page}{1}
\begin{refsegment}
\vspace*{0.5cm}
\begin{center}
\huge\bf
\input{title}
\end{center}
\vspace*{0.5cm}
\input{authors}
\newpage
\input{main}
\input{main_statement}
\printbibliography[segment=1]
\end{refsegment}

\singlespacing
\nolinenumbers
\clearpage
\renewcommand{\figurename}{Figure}
\renewcommand{\tablename}{Table}
\renewcommand{\thefigure}{S\arabic{figure}}
\renewcommand{\thetable}{S\arabic{table}}
\renewcommand{\theequation}{S\arabic{equation}}
\renewcommand{\thesection}{S\arabic{section}}
\setcounter{table}{0}
\setcounter{figure}{0}
\setcounter{page}{1}
\setcounter{section}{0}
\setcounter{equation}{0}
\begin{refsegment}
\begin{center}
\bf
{\Huge Supplementary Information}\\
\vspace*{1.0cm}
\Large
\input{title}
\end{center}
\input{authors}
\newpage
\tableofcontents
\newpage
\input{si_conf}
\input{si_omega}
\input{si_fpi}
\input{si_phys}

\input{si_analysis}
\input{si_win}

\input{si_finvol}
\input{si_finvol_pheno}
\input{si_ibcheck}

\input{si_tail}
\printbibliography[segment=2]
\end{refsegment}

\end{document}

%% file: figures/values.tex
\def\valuesLightSd{47.85(5)(13)[14]}
\def\valuesDiscoSd{-0.0014(22)(101)[104]}
\def\valuesLightId{206.92(37)(34)[50]}

\def\valuesLightIdFvNomon{206.48(50)(74)[90]}
\def\valuesLightIdFvEred{1-0.50/0.90=44\%}
\def\valuesDiscoId{-1.049(21)(24)[32]}

\def\valuesLightIdFvNomon{206.48(50)(74)[90]}
\def\valuesLightIdFvEred{1-0.50/0.90=44\%}
\def\valuesLightIdFvNomon{206.48(50)(74)[90]}
\def\valuesLightIdFvEred{1-0.50/0.90=44\%}
\def\valuesLightLdmilc{97.57(1.76)(1.17)[2.11]}
\def\valuesLipdiLdbody{365.6(2.4)(2.0)[3.1]}
\def\valuesLipdiBody{619.6(2.3)(2.1)[3.1]}
\def\valuesLightLdmilc{97.57(1.76)(1.17)[2.11]}
\def\valuesLightLd{404.6(2.7)(2.5)[3.7]}
\def\valuesLightLdFv{386.1(2.7)(2.5)[3.6]}
\def\valuesLightLdFvNomon{383.2(4.0)(4.1)[5.7]}
\def\valuesLightLdbodyFv{366.4(1.8)(2.2)[2.8]}
\def\valuesLightLdFvEred{1-3.6/5.7=37\%}
\def\valuesLightLdbodyEred{1-2.8/3.6=22\%}
\def\valuesLightLdFvDiff{2.8}
\def\valuesLightLdFv{386.1(2.7)(2.5)[3.6]}
\def\valuesLightLdFvNomon{383.2(4.0)(4.1)[5.7]}
\def\valuesLightLdFvEred{1-3.6/5.7=37\%}
\def\valuesLightLdFvDiff{2.8}
\def\valuesLightLdFvNomon{383.2(4.0)(4.1)[5.7]}
\def\valuesLightLdbodyFv{366.4(1.8)(2.2)[2.8]}
\def\valuesLightLdFvEred{1-3.6/5.7=37\%}
\def\valuesLightLdbodyEred{1-2.8/3.6=22\%}
\def\valuesLightLdFvDiff{2.8}
\def\valuesLightFv{640.1(2.8)(2.5)[3.7]}
\def\valuesLightFvNature{633.7(2.1)(4.2)[4.7]}
\def\valuesLightFvThisvsnatDiff{6.4}
\def\valuesLightFvThisvsnatErru{4.5}
\def\valuesLightFvThisvsnatErrc{3.8}
\def\valuesLightFvThisvsnatSigu{1.4}
\def\valuesLightFvThisvsnatSigc{1.7}
\def\valuesLightFvNature{633.7(2.1)(4.2)[4.7]}
\def\valuesLightFvThisvsnatDiff{6.4}
\def\valuesLightFvThisvsnatErru{4.5}
\def\valuesLightFvThisvsnatErrc{3.8}
\def\valuesLightFvThisvsnatSigu{1.4}
\def\valuesLightFvThisvsnatSigc{1.7}
\def\valuesStrangeSd{9.04(3)(6)[7]}
\def\valuesStrangeId{27.04(11)(8)[14]}
\def\valuesStrangeLdbody{16.89(9)(7)[11]}
\def\valuesStrangeBody{53.12(14)(11)[18]}
\def\valuesCharmSd{11.71(15)(21)[26]}
\def\valuesCharmId{2.95(12)(17)[21]}
\def\valuesCharmLdbody{0.0142(17)(22)[28]}
\def\valuesCharmBody{14.68(24)(29)[38]}
\def\valuesPhenoTail{27.59(17)(9)[26]}

\def\valuesPhenoTailnew{27.59(26)(45)[52]}
\def\valuesPhenoTailnew{27.59(26)(45)[52]}
\def\valuesTotalId{236.29(41)(39)[57]}
\def\valuesTotalLd{409.5(2.6)(2.2)[3.4]}
\def\valuesTotalId{236.29(41)(39)[57]}
\def\valuesTotalLd{409.5(2.6)(2.2)[3.4]}
\def\valuesTotal{715.1(2.5)(2.3)[3.4]}
\def\valuesTotalerr{3.4}
\def\valuesTotalprec{0.48}
\def\valuesTotalerr{3.4}
\def\valuesTotalprec{0.48}
\def\valuesImprovedprl{5.5}
\def\valuesImprovednat{1.6}

\def\valuesWindowbabarvsthis{3.5}
\def\valuesWindowkloevsthis{6.2}
\def\valuesWindowcmdvsthis{1.3}
\def\valuesWindowtauvsthis{2.7}
\def\valuesAmuthiswork{11659205.2(3.6)}
\def\valuesAmuthisworkPpm{0.31}
\def\valuesAmuthisworkPpm{0.31}
\def\valuesAmuexpvsthis{0.5}

\def\valuesHvpwhitepapervsthisImp{1.8}
\def\valuesHvpwhitepapervsthisImp{1.8}
\def\valuesHvpthisvsnatDiff{7.6}
\def\valuesHvpthisvsnatErru{5.2}
\def\valuesHvpthisvsnatErrc{4.5}
\def\valuesHvpthisvsnatSigu{1.5}
\def\valuesHvpthisvsnatSigc{1.7}
\def\valuesWnullFromfpi{0.17264(36)(33)[48]}
\def\valuesWnullFromfpiQed{-0.00027(10)(23)[25]}
\def\valuesWnullFromfpiQcd{0.17284(34)(25)[42]}
\def\valuesWnullFromfpiQcdNomon{0.17279(37)(40)[54]}
\def\valuesWnullFromfpiQcdEred{1-0.00042/0.00054=22\%}
\def\valuesWnullFromfpiErr{2.8}
\def\valuesWnullFromfpiSys{(33)= (27)_\mathrm{qcd}(18)_\mathrm{qvv}(9)_\mathrm{qss}(6)_{V_{ud}}(3)_\mathrm{exp}}
\def\valuesWnullFromfpiQed{-0.00027(10)(23)[25]}
\def\valuesWnullFromfpiQcd{0.17284(34)(25)[42]}
\def\valuesWnullFromfpiQcdNomon{0.17279(37)(40)[54]}
\def\valuesWnullFromfpiQcdEred{1-0.00042/0.00054=22\%}
\def\valuesWnullFromfpiQcdNomon{0.17279(37)(40)[54]}
\def\valuesWnullFromfpiQcdEred{1-0.00042/0.00054=22\%}
\def\valuesWnullFromfpiErr{2.8}
\def\valuesWnullFromfpiSys{(33)= (27)_\mathrm{qcd}(18)_\mathrm{qvv}(9)_\mathrm{qss}(6)_{V_{ud}}(3)_\mathrm{exp}}
\def\valuesWnullFromom{0.17245(22)(46)[51]}
\def\valuesWnullFromomQed{0.00010(3)(2)[3]}
\def\valuesWnullFromomQcd{0.17235(22)(46)[51]}
\def\valuesWnullFromomQed{0.00010(3)(2)[3]}
\def\valuesWnullFromomQcd{0.17235(22)(46)[51]}

%% file: title.tex
Hybrid calculation of hadronic vacuum polarization in muon g-2 to \valuesTotalprec\%

%% file: authors.tex
\noindent
A.~Boccaletti$^{1,2}$,
Sz.~Borsanyi$^{1}$,
A.~Cotellucci$^{2}$,
M.~Davier$^{3}$,
Z.~Fodor$^{4,5,1,2,6,7,*}$,
F.~Frech$^{1}$,
A.~G\'erardin$^{8}$,
D.~Giusti$^{2,9}$,
A.Yu.~Kotov$^{2}$,
L.~Lellouch$^{8}$,
Th.~Lippert$^{2}$,
A.~Lupo$^{8}$,
B.~Malaescu$^{10}$,
S.~Mutzel$^{8,11}$,
A.~Portelli$^{12,13}$,
A.~Risch$^{1}$,
M.~Sj\"o$^{8}$,
F.~Stokes$^{2,14}$,
K.K.~Szabo$^{1,2}$,
B.C.~Toth$^{1,2}$,
G.~Wang$^{8}$,
Z.~Zhang$^{3}$
\vspace*{1cm}

\noindent
$^{ 1}$\ Department of Physics, University of Wuppertal, D-42119 Wuppertal, Germany\\
$^{ 2}$\ J\"ulich Supercomputing Centre, Forschungszentrum J\"ulich, D-52428 J\"ulich, Germany\\
$^{ 3}$\ IJCLab, Universit\'e Paris-Saclay et CNRS/IN2P3, Orsay, 91405, France\\
$^{ 4}$\ Physics Department, Pennsylvania State University, University Park, PA 16802, USA\\
$^{ 5}$\ Institute for Computational and Data Sciences, Pennsylvania State University, University Park, PA 16802, USA\\
$^{ 6}$\ Institute for Theoretical Physics, E\"otv\"os University, H-1117 Budapest, Hungary\\
$^{ 7}$\ University of California, San Diego, 9500 Gilman Drive, La Jolla, CA 92093, USA\\
$^{ 8}$\ Aix Marseille Univ, Universit\'e de Toulon, CNRS, CPT, IPhU, Marseille, France\\
$^{ 9}$\ Fakult\"at f\"ur Physik, Universit\"at Regensburg, 93040, Regensburg, Germany\\
$^{10}$\ LPNHE, Sorbonne Universit\'e, Universit\'e Paris Cit\'e, CNRS/IN2P3, Paris, 75252, France\\
$^{11}$\ Laboratoire de Physique de l’Ecole Normale Sup\'erieure, Mines Paris - PSL, CNRS, Inria, PSL Research University, Paris, France\\
$^{12}$\ School of Physics and Astronomy, University of Edinburgh, Edinburgh EH9 3JZ, United Kingdom\\
$^{13}$\ RIKEN Center for Computational Science, Kobe 650-0047, Japan\\
$^{14}$\ Special Research Centre for the Subatomic Structure of Matter, Department of Physics, University of Adelaide, South Australia 5005, Australia\\

%% file: main.tex
{\bf For fifty years, the standard model of particle physics has been hugely
successful in describing subatomic phenomena. In the last quarter century, this
was challenged by a mismatch between its predictions and precision
measurements of the anomalous magnetic moment of the muon, $a_\mu$.  This
disagreement was eventually reconciled. First, through a determination in an
ab-initio lattice calculation~\cite{Borsanyi:2020mff} of the most uncertain
theoretical contribution: the leading-order hadronic vacuum polarization,
$\amulohvp$, and subsequently by experimental
results~\cite{CMD-3:2023alj} and updates of the reference standard-model
predictions using lattice results for
$\amulohvp$~\cite{Aliberti:2025beg}.  Here, we present a new calculation
for this crucial quantity, obtaining
$\amulohvp=\valuesTotal\times10^{-10}$. This reduces the uncertainty by a
factor of $\valuesImprovednat$ compared to our earlier
computation~\cite{Borsanyi:2020mff}. We adopt a hybrid approach that includes a
small, long-distance contribution from experiments in a low-energy regime where
they all agree. Our approach combines the strengths of experimental and lattice
data in different energy ranges, achieving better precision than with either
alone. Our lattice QCD simulations are performed on finer lattices than in
Ref.~\cite{Borsanyi:2020mff}, allowing for an even more accurate continuum
extrapolation. Combined with the calculations of the other standard-model
contributions summarized in Ref.~\cite{Aliberti:2025beg}, our result leads to a
prediction that differs from the recent measurement of $a_\mu$
\cite{Muong-2:2025xyk} by only $\valuesAmuexpvsthis$ standard deviations. This
provides a remarkable validation of the standard model to 11 digits.}

\bigskip

The muon is a short-lived elementary particle with spin 1/2 and a mass 207
times larger than that of the electron.  Both particles create a magnetic field
around them, characterized by a magnetic dipole moment.  This moment is
proportional to the spin and charge of the particle, and inversely proportional
to twice its mass.  Dirac's relativistic quantum mechanics predicts that the
constant of proportionality, $g_\mu$, known as the Landé factor, is precisely
2.  Relativistic quantum field theory introduces additional small
corrections induced not only by all particles and interactions of the Standard
Model (SM), but also potentially by yet undiscovered ones.
Because muons are more massive than
electrons, quantum corrections associated with heavy particles are generically
much larger for the former than for the latter~\cite{Jegerlehner:2009ry}.  This
increased sensitivity to the effects of possible unknown particles is the
reason for the current focus on the muon.  The corrections to $g_\mu$ are
commonly called the anomalous magnetic moment and are quantified as
$a_\mu=(g_\mu-2)/2$.

When calculating $a_\mu$, the uncertainty comes almost exclusively from the
strong interaction, described in the SM by quantum chromodynamics (QCD).  In
particular, the dominant source of uncertainty comes from hadronic vacuum
polarization (HVP) at leading order in the fine-structure constant (LO-HVP).
More generally, HVP induces a modification in the propagation of a virtual
photon in the vacuum, caused by the strong interaction.

Here we present a calculation of this LO-HVP contribution to $a_\mu$
($\amulohvp$) with unprecedented accuracy.  To that end we apply numerical
lattice quantum field theory techniques that allow QCD predictions to be made in
the highly nonlinear regime that is relevant here. Mathematically, QCD is a
generalized version of quantum electrodynamics (QED). However, QCD predicts physical phenomena that are drastically different from those described by QED. The Euclidean Lagrangian for a quark of mass $m$ and charge $q$ (in
units of the positron charge, $e$), subject to strong and electromagnetic
interactions, can be written as ${\cal L} = 1/(4e^2) F_{\mu\nu}F_{\mu\nu}
+ 1/(2g^2) \Tr\,G_{\mu\nu}G_{\mu\nu} + {\bar \psi}[\gamma_\mu (\partial_\mu +
iqA_\mu + iG_\mu) + m ] \psi$, where $F_{\mu\nu}=\partial_\mu A_\nu -
\partial_\nu A_\mu$, $G_{\mu\nu}=\partial_\mu G_\nu - \partial_\nu G_\mu
+i[G_\mu,G_\nu]$ and $g$ is the QCD coupling constant.  The fermionic quark
fields $\psi$ have an additional ``colour'' index in QCD, which runs from 1 to
3.  Different ``flavours'' of quarks are represented by independent fermionic
fields, with different masses and charges.  In QED, the gauge potential $A_\mu$
is a real valued field, whereas in QCD, $G_\mu$ is a 3$\times$3 traceless Hermitian matrix field
acting in ``colour'' space.  In the present work we include both QCD and QED as
well as four nondegenerate quark flavours (up, down, strange and charm) in a
fully-dynamical, staggered-fermion formulation.  We also consider the tiny
contribution of the bottom quark. Its error is subdominant, and we repeat
the treatment of our earlier analysis~\cite{Borsanyi:2020mff}.

To calculate the LO-HVP contribution to $a_\mu$, we start with the
zero-three-momentum, two-point function of the quark electromagnetic current in
Euclidean time $t$~\cite{Bernecker:2011gh}.  In this so-called time-momentum
representation, it is given by
\begin{equation}
    \label{eq:main_gdef}
    G(t)= -\frac{1}{3e^2}\sum_{\mu=1,2,3}\int d^3x
    \langle J_\mu(\vec{x},t) J_\mu(0)\rangle\ ,
\end{equation}
where $J_\mu$ is the quark electromagnetic current with $J_\mu/e= \frac{2}{3}
\bar{u}\gamma_\mu u - \frac{1}{3} \bar{d}\gamma_\mu d - \frac{1}{3}
\bar{s}\gamma_\mu s + \frac{2}{3} \bar{c} \gamma_\mu c$.  $u,d,s$ and $c$ are
the up, down, strange and charm quark fields. The angle brackets stand for
the QCD+QED expectation value to order $e^2$.  It is convenient to decompose
$G(t)$ into light ($u$ and $d$), strange, charm and disconnected components,
which have very different statistical and systematic uncertainties.  Performing
a weighted integral of the one-photon-irreducible part, $G_\mathrm{1\gamma
I}(t)$, of $G(t)$ from $t=0$ to infinity yields the LO-HVP contribution to
$a_\mu$~\cite{Bernecker:2011gh}.  The weight is a known kinematic function,
$K(t
m_\mu)$~\cite{Lautrup:1971jf,deRafael:1993za,Blum:2002ii,Bernecker:2011gh}.
Thus:
\begin{equation}
    \label{eq:amulohvp_def}
    a_{\mu}^\mathrm{LO-HVP}=\alpha^2
    \int_0^\infty dt\ K(t m_\mu)\ G_\mathrm{1\gamma I}(t)\ ,
\end{equation}
where $\alpha$ is the fine-structure constant at vanishing recoil and $m_\mu$
is the mass of the muon.

\begin{figure}[t]
    \centering
    \includegraphics{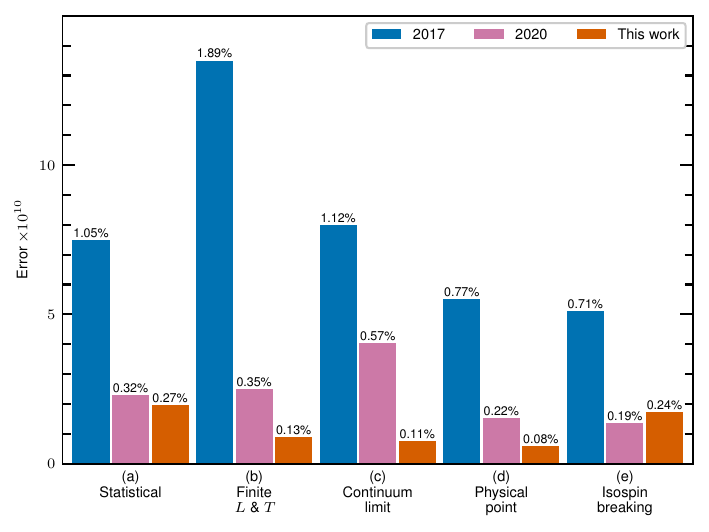}
    \caption
    {
	Main uncertainties and their reduction in our successive lattice
	calculations of $\amulohvp$.  Their sources are labelled (a-e) in the
	text and are given a short descriptive title below the bars in the
	plot. Their approximate size relative to the total LO-HVP contribution
	obtained in the present work is also shown. The blue bars on the left
	of each group correspond to our 2017 result~\cite{Borsanyi:2017zdw},
	the pink ones in the middle to our 2020 findings
	\cite{Borsanyi:2020mff} and the orange ones on the right to the work
	presented here. The isospin-breaking uncertainty (e) in this work is slightly
	larger than in 2020 due to changes in the way we set the physical scale.
	We moved from using the $\Omega^-$ baryon to the pion decay rate, which
	reduced other uncertainties, but increased the isospin-breaking uncertainty.
	Note: the statistical error (a) refers to that of
	the isospin-symmetric contribution in finite volume. The finite-size
	(b) and isospin-breaking (e) errors also contain statistical
	components of 0.08\% and 0.16\%, respectively.
    }
    \label{fig:err_improve}
\end{figure}

Reducing the uncertainty in the calculation of $\amulohvp$ to below half a
percent is a major challenge.  In particular, a number of contributions to this
uncertainty must be controlled.  They are (a) statistical uncertainties; (b)
those associated with the finite spatial size $L$ and time $T$ of the lattice;
(c) with the extrapolation to the continuum limit; (d) with fixing the five
parameters of four-flavour QCD; (e) with isospin-symmetry breaking. The progress made
in our successive lattice calculations of $\amulohvp$ is illustrated in
\fig{fig:err_improve}, where those contributions to the uncertainty are shown.
In the present work we focus on reducing the two largest ones in our
2020 calculation, which are (c) and (b).  We discuss all these contributions
(a-e) in detail now.

\noindent {\it ad} (a). Statistical uncertainties in the light-quark-connected
and disconnected contributions to the correlation function of
\eq{eq:main_gdef}, associated with the stochastic evaluation of the QCD and QED
path integrals, increase exponentially at large Euclidean times $t$.  In
addition to the many improvements made in~\cite{Borsanyi:2020mff}, to reduce
those uncertainties further we use mock analyses to determine which ensembles
required more statistics.  In particular, we increase the statistics on the
lattices which have the smallest lattice spacings and which are critical for
controlling the necessary continuum extrapolations.  Moreover, to control the
statistical uncertainties at large $t$ we replace the lattice calculation of
the contribution to $\amulohvp$ from $G(t)$ above $t\ge 2.8\,\fm$ by a
state-of-the-art, data-driven determination, via the HVPTools
setup~\cite{Davier:2010rnx,Davier:2010nc,Davier:2017zfy,Davier:2019can}.\footnote{
Such a combination was originally proposed in \refcite{RBC:2018dos}. There,
however, lattice results were replaced by $\epemtohad$ data above a much earlier
Euclidean time, $t\ge 1\,\fm$.}
Here and in the remainder of the paper the expression ``data-driven" refers
to predictions based on measurements of the hadron spectrum in
$e^+e^-$ annihilation and $\tau$-decay experiments.
Before combining the two results, we verify that the lattice and the
data-driven determinations of part of this long-distance ``tail'' contribution
agree within errors.  We compute this tail
contribution using the most precise measurements of the two-pion spectrum by
BaBar~\cite{BaBar:2009wpw,BaBar:2012bdw},
KLOE~\cite{KLOE:2008fmq,KLOE:2010qei,KLOE:2012anl,KLOE-2:2017fda} and
CMD-3~\cite{CMD-3:2023alj}, as well as the one obtained from hadronic $\tau$
decays~\cite{Davier:2009ag,Davier:2013sfa}. These experiments almost fully cover
the relevant energy range. For estimating the uncertainty of the tail observable, we
also use other experiments with partial coverage. The two-pion spectra of these experiments are supplemented by
the contributions from all of the other hadronic final states, as described in
\refcite{Davier:2023fpl}.  The tail contribution is dominated by centre-of-mass
energies below the $\rho$-meson peak, a region where all the measurements agree
very well. The tail only accounts for less than $5\%$ of our final,
lattice-dominated result for $\amulohvp$.
The Supplementary Information describes our
determination of this contribution and further justifies its use in our
calculation.

\noindent {\it ad} (b). Finite $L$ and $T$ corrections gave the largest
contribution to the error in 2017~\cite{Borsanyi:2017zdw}. Even in our 2020
calculation~\cite{Borsanyi:2020mff} it was still a significant source of
uncertainty.  Here our determination of the tail contribution using a
data-driven approach reduces those corrections by a factor of about two and the
associated uncertainties by roughly three.  We compute those corrections using the
dedicated simulations of \refcite{Borsanyi:2020mff}, supplemented by next-to-next-to-leading order
chiral perturbation theory for distances beyond
$11\,\fm$~\cite{Borsanyi:2020mff,Aubin:2019usy}.  Those results are checked
against nonperturbative analytical approaches to finite-volume
corrections~\cite{Meyer:2011um,Lellouch:2000pv,Luscher:1990ux,Hansen:2019rbh,
Hansen:2020whp} that we complement with experimental $\pi^+\pi^-$ cross-section
data below $1.3\,\gev$. Details are given in the Supplementary Information.

\noindent {\it ad} (c). The continuum extrapolation of the isovector
contribution to $\amulohvp$ was the largest source of uncertainty in our 2020
computation~\cite{Borsanyi:2020mff} and we have dedicated significant resources
to further control it.  The uncertainties were mainly due to long-distance,
taste-breaking effects that are present in staggered-fermion computations.
Here we add a new, finer lattice spacing.  The corresponding simulations
have a numerical cost close to the one required for the full 2020 computation.
In \refcite{Borsanyi:2020mff} the smallest lattice spacing was $0.064\,\fm$.
The new lattice spacing is $0.048\,\fm$.  Since the leading discretization
effects are proportional to the square of the lattice spacing, results at this
new lattice spacing have cutoff effects reduced by a factor of nearly two.  We
further account for the fact that different $t$-regions in $G(t)$ have
different cut-off effects by dividing the integral of \eq{eq:amulohvp_def} into
four $t$ intervals delimited by sigmoid functions. Such intervals or
``windows'' were first proposed in \refcite{RBC:2018dos}.  The first
window corresponds to the Euclidean-time interval $0.0$ to $0.4\,\fm$, known
as the short-distance (SD) window~\cite{Aoyama:2020ynm,RBC:2018dos} and denoted
$a_{\mu,00-04}^{\text{LO-HVP}}$ here.  We use three more intervals between
$0.4$ and $2.8\,\fm$ (separated at $2.0$ and $2.4\,\fm$) because this choice
yields a reduced uncertainty on the final result for $\amulohvp$.  We carry out
the continuum extrapolation in those windows separately. We then add the
individual extrapolated results to obtain the contribution to $\amulohvp$ from
the Euclidean time interval from $0$ to $2.8\,\fm$, taking correlations into
account.
The uncertainty on the light connected contribution is decreased by the new
ensembles by 37\%, and by using the data-driven approach to compute the tail by
an additional 22\%.
The whole procedure is detailed in the Supplementary Information.

\noindent {\it ad} (d). We improve the determination of the physical point,
which is now based on a very precise computation of the muonic decay rate of the
charged pion. As a crosscheck, we also perform the determination using the mass
of the $\Omega^-$ baryon as input, and find a good agreement between the two
approaches. The uncertainty associated with the physical point determination
was already small in \refcite{Borsanyi:2020mff} and is even smaller here.  For
details see the Supplementary Information.

\noindent {\it ad} (e). The uncertainties on the isospin-symmetry breaking
contributions obtained in \refcite{Borsanyi:2020mff} were already sufficiently small to
reach the precision sought here. Our error on this contribution is now slightly
increased: the isospin-breaking error on the pion decay rate is larger than it
was on the $\Omega^-$ baryon. Also we perform a variety of crosschecks that
confirm our earlier results on the isospin-breaking contributions. Our current
uncertainty details are given in the Supplementary Information.

By far the largest contributions to the various windows considered in this work
come from connected light-quark diagrams.  We focus on these here and discuss
the other contributions in the Supplementary Information.

\begin{figure}[t]
    \centering
    \includegraphics{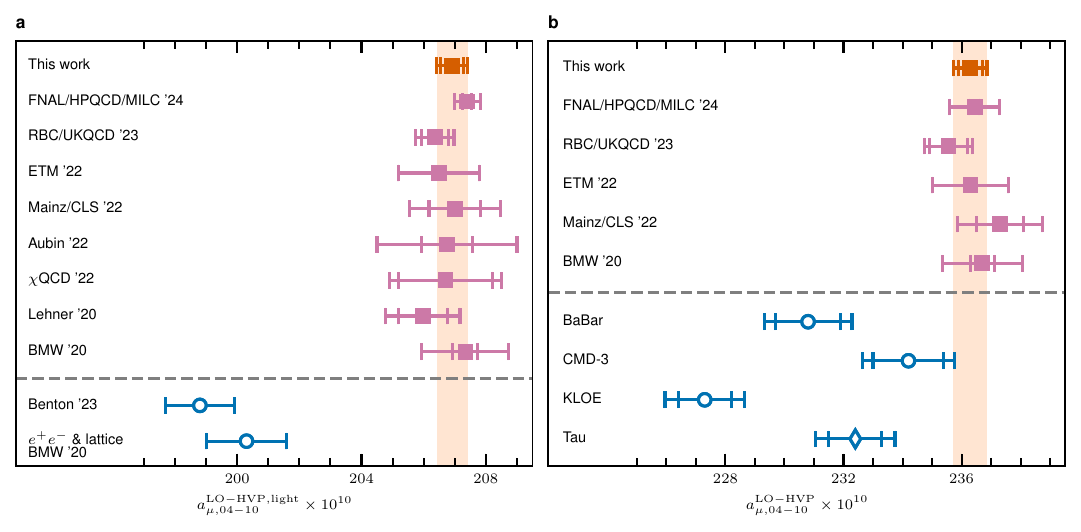}
    \caption
    {
	Comparison of our intermediate-window results with others in the
	literature. {\bf a.} Light contribution to the ID window,
	$a_{\mu,04-10}^{\text{LO-HVP},\mathrm{light}}$.  Our result is the orange
	square, the pink squares correspond to other lattice computations:
	Fermilab Lattice/HPQCD/MILC '24 \cite{MILC:2024ryz}, RBC-UKQCD '23
	\cite{RBC:2023pvn}, ETM '22 \cite{ExtendedTwistedMass:2022jpw}, Mainz
	'22 \cite{Ce:2022kxy}, Aubin et al '22 \cite{Aubin:2022hgm}, $\chi$QCD
	'22 \cite{Wang:2022lkq}, Lehner and Meyer '20 \cite{Lehner:2020crt},
	and our previous result BMW '20 \cite{Borsanyi:2020mff}.  The blue
	circles denote data-driven determinations of Benton et al '23
	\cite{Benton:2023dci} and of BMW '20 \cite{Borsanyi:2020mff}.  These
	two results are based on the KNT'19 data
	compilation~\cite{Keshavarzi:2018mgv,Keshavarzi:2019abf}.  {\bf b.}
	Full ID window, $a_{\mu,04-10}^\text{LO-HVP}$.  Here, in the
	data-driven case, we show results~\cite{Davier:2023fpl} that use the
	measurements of the two-pion spectrum obtained in individual
	electron-positron annihilation experiments and in $\tau$-decays, as
	explained in \refcite{Davier:2023fpl}. The error bars correspond to the
	standard error of the mean (SEM).
    }
    \label{fig:wincomp}
\end{figure}

For the connected contribution of the light $u$ and $d$ quarks to the
intermediate-distance (ID) window we find
$a_{\mu,04-10}^{\text{LO-HVP},\mathrm{light}}=\valuesLightId\times 10^{-10}$,
where the first and second numbers in parentheses refer to the statistical and
systematic uncertainties, respectively, and the number in square brackets is
their quadrature sum, the total uncertainty. As shown in the panel a.\ of
\fig{fig:wincomp} our result agrees with eight other lattice
calculations of this
quantity~\cite{Borsanyi:2020mff,Lehner:2020crt,Wang:2022lkq,Aubin:2022hgm,Ce:2022kxy,ExtendedTwistedMass:2022jpw,FermilabLatticeHPQCD:2023jof,RBC:2023pvn},
including our previous determination, within less than a standard deviation.

On the other hand, our new result for
$a_{\mu,04-10}^{\text{LO-HVP},\mathrm{light}}$ differs from the data-driven one
presented in \refcite{Borsanyi:2020mff} by $4.3\sigma$. This number was
obtained by using the total result $a_{\mu,04-10}^{\text{LO-HVP}}$ from the
data-driven approach and subtracting all but the light-connected contributions
measured in our 2020 lattice simulations. There is another published result
using the data-driven approach by Benton et al ~\cite{Benton:2023dci}. These
two results for the light-connected ID window are the only data-driven ones
published. They are both based on the KNT data
compilation~\cite{Keshavarzi:2018mgv,Keshavarzi:2019abf} that does not include
the more recent CMD-3 measurement nor the ones from $\tau$ decays. Their
difference with our new result, as shown in the left panel of
\fig{fig:wincomp}, reinforces the disagreement between the lattice and
data-driven determinations found in \refcite{Borsanyi:2020mff} which was a
first strong indication that the lattice~\cite{Borsanyi:2020mff} and reference
predictions for $\amulohvp$~\cite{Aoyama:2020ynm} could not both be correct.

Note, that the exact value of $a_{\mu,04-10}^{\text{LO-HVP},\mathrm{light}}$
depends on the scheme used to define the isospin-symmetric limit of QCD. Our
scheme, originally defined in \cite{Borsanyi:2020mff}, is specified in
the Supplementary Information. In \refcite{RBC:2023pvn} it
is shown that the difference between the value of
$a_{\mu,04-10}^{\text{LO-HVP},\mathrm{light}}$ obtained in the RBC-UKQCD scheme
and in our scheme is approximately $0.10(24)\times 10^{-10}$, smaller than even
our present uncertainties. The differences with other schemes used by the other
collaborations are probably on the same level. However we emphasize that this
scheme dependence in no way affects our final result for $\amulohvp$, nor for
the full value of $a_{\mu,04-10}^{\text{LO-HVP}}$ that includes all flavour,
isospin-breaking contributions. Both are unambiguous physical quantities.

In panel b.\ of \fig{fig:wincomp} we display a comparison of our result
for the full ID window contribution,
$a_{\mu,04-10}^{\text{LO-HVP}}=\valuesTotalId$, with the five other lattice
determinations of that quantity. Here the results do not depend on any scheme
choice and agreement is still excellent. Also plotted are the individual
data-driven results~\cite{Davier:2023fpl} obtained using the same data sets as
for computing the central value of the tail. Those results display significant tensions which forbid
an overall comparison between the lattice and data-driven approaches. However,
important progress is being made on understanding the sources of those
differences and we expect that the situation on the data-driven side will be
clarified soon. The differences may be partly due to the treatment of radiative
corrections, as explained in \refscite{BaBar:2023xiy,Davier:2023fpl}. While the
difference of our lattice result with the one obtained using KLOE's
measurement~\cite{KLOE:2008fmq,KLOE:2010qei,KLOE:2012anl,KLOE-2:2017fda} is
$\valuesWindowkloevsthis\sigma$, it reduces to $\valuesWindowbabarvsthis\sigma$
for the BaBar measurement~\cite{BaBar:2009wpw,BaBar:2012bdw} and even to
$\valuesWindowcmdvsthis\sigma$ for the one by CMD-3~\cite{CMD-3:2023alj}.
Compared to the determination obtained via $\tau$
decays~\cite{Davier:2009ag,Davier:2013sfa}, the difference is
$\valuesWindowtauvsthis\sigma$. With an alternative evaluation of the
$\tau$-data~\cite{Masjuan:2023qsp} the difference is even smaller.
These numbers illustrate the known discrepancies between measurements at
energies around the $\rho$-meson peak. Note, that these contributions are
highly suppressed in the tail observable. Nevertheless, we take into account
these discrepancies by performing the analysis of the tail with and without the
most extreme experiments. The associated uncertainty is an order-of-magnitude
below our final error on $\amulohvp$. Details can be found in the
Supplementary Information.

Our result for the light-connected contribution to the SD window,
$a_{\mu,00-04}^{\text{LO-HVP},\mathrm{light}}=\valuesLightSd\times 10^{-10}$,
is in excellent agreement with five other lattice computations of this
quantity~\cite{ExtendedTwistedMass:2022jpw,RBC:2023pvn,Kuberski:2024bcj,MILC:2024ryz,Spiegel:2024dec}.
We also consider the window observable proposed in \refcite{Aubin:2022hgm},
from $1.5$ to $1.9~\mathrm{fm}$, and we obtain
$a_{\mu,15-19}^{\text{LO-HVP},\mathrm{light}}=\valuesLightLdmilc\times
10^{-10}$.  Again we find a good agreement with the two other computations of
this quantity \cite{Aubin:2022hgm,FermilabLatticeHPQCD:2023jof}. A more
detailed comparison of our results for the above windows is provided in the
Supplementary information.

Now, summing the connected-light and disconnected contributions obtained in our
four chosen Euclidean-time intervals, and combining them with all the other
required contributions, including the data-driven tail, we obtain
$\amulohvp=\valuesTotal\times 10^{-10}$, as detailed in the Supplementary
Information.  This result agrees with our earlier 2017 and 2020 determinations,
but reduces uncertainties by a factor of $\valuesImprovedprl$ compared to the
former and of $\valuesImprovednat$ to the latter.  The difference between our
current and the 2020 result
is $\valuesHvpthisvsnatDiff\times 10^{-10}$, with an uncertainty
of $\valuesHvpthisvsnatErru\times 10^{-10}$, indicating that the new
result is $\valuesHvpthisvsnatSigu\sigma$ higher. To obtain that
result, we assume zero correlation among some of the systematics. When
assuming full correlation, the uncertainty becomes
$\valuesHvpthisvsnatErrc\times 10^{-10}$, and in this case, the new
result is $\valuesHvpthisvsnatSigc\sigma$ higher.

\begin{figure}[t]
    \centering
    \includegraphics{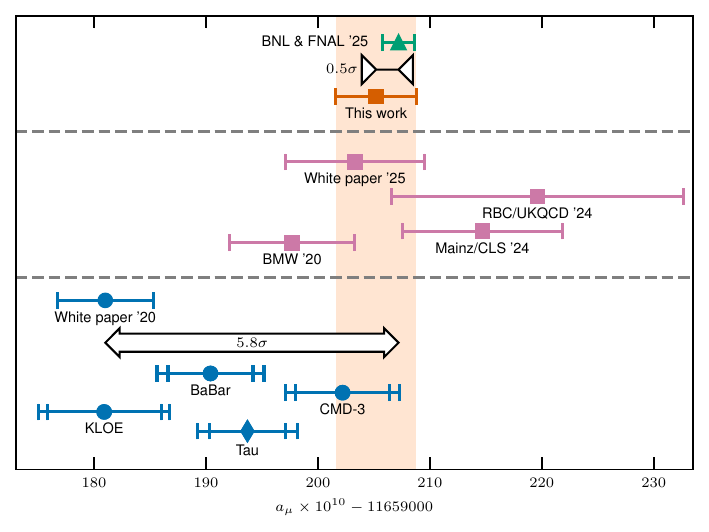}
    \caption
    {
	Comparison of standard-model predictions for the muon anomalous
	magnetic moment with its measured value. \emph{Top panel:}
	world-average measurement of $a_\mu$~\cite{Muong-2:2025xyk} and the
	standard-model prediction of this work. The latter is denoted by the
	orange band and is obtained by adding the value of $\amulohvp$ computed
	here to the results for all of the other contributions summarized in
	\cite{Aliberti:2025beg}. \emph{Middle panel:} predictions using recent
	lattice computations for $\amulohvp$, RBC-UKQCD
	\cite{RBC:2018dos,RBC:2023pvn,RBC:2024fic}, Mainz
	\cite{Djukanovic:2024cmq} and our previous computation
	\cite{Borsanyi:2020mff}. The muon $g-2$ Theory Initiative combination
	from 2025 \cite{Aliberti:2025beg}, which is obtained using lattice
	results for $\amulohvp$, is labelled ``White paper '25''. \emph{Bottom
	panel:} predictions using the data-driven approach for $\amulohvp$
	including the most precise measurements of the two-pion spectrum in
	electron-positron annihilation and $\tau$-decay
	experiments~\cite{Davier:2023fpl}. These correspond to
	BaBar~\cite{BaBar:2009wpw,BaBar:2012bdw},
	KLOE~\cite{KLOE:2008fmq,KLOE:2010qei,KLOE:2012anl,KLOE-2:2017fda} and
	CMD-3~\cite{CMD-3:2023alj} for $e^+e^-$ annihilation and Tau for $\tau$
	decays~\cite{Davier:2009ag,Davier:2013sfa}. The earlier Theory
	Initiative combination from 2020 \cite{Aoyama:2020ynm}, which is
	obtained using the data-driven results, is labelled ``White paper
	'20''.
	Note, all standard-model predictions include non-HVP contributions from
	``White paper '25'', except for ``White paper '20''.
	The error bars are SEM.
    }
    \label{fig:amucomp}
\end{figure}

Adding our determination of $\amulohvp$ to the other standard-model
contributions compiled in \refcite{Aliberti:2025beg} yields
$a_\mu=\valuesAmuthiswork\times 10^{-10}$. In \fig{fig:amucomp} we compare this
result with the world average of the direct measurements of the magnetic moment
of the muon~\cite{Muong-2:2025xyk}. Our prediction differs from that
measurement by $-\valuesAmuexpvsthis\sigma$. Also given are the muon $g-2$
Theory Initiative combinations from the years 2020 \cite{Aoyama:2020ynm} and
2025 \cite{Aliberti:2025beg}, where the $\amulohvp$ contribution was obtained
solely from the data-driven and solely from the lattice approach, respectively.
In addition to these combinations, we also provide individual results in both approaches.
As the figure shows, some of the data-driven results
are in serious tension both with our and the lattice-only estimates.
Our $\amulohvp$ is in good agreement with the latest Theory Initiative
combination and our uncertainty is a factor of $\valuesHvpwhitepapervsthisImp$ smaller.

In the near future we expect
more data for the $\epemtopippim$ cross section
~\cite{Colangelo:2022jxc}. Beyond consolidating our current
understanding~\cite{BaBar:2023xiy,Davier:2023fpl} of the tensions in the
measurements of that cross section, this new data should improve the
data-driven determination of $\amulohvp$.  In addition, the possibility of
directly measuring HVP in the spacelike region is being investigated by the
MUonE collaboration~\cite{Abbiendi:2016xup}.  Finally, combinations of lattice
and data-driven results, beyond the simple one presented here, ought to be
pursued, following e.g.\ the methods put forward in \cite{Davier:2023cyp}.
Investigations along all of those lines are underway.

The precise measurement and standard-model prediction for the muon anomalous
magnetic moment reflect significant scientific progress.  Experimentally,
Fermilab's ``Muon $g-2$'' collaboration has measured $a_\mu$ to $0.127$~ppm
\cite{Muong-2:2025xyk}. Furthermore, there is the ``muon $g-2$/EDM'' experiment
under development at KEK's J-PARC~\cite{Abe:2019thb} to measure this
quantity using a completely new and independent experimental approach.
On the theoretical side, physicists from around the
world have performed complex calculations  (see e.g.\
\refcite{Aliberti:2025beg}), some based on additional precise measurements,
incorporating all aspects of the standard model and many quantum field theory
refinements. It is remarkable that the electromagnetic, electroweak and strong interactions, which require
very different computational tools, can be combined into a single calculation
with such precision. The result for $\amulohvp$ presented here, combined with
other contributions to $a_\mu$ summarized in \refcite{Aliberti:2025beg},
provides a standard-model prediction with a precision of
$\valuesAmuthisworkPpm$~ppm. At such a level of precision, the agreement found
between experiment and theory, to within less than one standard deviation, is a
remarkable success for the standard model and, from a broader perspective, for
renormalized quantum field theory.

%% file: main_statement.tex
\bigskip

{\bf Data availability.} The datasets for the continuum extrapolation tables
are publicly available from {\tt doi:10.5281/zenodo.17880027}. Those for other
figures and tables are available from the corresponding author on request.

{\bf Code availability.} A CPU-code, which was used for configuration
production and measurements, can be obtained from the corresponding author upon
request, subject to possible export control constraints. The Wilson flow
evolution code, which was used to determine $w_0$, can be downloaded from {\tt
arxiv.org/abs/1203.4469}.

{\bf Acknowledgements:} We warmly thank A.~El-Khadra,
M.~Hansen, E.~Jenkins, L.~Jin, Ch.~Lehner, A.~Manohar, H.~Meyer, D.~Nogradi, A.~Patella, A.~Ramos, R.~Sommer,
P.~Stoffer and H.~Wittig for enlightening discussions, and K.L.~Kelley and I.~Frankel for insight
on the scale setting observables.  We are grateful to A.~Keshavarzi, D.~Nomura
and T.~Teubner for sharing their KNT19 compilations of
$e^+e^-\to\text{hadrons}$ cross sections and for correspondence. The authors
gratefully acknowledge the Gauss Centre for Supercomputing (GCS) e.V. ({\tt www.gauss-centre.eu}),
GENCI ({\tt www.genci.fr}, grant 502275), EuroHPC Joint Undertaking
(grants EXT-2023E02-063, EXT-2024E02-109) and Australian National Computational Merit Allocation
Scheme for providing computer time on the GCS supercomputers SuperMUC-NG at
Leibniz Supercomputing Centre in München, HAWK and HUNTER at the High Performance
Computing Center in Stuttgart and JUWELS, JURECA and JUPITER at Forschungszentrum
Jülich, as well as on the GENCI supercomputers Joliot-Curie/Irène Rome at TGCC,
Jean-Zay V100 at IDRIS, Adastra at CINES, on the EuroHPC JU flagship supercomputers Leonardo
at CINECA and LUMI at CSC, Gadi at NCI and Setonix at PSC.
This work received funding from
the French National Research Agency under contract ANR-22-CE31-0011, from the
Excellence Initiative of Aix-Marseille University -- A*Midex, a French
``Investissements d’Avenir" program under grants AMX-18-ACE-005,
AMX-22-RE-AB-052, and from grants NW21-024-A and BMBF-05P21PXFCA. AB, ZF, DG
and AYK are supported by ERC-MUON-101054515. ZF is also partially supported by
NW21-024-B and DOE-0000278885. AP is partly supported by UK STFC Grant
ST/X000494/1 and by long-term Invitational Fellowship L23530 from the Japan
Society for the Promotion of Science. FS is supported by a Ramsay Fellowship
from the University of Adelaide.

{\bf Author contributions:} Code development: SB, KKS, BCT, GW.  Runs and data
management: KKS, GW.  Autocorrelations, crosschecks: FF, SM, AR.  Meson masses:
FF, FS, KKS, GW.  Omega masses: FF, FS, GW.  Pion decay rate: AC, FF, DG, LL,
AP, FS, KKS, BCT. Analysis strategy: ZF, LL, SM, FS, KKS, BCT.  Short distance
window: AYK, LL, SM.  Intermediate and long distance observables: AYK, FS, KKS,
BCT. Strange/charm contributions: AYK, BCT. Finite-size effects from lattice:
KKS, BCT. Finite-size effects from data: AB, DG, LL, AL, MS. Isospin-breaking
contributions: AG, LL, AP, FS, KKS, BCT. Data driven approach: MD, ZF, LL, BM,
ZZ. Acquisition of computer resources: SB, ZF, DG, LL, TL, FS, KKS, BCT. Main
paper text: ZF, LL, FS. Coordination: ZF, LL, FS, KKS.

{\bf Competing interests.} The authors declare no competing interests.

{\bf Additional information.} Supplementary Information is available for
this paper. Correspondence and requests for materials should be addressed to
Zoltan Fodor ({\tt fodor@bodri.elte.hu}). Reprints and permissions information
is available at www.nature.com/reprints.

%% file: si_conf.tex
\section{Configurations and measurements}
\label{se:si_conf}

\subsection{Action and ensembles}
\label{se:action}

We perform simulations using a {\tt 4stout} lattice action, given by the
tree-level Symanzik gauge action~\cite{Luscher:1984xn} and a one-link
staggered fermion action. Where the gauge link appears in the fermion action,
we apply four steps of stout smearing~\cite{Morningstar:2003gk} with a
smearing parameter of $\rho = 0.125$.  Some of the measurements are performed
on supercomputers with multi-GPU compute nodes, here we use the
highly-optimized QUDA \cite{Clark:2009wm,Babich:2011np} and Qlattice \cite{web:qlattice} software
suites.

\begin{figure}
    \centering
    \includegraphics{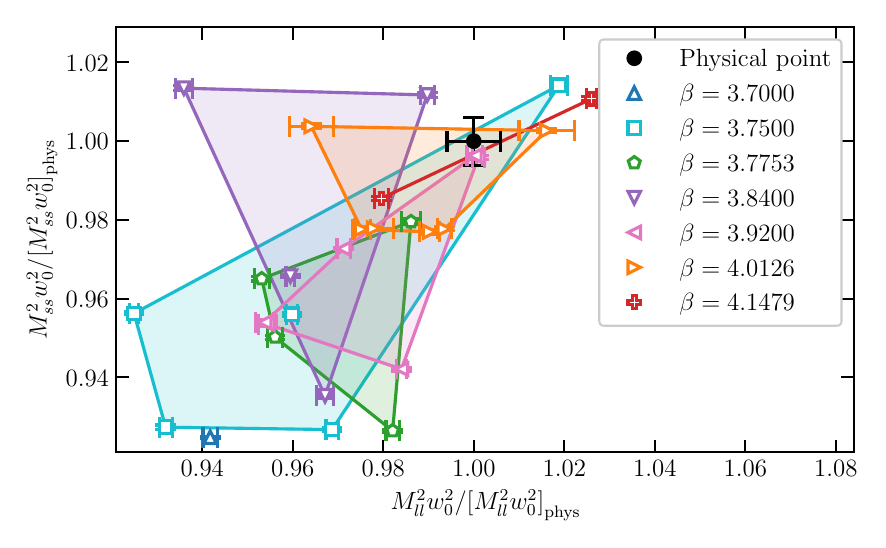}
    \caption
    {
	\label{fig:landscape} Landscape of our ensembles.  The horizontal and
	vertical axes are the squared pseudo-scalar masses, $M_{ll}^2$ and
	$M_{ss}^2$, in units of the $w_0$-scale, both normalized to the central
	value of their respective physical point.  Different colours denote different
	lattice spacings. The black point denotes the (isospin-symmetric) physical
	point, with error bars corresponding to the uncertainties from our
	determination of the $M_{ll},M_{ss}$ and $w_0$ parameters in physical
	units. (Figures in this work were produced with the aid of Matplotlib~\cite{Hunter:2007}.)
    }
\end{figure}

\begin{table}[p]
    \centering
    \begin{tabular}{c|c|c|c|c|c|c}
	$\beta$&$a$ [fm]&$L/a \times T/a $&tag& $am_s$&$m_s/m_l$&\#confs\\
	\hline\hline
	3.7000&0.1315&$48\times 64$&dir00&0.057291&27.899&904\\
	\hline
	3.7500&0.1191&$56\times 96$&dir00&0.049593&28.038&315\\
	&&&dir01&0.049593&26.939&516\\
	&&&dir02&0.051617&29.183&504\\
	&&&dir03&0.051617&28.038&522\\
	&&&dir05&0.055666&28.083&215\\
	\hline
	3.7753&0.1116&$56\times 84$&dir00& 0.047615&27.843&510\\
	&&&dir01&0.048567&28.400&505\\
	&&&dir02&0.046186&26.469&507\\
	&&&dir03&0.049520&27.852&385\\
	\hline
	3.8400&0.0952&$64\times 96$&dir00&0.043194&28.500&510\\
	&&&dir02b&0.043194&30.205&436\\
	&&&dir04&0.040750&28.007&1503\\
	&&&dir05&0.039130&26.893&500\\
	\hline
	3.9200&0.0787&$80\times 128$&dir02& 0.032440&27.679&506\\
	&&&dir04&0.034240&27.502&512\\
	&&&dir01b&0.032000&26.512&1001\\
	&&&dir02b&0.032440&27.679&327\\
	&&&dir03b&0.033286&27.738&1450\\
	&&&dir04b&0.034240&27.502&500\\
	\hline
	4.0126&0.0640&$96\times 144$&phys1&0.026500&27.634&446\\
	&&&phys2&0.026500&27.124&551\\
	&&&phys1b&0.026500&27.634&2248\\
	&&&phys2b&0.026500&27.124&1000\\
	&&&phys3&0.027318&27.263&985\\
	&&&phys4&0.027318&28.695&1750\\
	\hline
	4.1479&0.0483&$128\times 192$&phys1&0.019370&27.630&2792\\
	&&&phys2&0.019951&27.104&2225\\
    \end{tabular}
    \caption
    {
	\label{tab:configs} List of the ensembles used in this work, with gauge
	coupling, lattice spacing, lattice size, ensemble tag, strange-quark
	mass, mass ratio of strange and light quarks and number of
	configurations. The numbers are rounded to the accuracy provided by the number of displayed digits.
    }
\end{table}

We use $2+1+1$ dynamical flavours with equal up and down quark masses
$m_u=m_d=m_l$. The mass
parameters for the light, $m_l$, and the strange quark, $m_s$, are chosen to
scatter around the physical point.  This can be seen in
Figure~\ref{fig:landscape}, where we show the landscape of our ensembles in the plane
of the light and strange quark masses. The charm mass parameter is set by the
ratio $m_c/m_s = 11.85$, which is taken from the $c\bar{c}$ current analysis in
\cite{McNeile:2010ji}. This value is within one per-cent of the most recent
lattice average from FLAG \cite{FlavourLatticeAveragingGroupFLAG:2021npn,EuropeanTwistedMass:2014osg,Chakraborty:2014aca,FermilabLattice:2018est,ExtendedTwistedMass:2021gbo}.  We
use seven different values for the gauge coupling parameter $\beta$.  The
corresponding lattice spacings and the set of ensembles at each
lattice spacing, together with the number of
configurations are listed in Table~\ref{tab:configs}. Most of the ensembles in
this study were also used in our earlier work on the magnetic moment of the
muon \cite{Borsanyi:2020mff}. Since then we added a finer lattice spacing,
corresponding to
$\beta= 4.1479$, with two different ensembles, which bracket the physical point
in both the light and the strange-quark mass. 

To set the scale and the physical point we use the Wilson-flow-based observable
$w_0$ and the masses of $q\bar{q}$ connected pseudo-scalar mesons with $q=u,d,s$.
The physical values of these observables were computed in our earlier
work \cite{Borsanyi:2020mff}, where we used the experimental values of the
hadron masses $\pi^0$, $K^+$, $K^0$ and $\Omega^-$ as inputs.  A more detailed
discussion about our setting of the physical point is given in
Section~\ref{se:si_phys}.

\begin{figure}
    \centering
    \includegraphics{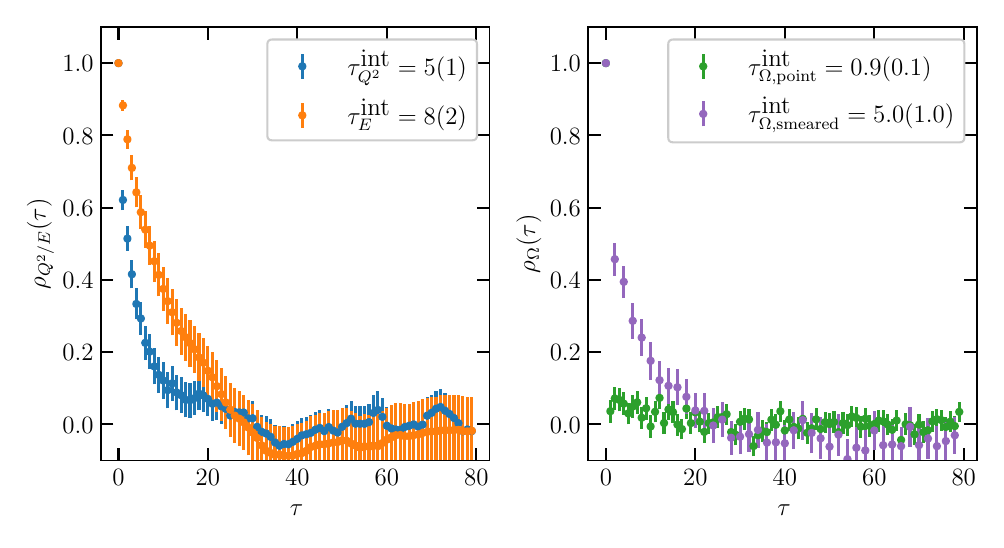}
    \caption
    {
	Left panel: normalized autocorrelation function of the energy density $E$
	and topological charge squared $Q^2$, both computed at a gradient-flow time of
	$w_0^2$ on one of the ensembles at $\beta=4.1479$. Right panel:
	normalized autocorrelation function of the $\Omega_\mathrm{VI}$ hadron
	correlator with point and smeared sources.  The correlator is taken at
	$t= 1.5~\mathrm{fm}$ time separation, which is located in the middle of
	the fitting window. The plots also include the integrated
	autocorrelation times, given in units of configurations.
    }
    \label{fig:ACT}
\end{figure}

The configurations are generated with a Rational Hybrid Monte Carlo algorithm
\cite{Clark:2006fx} including force gradient \cite{Yin:2011np}, multiple
timescales \cite{Sexton:1992nu} and Hasenbusch preconditioning
\cite{Hasenbusch:2001ne}. The configurations are separated by 10 unit length
trajectories. In Figure~\ref{fig:ACT} we show normalized autocorrelation
functions, denoted here by $\rho(\tau)$, for the topological charge squared
\cite{Schaefer:2010hu}, smeared energy density and Omega propagator.  The
charge is computed using the standard clover discretization of the topological
charge density at a gradient-flow time of $w_0^2$, which corresponds to a
smearing radius of about $0.5\,\fm$.  The same flow-time is used for the
smeared energy density. The definition of $\Omega_\mathrm{VI}$ operator is
given in Section~\ref{se:si_omega}.  The autocorrelation function and its error
are computed using the {\tt pyerrors} package \cite{Joswig:2022qfe}. The
integrated autocorrelation times are also given in Figure~\ref{fig:ACT}, the
largest of which corresponds to the smeared energy density.

We use jackknife resampling to calculate the statistical errors. To suppress
the auto-correlation between data from subsequent configurations we introduce a
blocking procedure. It is very convenient to use an equal number of blocks for
all ensembles. In this work we use 48 blocks. With this choice we have
typically 10 configurations or more in a block, which is larger than the
autocorrelation time of any of the quantities we consider even on our finest
ensembles, as shown in Figure~\ref{fig:ACT}. For the blocks we apply the
delete-one principle, resulting in 48 jackknife samples plus the full sample.

Topological properties of QCD were investigated with the {\tt 4stout} action in
Ref.~\cite{Borsanyi:2016ksw}, where the topological susceptibility $\chi$ was
computed for a wide range of lattice spacings.  The continuum extrapolation
of $\chi$ is notoriously difficult, because of the absence of exact zero modes
of the staggered Dirac operator at finite lattice spacing.  The behaviour
towards the continuum can be much improved by rescaling $\chi$ with the square
of the ratio of the Goldstone and taste singlet pion masses.  On our finest
lattice we find $\chi=0.0358(29)$~fm$^4$ for the unimproved and
$\chi=0.0299(24)$~fm$^4$ for the improved susceptibility.  These numbers nicely
fit on the continuum extrapolation curves presented in Figure~S1 of
Ref.~\cite{Borsanyi:2016ksw}. In that work it was found, that the continuum
extrapolated value agrees well with the prediction of chiral perturbation
theory.

\subsection{Taste violation}
\label{se:tavi}

\begin{figure}
    \centering
    \includegraphics{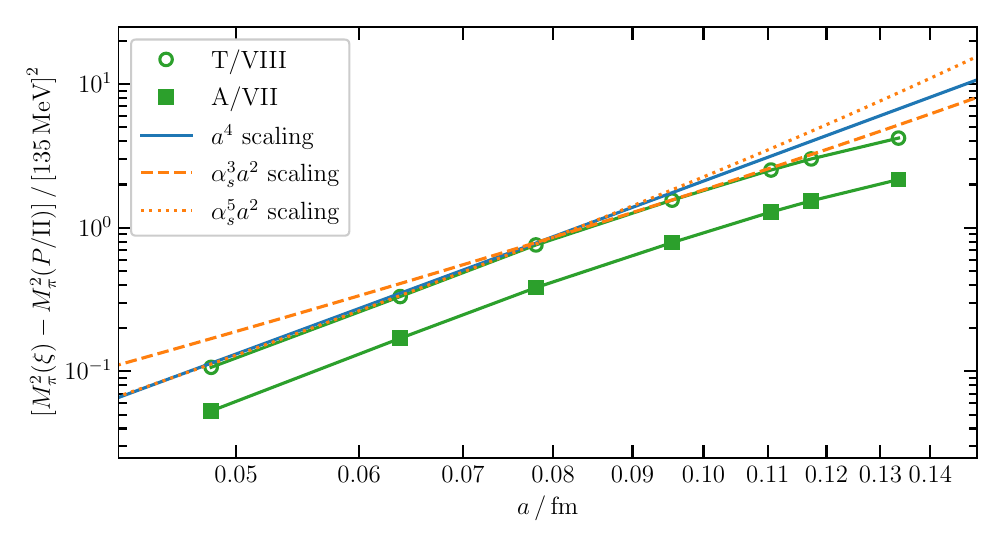}
    \caption
    {
	\label{fig:taste_violation} Taste violation as a function of lattice
	spacing, for axial-vector $A$ and tensor $T$ tastes. The Roman-number
	labelling of the tastes is taken from \cite{Ishizuka:1993mt}, e.g.\ the
	pseudo-scalar taste $P$ corresponds to $\text{II}$.
    }
\end{figure}
An important aspect of these computations is the lattice artefact related to
the taste symmetry violation of staggered fermions. This makes pseudo-scalar
mesons heavier than in the continuum, depending on their taste quantum number.
In particular the masses of pions and eta mesons built from light quark flavours are
given as:
\begin{equation}
    \label{eq:pieta}
    M^2_{\pi}(\xi)= M^2_{ll} + \Delta_{KS}(\xi)
    \quad\text{and}\quad
    M^2_{\eta}(\xi)= M^2_{ll} + \Delta_{KS}(\xi) + \tfrac{1}{2}\delta_{KS}(\xi)\ ,
\end{equation}
where $M^2_{ll}$ is the squared mass of the usual pseudo-Goldstone pion.  The
pions have only connected contractions, whereas the etas also contain
disconnected contributions. The $\xi$ stands for one of the sixteen meson
tastes of the taste $SU(4)$ group. In practical lattice simulations these group
into multiplets of the $SO(4)$ group to a very good approximation, so the meson
tastes are conventionally labelled by $P,A,T,V,I$. The $P$ taste corresponds to
the pseudo-Goldstone pion, for which $\Delta_{KS}(P)$ and $\delta_{KS}(P)$ are
zero. 

We compute the taste-symmetry breaking terms $\Delta_{KS}(\xi)$ by comparing
the masses of different meson tastes on our ensembles. For this purpose we use
an average ``flavour-symmetric'' valence quark mass of $\tfrac{1}{3}(2m_l +
m_s)$.  In Figure~\ref{fig:taste_violation} we show this comparison for the $A$
and $T$ tastes as a function of the lattice spacing. We observe a decrease with
approximately the fourth power of the lattice spacing towards our finer
lattices. This is much faster than the $\alpha_s a^2$ expected from naive
scaling, where $\alpha_s(a)$ is the strong coupling constant at the lattice cutoff
scale. The falloff is actually consistent with an $\alpha_s^n(a) a^2$ type
behaviour with $n=3$ on the coarser and with $n=5$ towards the finer lattices.

The taste-symmetry breaking terms $\delta_{KS}(\xi)$ are related to the eta
mesons.  $\delta_{KS}(T)=0$ in the $SO(4)$ approximation, $\delta_{KS}(I)$
arises from the chiral anomaly, and $\delta_{KS}(A)$ and $\delta_{KS}(V)$ are
called hairpin parameters. These have never been computed directly, since the
disconnected contributions are computationally demanding. Here we assume that
these artefacts decrease with the same rate as $\Delta_{KS}(\xi)$; the same
assumption was made in Ref.~\cite{FermilabLattice:2018zqv}. 

%% file: si_omega.tex
\section{Scale setting with the omega baryon mass}
\label{se:si_omega}

\subsection{Omega propagator measurements}

To extract the mass of the positive-parity, ground-state $\Omega$ baryon, we
use a similar strategy as in our previous work \cite{Borsanyi:2020mff}. In
particular we consider three different operators~\cite{Golterman:1984dn,
Ishizuka:1993mt, Bailey:2006zn}:
\begin{eqnarray}
    \begin{aligned}
    \label{eq:omegaop}
	\Omega_{\rm{VI}}(t) &= \sum_{x_k even} \epsilon_{abc} \left[ S_1 \chi_a S_{12} \chi_b S_{13} \chi_c - S_2 \chi_a S_{21} \chi_b S_{23} \chi_c + S_3 \chi_a S_{31} \chi_b S_{32} \chi_c \right](x)\ , \\
	\Omega_{\rm{XI}}(t) &= \sum_{x_k even} \epsilon_{abc} \left[ S_1 \chi_a S_{2} \chi_b S_{3} \chi_c \right](x)\ ,\\
	\Omega_{\rm{Ba}}(t) &= [2\delta_{\alpha 1}\delta_{\beta 2} \delta_{\gamma 3} -  \delta_{\alpha 3}\delta_{\beta 1} \delta_{\gamma 2}- \delta_{\alpha 2}\delta_{\beta 3} \delta_{\gamma 1} + (\cdots \beta \leftrightarrow \gamma \cdots)] \\
	&\sum_{x_k even} \epsilon_{abc} \left[ S_1 \chi_{a \alpha} S_{12} \chi_{b \beta} S_{13} \chi_{c \gamma} - S_2 \chi_{a \alpha} S_{21} \chi_{b \beta} S_{23} \chi_{c \gamma} +  S_3 \chi_{a \alpha} S_{31} \chi_{b \beta} S_{32} \chi_{c \gamma} \right](x)\ .
    \end{aligned}
\end{eqnarray}
Here, $\chi_a(x)$ is the strange-quark field with colour index $a$ and $\chi_{a
\alpha}(x)$ with $\alpha = 1, 2, 3$ the additional ``flavour'' index introduced
by Bailey~\cite{Bailey:2006zn}.  The operator $S_\mu$ performs a symmetric,
gauge-covariant shift in direction $\mu$, while $S_{\mu \nu} \equiv S_\mu
S_\nu$.  $\Omega_{\rm{VI}}$ and $\Omega_{\rm{XI}}$ couple to two different
tastes of the $\Omega$ baryon and $\Omega_{\rm{Ba}}$ only couples to a single
taste. In the continuum limit the masses of these states become degenerate. In
our analyses we include all three in order to assign a systematic to the
choice of the operator.

Beside point sources we also use smeared ones to construct the propagator.
Smearing changes the excited state contamination and including smeared
propagators in the analysis makes a more reliable extraction of the ground
state possible. These propagators are constructed by applying spatial Wuppertal
smearing~\cite{Gusken:1989ad} on a source $\psi$ vector
\begin{eqnarray}
    \label{eq:wptl}
    [\hat{W} \psi]_x = (1 - \sigma) \psi_x +
    \frac{\sigma}{6} \sum_{\mu = 1, 2, 3}
    \left( U_{\mu, x}^\mathrm{3d} U_{\mu, x + \mu}^\mathrm{3d} \psi_{x + 2 \mu} +
    U_{\mu, x - \mu}^{\mathrm{3d},\dagger} U_{\mu, x - 2 \mu}^{\mathrm{3d},\dagger} \psi_{x- 2 \mu} \right)
\end{eqnarray}
with smearing parameter $\sigma=0.5$. The spatial derivatives involve two hops
in order to preserve the staggered symmetries of the operators. They also
include a smeared-gauge field $U^\mathrm{3d}$ obtained by applying
$N_\mathrm{3d}$ spatial stout smearing steps with smearing parameters $0.125$.
The smeared source is then obtained by applying the Wuppertal-smearing
procedure $N_\mathrm{Wptl}$ times on a point source.  The number of smearing
steps, see Table~\ref{tab:omega_parameters}, depends on the $\beta$ in such a
way to keep the effective smearing radius of the procedure approximately
constant in physical units.

\begin{table}
    \centering
    \begin{tabular}{  c || c | c || c | c | c | c | c || c  }
	$\beta$  & $N_{\rm{Wptl}}$ & $N_{\rm{3d}}$ & $t_p$ & $t_a$ & $t_b$ & range \#1 & range \#2 & \# pt, sm sources \\
	\hline
	3.7000 &  24 &  32 & 1 & 4 & 7  &  7$\dots$15 & 8$\dots$15  &   28928, 229376  \\
	\hline
	3.7500 &  30 &  40 & 1 & 4 & 7  &  8$\dots$18 & 9$\dots$18  &   66208, 530176  \\
	\hline
	3.7553 &  34 &  46 & 1 & 4 & 7  &  9$\dots$19 & 10$\dots$19 &   61024, 488192  \\
	\hline
	3.8400 &  46 &  62 & 2 & 4 & 9  & 10$\dots$20 & 11$\dots$20 &  125440, 2807552 \\
	\hline
	3.9200 &  67 &  90 & 2 & 6 & 9  & 12$\dots$25 & 13$\dots$25 &  137472, 3038720 \\
	\hline
	4.0126 & 101 & 135 & 3 & 6 & 9  & 15$\dots$30 & 16$\dots$30 &  223360, 4235520 \\
	\hline
	4.1479 & 178 & 238 & 5 & 6 & 11 & 19$\dots$40 & 21$\dots$40  & 160544, 2068736 \\
	\hline
    \end{tabular}
    \caption
    {
	Parameters of our procedure to obtain $\Omega$ mass: number of
	Wuppertal and stout smearing steps; the parameters $t_p$, $t_a$ and
	$t_b$ of the GEVP procedure and two different fit ranges, given by
	start and end points $t_\mathrm{min}$ and $t_\mathrm{max}$.
	Definitions are given in the text. In the last column the total number of
	measurements with point and smeared sources are given.
    }
    \label{tab:omega_parameters}
\end{table}
To enhance the signal, we calculate the $\Omega$ propagators using $256$ or
$512$ smeared and $32$ point source fields per gauge configuration; the total
number of measurements for each $\beta$ are given in the last column of
Table~\ref{tab:omega_parameters}. For each source field we select a random time
slice, which, in turn, is populated with eight independent $\mathbb{Z}_3$
random point sources at $(0, 0, 0)$, $(L/2, 0, 0)$, $\cdots$ and $(L/2, L/2,
L/2)$. The idea with eight sources was originally proposed in \cite{Billoire:1985yn}.

\subsection{Omega propagator fits}

Our mass extraction procedure combines the Generalized Eigenvalue Problem
(GEVP) approach, see Ref.~\cite{Blossier:2009kd} and references therein, with
the Generalized Pencil-of-Function approach proposed in
Ref.~\cite{Aubin:2010jc}. We first apply a folding transformation to the
original hadron propagator $H_t$:
\begin{eqnarray}
    H_t \rightarrow
    \left\{
	\begin{matrix*}[l]
	    \; \frac{1}{2} \left[ H_t + (-1)^{t+1} H_{T - t} \right] & 0<t<\frac{T}{2} \\
	    \; H_t & t=0 \; {\rm{or}} \; t=\frac{T}{2}
	\end{matrix*}
	\right.\ ,
\end{eqnarray}
where the staggered phase factor $(-1)^{t+1}$ ensures the parity is consistent
between the forward and backward-propagating states in the folding. Then for
each time slice $t$ we construct the following $6\times6$ matrix:
\renewcommand{\arraystretch}{1.5}
\begin{equation}
    \mathbf{H}(t) =
    \left(
    \begin{array}{cc|cccc}
	H_{t+2 t_p+0}^{pp} & H_{t + 2 t_p+1}^{pp} & H_{t+t_p+0}^{ps} & H_{t+t_p+1}^{ps} & H_{t+t_p+2}^{ps} & H_{t+t_p+3}^{ps}\\
	H_{t+2 t_p+1}^{pp} & H_{t + 2 t_p+2}^{pp} & H_{t+t_p+1}^{ps} & H_{t+t_p+2}^{ps} & H_{t+t_p+3}^{ps} & H_{t+t_p+4}^{ps}\\
	\hline
	H_{t+t_p+0}^{sp}   & H_{t+ t_p+1}^{sp}    & H_{t+0}^{ss}     & H_{t+1}^{ss}     & H_{t+2}^{ss}     & H_{t+3}^{ss}    \\
	H_{t+t_p+1}^{sp}   & H_{t+ t_p+2}^{sp}    & H_{t+1}^{ss}     & H_{t+2}^{ss}     & H_{t+3}^{ss}     & H_{t+4}^{ss}    \\
	H_{t+t_p+2}^{sp}   & H_{t+ t_p+3}^{sp}    & H_{t+2}^{ss}     & H_{t+3}^{ss}     & H_{t+4}^{ss}     & H_{t+5}^{ss}    \\
	H_{t+t_p+3}^{sp}   & H_{t+ t_p+4}^{sp}    & H_{t+3}^{ss}     & H_{t+4}^{ss}     & H_{t+5}^{ss}     & H_{t+6}^{ss}
    \end{array}
    \right)\ ,
\end{equation}
where $\mathrm{type}$ in $H_t^\mathrm{type}$ denotes the different source-sink
combinations, such as point-point, smear-point, point-smear and smear-smear,
labelled with $pp$, $sp$, $ps$ and $ss$.  We introduce an additional time shift
$t_p$ in case of the point source operator to suppress its excited states. The
time shifts between different rows and columns are needed to fully resolve the
negative parity states, which have a negative amplitude. 

\renewcommand{\arraystretch}{1.0}
\begin{figure}[t]
    \centering
    \includegraphics{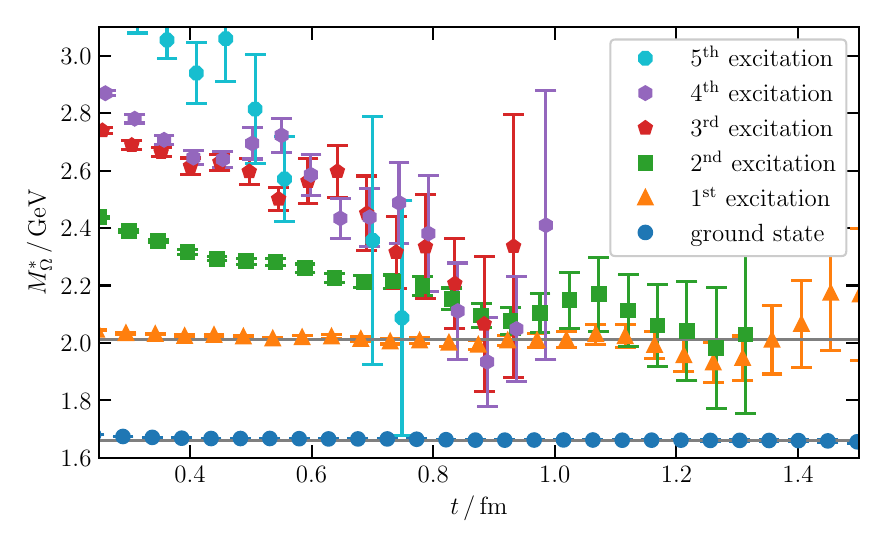}
    \caption
    {
	$\Omega$ baryon effective masses extracted from the GEVP for five
	excited states and the ground state on
	an ensemble at $\beta=4.1479$. The blue band through the ground-state points is
	obtained from a single exponential fit to the ground state propagator.
	The grey horizontal line through the first excited state corresponds to the mass of
	the recently observed excited $\Omega$ baryon by the Belle
	experiment~\cite{Belle:2018mqs}. To convert the numbers into physical
	units we used $a=0.0483\,\fm$.
    }
    \label{fig:omega_gevp_spectrum} 
\end{figure}

\renewcommand{\arraystretch}{1.0}
\begin{figure}[t]
    \centering
    \includegraphics{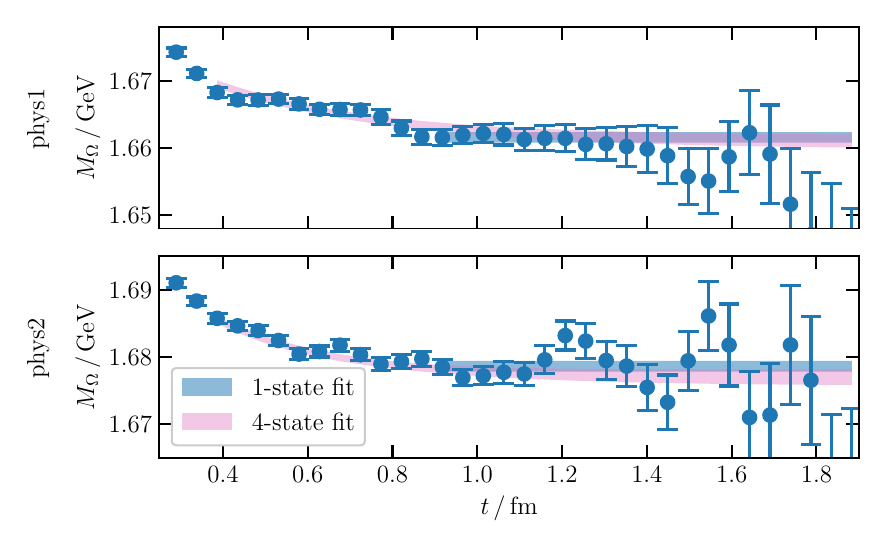}
    \caption
    {
	$\Omega$ baryon ground state effective masses extracted from the GEVP
	for the two ensembles at our finest lattice spacing, $\beta=4.1479$.
	The blue horizontal band shows the ground state mass obtained from a
	single exponential fit; the pink, curved band corresponds to the
	effective mass of a fit with four exponentials.
    }
    \label{fig:omega_gevp_fits} 
\end{figure}

For a given $t_a$ and $t_b$, let $\lambda(t_a , t_b)$ be an eigenvalue and
$v(t_a,t_b)$ an eigenvector of the following GEVP:
\begin{eqnarray}
    \mathbf{H}(t_a) v(t_a, t_b) = \lambda(t_a, t_b) \mathbf{H}(t_b) v(t_a, t_b)\ .
\end{eqnarray}
The ground state corresponds to the largest eigenvalue. The propagator
corresponding to the eigenvector $v$ is given by vector-matrix-vector product:
\begin{equation}
    P(t;t_a,t_b)= v^\dagger(t_a, t_b) \mathbf{H}(t) v(t_a, t_b)\ .
\end{equation}
This correlation function can be fitted to an exponential function $A\exp(-Mt)$
with amplitude $A$ and mass $M$. Note, that backward propagating states have
negligible contribution for the time-slices that we work with.  From $P(t)$ one
can also construct an effective mass in the standard way. The tuneable
parameters of the procedure are $t_a$ and $t_b$ for specifying the GEVP, as
well as the fit range $[t_\mathrm{min},t_\mathrm{max}]$ for the exponential fitting.

In Figure~\ref{fig:omega_gevp_spectrum} we show the effective masses obtained with the
GEVP procedure for an ensemble at $\beta= 4.1479$. This plot
shows the excited states. The phenomenological interpretation of these is
non-trivial, since they are hadron resonances decaying into scattering states.
As an illustration we show the excited, negative-parity $\Omega$ baryon from
the Belle experiment~\cite{Belle:2018mqs} with a mass of
$2012.4(7)(6)~\mathrm{MeV}$. As shown in Figure~\ref{fig:omega_gevp_fits},
the ground state is well resolved with a per-mill level of precision.

We perform the exponential fit taking into account the correlations
between different timeslices in the $P(t)$ propagator. For that we construct
the covariance matrix with $200$ jackknife samples, instead of our usual choice
of $48$ samples, to stabilize its inversion. The
smaller block-size should not be problematic with the $\Omega$, given its
autocorrelation time provided in Section~\ref{se:si_conf}. When inverting the
covariance matrix we apply a singular value decomposition and regulate the
smallest eigenvalues of the correlation matrix as described in Appendix A.1 of
Ref.~\cite{Bouchard:2014ypa}.  In an uncorrelated fit or in a correlated fit
with regulated eigenvalues the statistical interpretation of the fit quality
becomes questionable. Recently an improved estimator for the fit quality was
given in \cite{Bruno:2022mfy}, which we call $Q$-value here.  It can be
employed both in the uncorrelated and correlated cases.

\begin{figure}[t]
    \centering
    \includegraphics{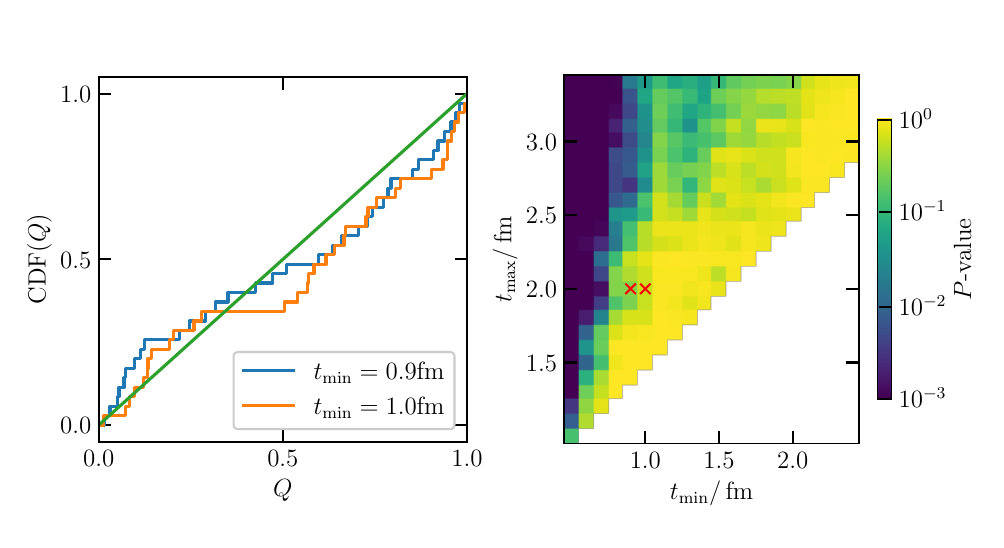}
    \caption
    {
	Left panel: cumulative distribution function (CDF) of $Q$-values
	\cite{Bruno:2022mfy} over all of our ensembles, where we show results
	for our two preferred fit-ranges for the $\Omega_\mathrm{VI}$
	operator. The first fit-range, shown with blue, is
	$[0.9,2.0]\,\fm$, whose Kolmogorov-Smirnov
	significance level is $P=0.27$, i.e.~the probability we get the observed
	CDF or anything worse is 27\%, assuming a uniform distribution. We also
	use a second fit-range $[1.0,2.0]\,\fm$, shown
	with orange, with $P=0.57$.  The green line represents the CDF of a
	hypothetical uniform distribution. Right panel: we show the
	Kolmogorov-Smirnov significance level for several different fit-ranges,
	$[t_\mathrm{min},t_\mathrm{max}]$, the brighter/darker colours stand for
	better/worse significance. Our final choices for the above two
	fit-ranges are shown by the red crosses.
    }
    \label{fig:omega_heatmap} 
\end{figure}
The time range in the exponential fit is going to be chosen by the $Q$-value of
the fit. For each operator in Equation \eqref{eq:omegaop} we compute the
$Q$-values on all of our ensembles for several different fit ranges
$[t_\mathrm{min},t_\mathrm{max}]$. In the left panel of
Figure~\ref{fig:omega_heatmap} we show the cumulative distribution function of
the $Q$-values over all of our ensembles for some selected fit-ranges and for
the $\Omega_\mathrm{VI}$ operator. For a given fit range the $Q$-values will
follow a uniform distribution if the ground state correlator $P(t)$ can be
described by a single exponential $A\exp(-Mt)$. To decide if the observed CDF
is uniform, we use a one-sided Kolmogorov-Smirnov test with the uniform
distribution of the $Q$-values as null-hypothesis. We vary the fit-ranges and
compute the Kolmogorov-Smirnov significance. These are shown in a heat-map
format in the right panel of Figure~\ref{fig:omega_heatmap}. We choose to work
with the fit-range $[0.9\,\rm{fm},2.0\,\rm{fm}]$.  To estimate the systematic
error related to the fit-range we also use one with a later start,
$[1.0\,\rm{fm},2.0\,\rm{fm}]$. Both ranges are plotted on the heatmap figure,
with red crosses.

\begin{figure}[p]
    \centering
    \includegraphics{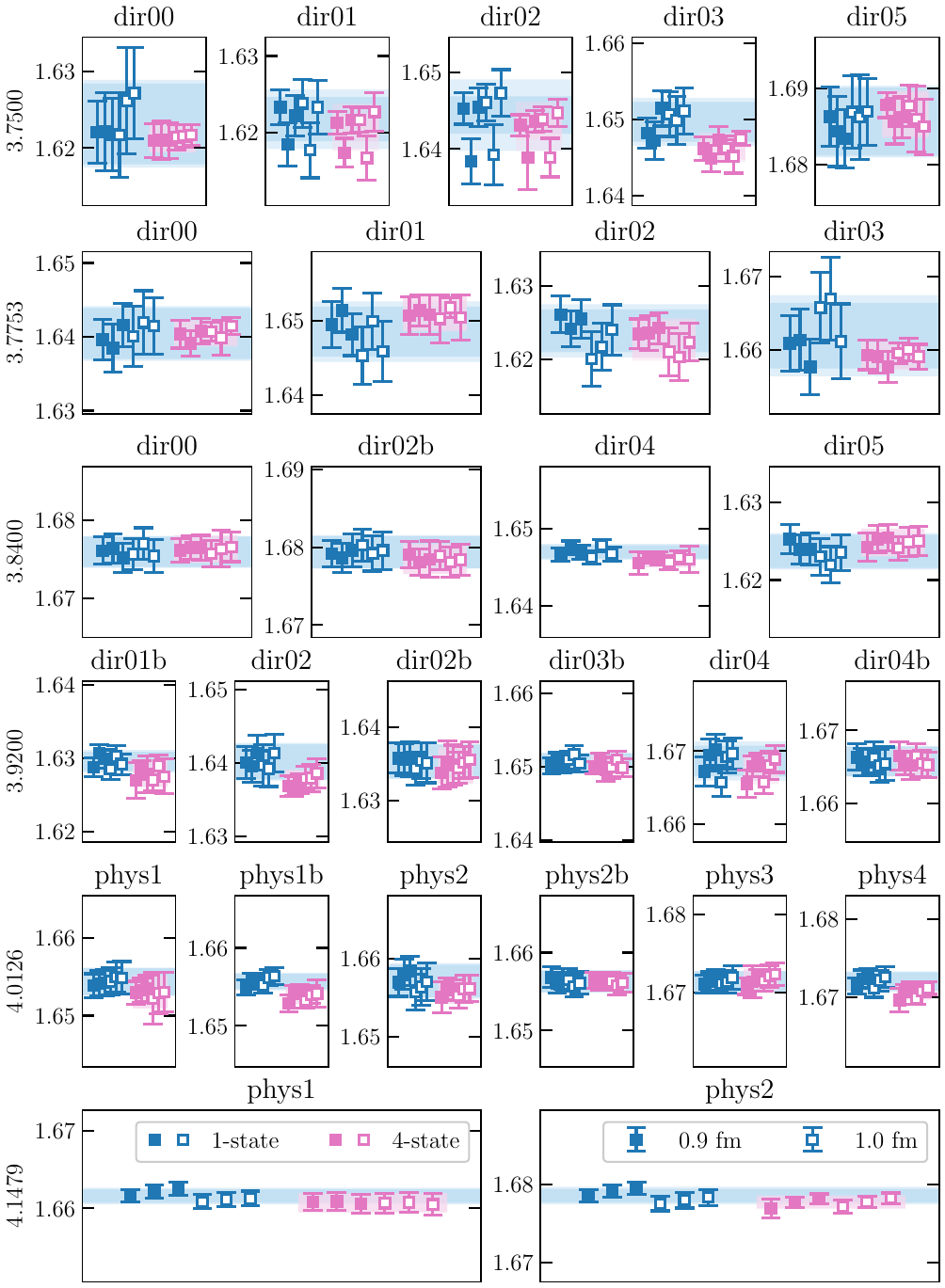}
    \caption
    {
	Results for the ground state $\Omega$ baryon mass on different
	ensembles, in units of GeV, using the lattice spacings from
	Table~\ref{tab:configs}. Each ensemble has six data points with the blue
	colour, which correspond to correlated single exponential fits to the
	$\Omega_{\rm{VI}}$, $\Omega_{\rm{XI}}$ and $\Omega_{\rm{Ba}}$
	propagators with fit starting point set at $0.9$~fm, followed by the
	same fits starting at $1.0$~fm.  The earlier/later fit ranges are
	displayed with filled/open symbols. In addition we have another six
	data points with pink colour corresponding to fits with four
	exponentials. The blue/pink bands correspond to averages of the
	blue/pink points.
    }
    \label{fig:omega_total}
\end{figure}

To investigate the excited state contamination of the ground state masses, we
perform correlated fits to the ground state propagator with four exponentials
instead of one. We introduce $100$~MeV wide priors on the excited states, for
the central values we use masses of $\Omega$ resonances taken from the Particle
Data Book.  This choice of priors assumes that the dominant contaminations come
from states near the resonance energies. This is motivated by the local nature
of our Omega operators, but ignores possible contributions from the
non-resonant scattering states discussed at the end of this section.  Results
of these fits are shown in the effective mass plot of
Figure~\ref{fig:omega_gevp_fits}, in case of our finest two ensembles. The
ground state mass is in good agreement between the single-state and four-state
fits.

The extracted ground state energies for all of our ensembles are given in
Figure~\ref{fig:omega_total}.  We reach better than $0.1$\% precision on our
finest lattice. We see an increase of the precision towards finer lattice
spacings, which can be partly explained by the increase in the number of
measurements with $\beta$. To confirm the robustness of our procedure, we
performed fits also using the previously discussed four-state fits. The results
of all these fits are shown in Figure~\ref{fig:omega_total}, and we observe a
good agreement across the different determination procedures.

The determination of the ground state baryon mass, as detailed in this Section,
has a caveat\footnote{We thank our referee for bringing this problem to
our attention.}. It does not account for the presence of non-resonant, scattering
states in the excited state spectrum, like $(\Xi,K)$, $(\Omega,\pi,\pi)$ and
other multi-hadron combinations. Though our single-hadron operators are
expected to couple weakly to these states, we cannot guarantee that the
scattering state contamination remains within our quoted precision. A similar
problem occurs in the case of nucleon propagators, where the $(\Delta,\pi)$ states
distort the ground state determination. This problem has been investigated in
chiral perturbation theory \cite{Bar:2015zwa,Tiburzi:2015tta}.

\subsection{Determination of $w_0$ using the Omega mass}

\begin{table}[p]
    \centering
    \includegraphics{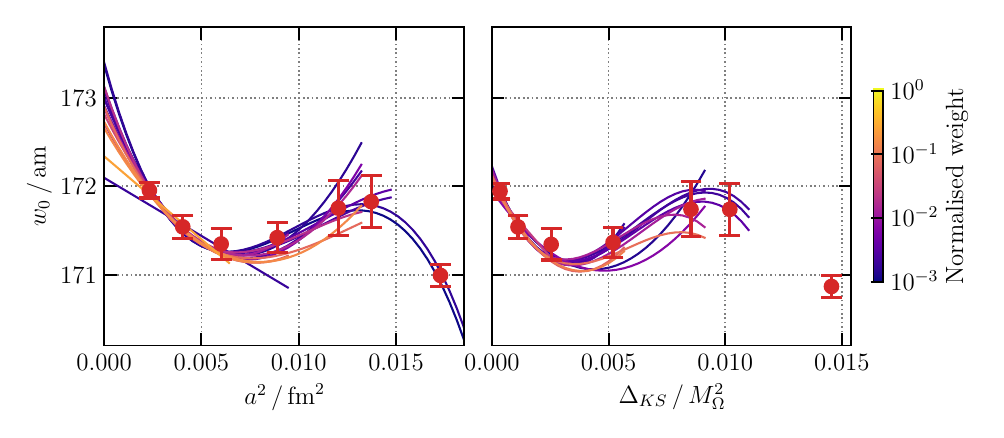}\\
    \includegraphics{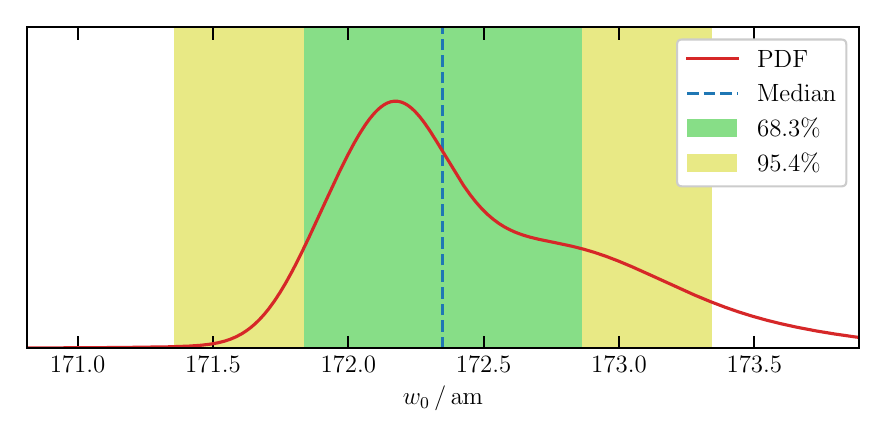}\\
    \input{figures/si_win/legacy/w0}
    \caption
    {
	Gradient-flow scale $w_0$ in attometers using Omega baryon mass
	as input. Continuum extrapolations as a function of $a^2$ and of
	$\Delta_{KS}$ are shown in the first row. The highest-weighted
	fit is a function of $a^2$, and at a nominal lattice spacing of
	$0.1\,\fm$, the $a^2$ term contributes $4.7\%$ of the total
	value, the $a^4$ term $5.5\%$, and the $a^3$ term $1.9\%$.
	The rest of the table shows the probability distribution
	function (PDF), and our error budget.  The plot conventions,
	including small offsets of the points, are described in more
	detail in the first part of Section~\ref{se:si_win:table}.
    }
    \label{ta:res_w0om}
\end{table}

In order to compute the physical value of the gradient-flow scale $w_0$ in full
QCD plus QED, we use the analysis strategy from our 2020 work. There we
parameterize the quark mass and electromagnetic coupling dependence of the $w_0
M_\Omega$ product as
\begin{equation}
    \label{eq:w0om}
    w_0 M_\Omega= A + B\ M_\Omega^{-2} M_{ud}^2 + C\ M_\Omega^{-2}(M_{us}^2 + M_{ds}^2 - M_{ud}^2)/2 + E\ e_v^2 + F\ e_v e_s + G\ e_s^2\ ,
\end{equation}
where $e_v$ and $e_s$ are the electromagnetic couplings of the valence and sea
quarks. We called these fits ``Type-I`` in our work. The procedure determines
the parameters $A,\dots,F$ by fitting measurements over several ensembles.  The
physical value of $[w_0 M_\Omega]_\mathrm{qcd+qed}$ is obtained by
substituting the physical values of the $M_{ud}$, $M_{us}$, $M_{ds}$ and
$M_\Omega$ in the above formula. Then dividing by the value $M_{\Omega} =
1672.45\,\text{MeV}$ \cite{ParticleDataGroup:2024cfk} yields the physical
value of the gradient flow scale $[w_0]_\mathrm{qcd+qed}$.

The parameters $A,B$ and $C$ can be determined from isospin-symmetric
measurements. This has already been done in our 2020 work, here we redo this
determination, where we add the $0.048$~fm lattice spacing into the analysis
and also increase the statistics on coarser ensembles. The current analysis
consists of $8640$ fits, differing in the fit ranges of the pseudoscalars and
the omega baryon, the cuts in the lattice spacing and the functional form of
the fit function. We also vary the expansion variable between a naive $a^2$ and
a $\Delta_{KS}(a)$ type dependence in the lattice spacing.  As opposed to our
earlier publication in 2020, where we used only quadratic polynomials of the
expansion variable, this time we also include cubic ones.  We have $8640$ fits
altogether, 83\% of which have $P$-value of at least $0.1$; Sample continuum
extrapolations, probability distribution function and error budget are shown in
Table~\ref{ta:res_w0om}. The $w_0$ values in the table are obtained by
dividing the $w_0M_\Omega$ product by the experimental value of the $\Omega$
baryon mass. The distribution has a narrower Gaussian peak at a lower value
corresponding to $\Delta_{KS}$ fits and a broader at a higher value
corresponding to $a^2$ fits.

From the above fits we obtain the value of $w_0$ in the isospin-symmetric
point, which is defined by setting the hadron masses to the physical values and
$E,F$ and $G$ to zero in Equation \eqref{eq:w0om}:
\begin{equation}
    [w_0]_\mathrm{qcd,typeI}= \valuesWnullFromomQcd~\mathrm{fm}\ ,
\end{equation}
where the first and second numbers in parentheses refer to the
statistical and systematic uncertainties, respectively, and the number
in square brackets is their quadrature sum, the total uncertainty.
Note, that the definition of the
QCD point is ambiguous, we therefore have indicated the choice of the QCD point
in the subscript.  For the $E,F$ and $G$ parameters we need to compute the
electromagnetic derivatives of the various hadron masses. This has been done in
our 2020 work and the fits yield the QED part:
\begin{equation}
    \label{eq:w0fromomqed}
    [w_0]_\mathrm{qed,typeI}= \valuesWnullFromomQed~\mathrm{fm}\ .
\end{equation}
Finally, for the sum of the QCD and QED contribution we get
\begin{equation}
    \label{eq:w0fromom}
    [w_0]_\mathrm{qcd+qed}= \valuesWnullFromom~\mathrm{fm}\ ,
\end{equation}
which is a scheme independent quantity and the main result of this section. We
will compare this value to other determinations in later sections of this
paper.

%% file: figures/si_win/legacy/w0.tex
\begin{tabular}{l|r|r}
    Median & \multicolumn{2}{c}{172.35 am}\\
    \hline
    Total error & 0.51 am & 0.30 \% \\
    Statistical error &  0.22 am & 0.13 \% \\
    Systematic error &  0.46 am & 0.27 \% \\
    \hline
    Pseudoscalar fits & 0.01 am & $<$ 0.01 \% \\
    Omega baryon fits & 0.24 am & 0.14 \% \\
    Physical value of $M_\Omega$ & 0.06 am & 0.03 \% \\
    Lattice spacing cuts & 0.09 am & 0.05 \% \\
    Order of fit polynomials & 0.17 am & 0.10 \% \\
    Continuum parameter ($\Delta_{KS}$ or $a^2$) & 0.30 am & 0.17 \% \\
\end{tabular}

%% file: si_fpi.tex
\section{Scale setting with the pion decay rate}
\label{se:si_fpi}

\subsection{Pion propagator measurements and fits}

\begin{figure}
    \centering
    \includegraphics{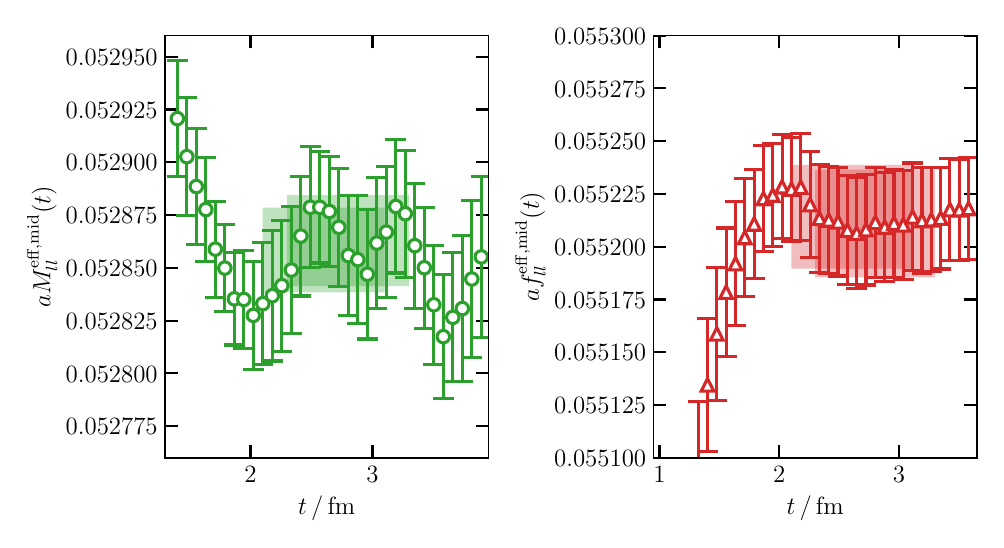}
    \caption
    {
	Midpoint effective pion mass and decay constant from one of the
	$\beta = 3.9200$ ensembles. The semi-transparent shaded bands
	show the plateaus, whose x-extents correspond to the two
	plateau regions ($1.9$~fm -- $2.9$~fm for the first, and
	$2.1$~fm -- $3.1$~fm for the second) and y-extents, to the
	uncertainties on the associated fit results.
    }
    \label{fig:plateau}
\end{figure}

On each ensemble we obtain the pion mass and decay constant from
the zero-momentum two-point function $G(t)$ of the flavour-non-singlet, pseudo-scalar density with pseudo-Goldstone staggered
taste. We use random wall sources and point sinks for the measurements.
On our finest lattice the integrated autocorrelation time is at most 10
trajectories for these propagators.

For sufficiently large \(t \ll T\), excited states are exponentially suppressed
and \(G(t)\) approaches the single-state form
\begin{equation}
    G(t) \simeq \frac{f_{ll}^2 M_{ll}^3}{4 m_l}
    \frac{\cosh\!\left(M_{ll}\left(t-\frac{T}{2}\right)\right)}{e^{M_{ll}\frac{T}{2}}\left(1 - e^{-M_{ll} T}\right)}\ ,
\end{equation}
where \(M_{ll}\) and \(f_{ll}\) are the mass and decay constant of the
pion composed of degenerate, light quarks of bare mass $m_l$.

We define an effective mass which asymptotically takes the value \(M_{ll}\). We
consider two different definitions: the local effective mass
\begin{equation}
    M^{\textrm{eff, loc}}_{ll}(t)= \frac{1}{\Delta}\cosh^{-1}\frac{G(t + \Delta) + G(t-\Delta)}{2G(t)}\ ,
\end{equation}
and the midpoint effective mass
\begin{equation}
    M^{\textrm{eff, mid}}_{ll}(t)= -\frac{1}{2\Delta}\left[\cosh^{-1}\frac{G(t + \Delta)}{G(T/2)} - \cosh^{-1}\frac{G(t - \Delta)}{G(T/2)}\right]\ ,
\end{equation}
and take the difference between the two as a source of systematic error.  The
parameter \(\Delta\) is chosen independently for each ensemble such that
\(\Delta/a\) is even (to minimize the effects of the oscillating parity
partners), and the physical value \(\Delta\) is close to either $0.2$~fm or
$0.4$~fm.  The difference between the two choices for \(\Delta\) is also taken
as a systematic error.

We can also define an effective decay constant
\begin{equation}
    f^{\textrm{eff}}_{ll}(t) =  \sqrt{
	\frac{8m_l^2G(t)}{M^{\textrm{eff}}_{ll}(t)^3}
	\frac{e^{M^{\textrm{eff}}_{ll}(t)\frac{T}{2}}(1-e^{-M^{\textrm{eff}}_{ll}(t)T})}{\cosh\!\left(M^{\textrm{eff}}_{ll}(t-\frac{T}{2})\right)}
    }\ ,
\end{equation}
where we take \(M^{\textrm{eff}}_{ll}\) to be the same combination of \(\textrm{loc}\) or
\(\textrm{mid}\) and \(\Delta\) value used
to extract the mass in a given analysis. Figure~\ref{fig:plateau} shows an
example of these effective mass and decay constants as they approach the
asymptotic region.

In order to extract the asymptotic values of the effective mass and decay
constant, we perform constant fits to their values in plateau regions chosen to
be at sufficiently large Euclidean time that excited state effects are
negligible. These plateau regions are selected with a method that is already
used to optimize the plateau regions for the effective mass of the
$\Omega$ baryon and makes use of the $Q$-value that is defined in
Ref~\cite{Bruno:2022mfy}. Based on this analysis, we select a plateau region of
$1.9$~fm -- $2.9$~fm. In order
to account for possible systematic errors arising from any residual excited
state contaminations, we also repeat the analysis with later plateaus at
$2.1$~fm -- $3.1$~fm and include the difference between
the two as a systematic error. In Figure~\ref{fig:plateau}, we show the fits to
these two plateau regions for an example effective mass and decay constant.

The plateau fits included here are all performed by minimizing the
uncorrelated \(\chi^2\), which provides improved fit stability when fitting to many
points~\cite{Michael:1993yj}. The effects of correlations between the data points
are taken into account by utilizing the $Q$-value~\cite{Bruno:2022mfy},
and by computing uncertainties in the fit values using the jackknife procedure.

\subsection{Finite-size effects}

In our analysis, we correct the lattice data for finite-volume effects with
continuum NNLO XPT \cite{Bijnens:2014dea}, where for the low-energy constants
we take the default values of the computer code in \cite{Bijnens:2014gsa}. For
our setup, we observe a good convergence of the chiral expansion: in a $6$~fm
box, the pion decay constant is smaller than its infinite-volume value by about
$0.11$\% and $0.12$\% in NLO and NNLO, respectively.  In our treatment of
systematic uncertainties, we estimate possible higher-order effects by using the
difference between NNLO and NLO results.

We also take into account the staggered artefacts of finite-size effects at NLO
\cite{Aubin:2003mg,Aubin:2003uc,Bailey:2011bkm,Bailey:2012jy}.  The staggered
XPT expressions depend on hairpin parameters, $\delta_{KS}(A)$ and
$\delta_{KS}(V)$, which are defined in Section~\ref{se:tavi}. We use two
different choices for these parameters in the analysis; their difference is
treated as a systematic error.  For the first choice, we take
$\delta_{KS}(A)=-1.76$ and $\delta_{KS}(V)=0.92$ from global fits to the
pseudoscalar data of the Fermilab-MILC collaboration
\cite{FermilabLattice:2018zqv}. As a second choice, we use
$\delta_{KS}(A)=-1.87$ and $\delta_{KS}(V)=0.00$, which best fits our lattices
with different volumes.  On our coarsest lattice, the finite-size
effect on the pion decay constant is increased from the above $0.11$\%
to about $0.15$\% due to the staggered artefacts.

Finite-time effects are also included at NLO, and are found to be on the order
of the difference between the NNLO and NLO finite-volume effects.

\subsection{Electromagnetic effects}

\begin{table}[t]
    \centering
    \begin{tabular}{c|c|c|c|c|c|c}
	$\beta$&$a$ [fm]&$L/a \times T/a $&tag& $am_s$&$m_s/m_l$&\#confs\\
	\hline\hline
	3.7000&0.1315&$24\times 48$&volume/24&0.057291&27.899&726\\
	      &      &$48\times 64$&volume/48&0.057291&27.899&300\\
	\hline
	3.7753&0.1116&$28\times 56$&dir00& 0.047615&27.843&887\\
	\hline
	3.8400&0.0952&$32\times 64$&dir00&0.043194&28.500&1110\\
	&&&dir02&0.043194&30.205&1072\\
	&&&dir04&0.040750&28.007&1036\\
	&&&dir05&0.039130&26.893&1035\\
	\hline
	\hline
	0.7300&0.1120&$56\times84$&phys2/56&0.06061&33.728&1305\\
    \end{tabular}
    \caption
    {
	List of the ensembles used to measure the electromagnetic sea-sea
	contribution to the decay rate. The action of the first seven ensembles
	is {\tt 4stout}, that of the last ensemble is {\tt 4hex}. Columns are
	as in Table~\ref{tab:configs}. In case of the {\tt 4hex} ensemble, our
	strange/light mass ratio is somewhat larger than physical, since
	there we tuned the light mass such that the taste-averaged pion mass
	takes approximately the physical value.
    }
    \label{tab:configs2} 
\end{table}

In this work, we also consider the electromagnetic corrections to the pion
decay rate. We restrict the analysis to the virtual photon exchange diagrams and
use the point-like meson approximation for the real photon emission
contribution. The structure-dependent corrections to the latter, for the
muonic decay rate we use in this work, are found to be
negligible \cite{Desiderio:2020oej,Frezzotti:2020bfa}\footnote{This statement
holds also beyond the electro-quenched approximation, as shown
recently in \cite{DiPalma:2025iud}.}. The virtual photon exchange contributions
have been computed on the lattice in the electro-quenched approximation in
\cite{DiCarlo:2019thl}. The novelty of our work is that we compute the leading
electro-unquenched contributions.  Together, the contributions can be
used to determine the muonic pion decay rate in lattice QCD.

There are two types of diagrams missing in \cite{DiCarlo:2019thl},
i.e.\ those where the photon is coupled to a sea-quark loop. In the
first, the photon
connects a sea quark to a valence quark or a lepton; in the second, the photon
connects two sea quarks; we call them sea-valence and sea-sea contributions,
respectively. The former is an $SU(3)$ flavour symmetry violating observable,
and as such we expect it to be about $20$\% of the size of the latter. This
flavour suppression of the sea-valence contributions compared to the sea-sea
one is seen in several observables in our 2020
work~\cite{Borsanyi:2020mff}. There, we call these contributions $F$ and $G$,
and the smallness of $F$ compared to $G$ is apparent in the case of hadron
masses in Figures~20 and 21 and, in the case of $a_\mu$, in Figures~25 and 26.
Also, one can estimate the
individual electromagnetic contributions using partially quenched XPT at
leading order $O(e^2p^2)$---see Refs.~\cite{Knecht:1999ag,Bijnens:2006mk} for
the necessary expressions. The contribution of the interaction between a sea
and a valence quark is free of low-energy constants, and we
find that it is a $-0.01$\% relative correction on the isospin-symmetric decay
rate. The diagram with the interaction between a sea quark and a lepton is
structure-dependent and only appears at $O(e^2p^4)$ in partially-quenched XPT.
The flavour suppression and the XPT estimate justify neglecting the sea-valence
contribution at our current level of precision. Therefore, in this work we
compute only the sea-sea contribution.

The sea-sea contribution to an observable expectation value $\langle O \rangle$
is given by the second derivative with respect to the sea electric charge:
\begin{equation}
    \label{eq:o02}
    \partial_{02}\langle O \rangle=
    \left\langle \left[O_0-\left \langle O_0 \right\rangle_U\right] \left\langle \frac{\mathrm{dets}_2^{''}}{\mathrm{dets}_0} \right\rangle_A \right\rangle_U\ ,
\end{equation}
where we use the formulae from Section~5 of our 2020 work
\cite{Borsanyi:2020mff} and replace our old notation $[\dots]^{''}_{02}$
by $\partial_{02}\dots$ for more clarity. Our definition of the partial
derivative keeps the bare quark masses constant, and this type of ``bare''
derivative has to be renormalized (see later). We also consider the second
derivative for the product of two or more expectation values, for example
\begin{equation}
\label{eq:ss-deriv-prod}
    \partial_{02} ( \langle A \rangle \langle B \rangle )=
    \partial_{02} \langle A \rangle \cdot \langle B \rangle_{0} +
     \langle A \rangle_{0} \cdot
    \partial_{02}\langle B \rangle
    \ ,
\end{equation}
where we use the fact that the first derivatives vanish at zero electromagnetic
charge. We also use this notation for observables that are more general
functions of expectation values,
\begin{equation*}
\partial_{02} f(\langle A \rangle , \langle B \rangle ) = \frac{\partial f(
  \langle A \rangle ,\langle B \rangle )}{\partial \langle A \rangle } \cdot
\partial_{02} \langle A \rangle + \frac{\partial f( \langle A \rangle ,\langle
  B \rangle )}{\partial \langle B \rangle }\cdot \partial_{02} \langle B
\rangle \ .
\end{equation*}

For the measurement of Equation \eqref{eq:o02}, we need the correlation, with respect to the gluon fields $U$, between
the isospin-symmetric observable $O_0$ and the second derivative of the
quark determinant $\langle \mathrm{dets}_2''/ \mathrm{dets}_0\rangle_A$ that
is averaged over photon fields $A$.  The fermion determinants are
measured on seven ensembles using the {\tt 4stout} action and on one ensemble
using the {\tt 4hex} action. The measurement algorithm is given in Section~7 of
\cite{Borsanyi:2020mff}, and the ensemble parameters can be found in
Table~\ref{tab:configs2}.

The electromagnetic sea-sea contribution to the pion decay rate only enters in
the quark part of the diagram; the leptons are unaffected. Therefore, the
correlation of the isospin-symmetric pion decay constant $f_{ll}$ with the
derivative of the quark determinant gives the sea-sea contribution. For our
computation, we also need the sea-sea contribution to the pion and kaon masses
and to the gradient flow scale. These have already been computed in our 2020
work \cite{Borsanyi:2020mff}.

We investigate finite-volume effects of the sea-sea contribution by utilizing
lattices with spatial extent around $3$~fm and around $6$~fm; see
Table~\ref{tab:configs2}. In the analysis, we assume a volume dependence that is
linear in the $1/L^2$ variable. This is justified by the fact that the sea-sea
contribution probes the internal structure of the pion and the leading
structure-dependent finite-volume effects are found to start at $1/L^2$ in the case
of the leptonic decay rates of pseudoscalar mesons \cite{Lubicz:2016xro}.

\subsection{Determination of $w_0$ using the pion decay rate - formulae}

Our objective is to determine the gradient-flow
scale $w_0$ from lattice QCD to leading order in isospin-breaking corrections, 
using the muonic decay rate of charged pions as input. In this
subsection, we provide the necessary formulae.  We start
by defining $F_{ud}$, the square root of the decay
rate $\Gamma$ of the pion, excluding $V_{ud}$ and kinematical factors:
\begin{equation}
\label{eq:Fud_def}
    F_{ud}^2= \Gamma(\pi\to\mu\bar\nu_\mu[\gamma])\left[ \tfrac{G_F^2}{8\pi}|V_{ud}|^2 M_{ud} m_\mu^2 \left( 1 - m_\mu^2/M_{ud}^2 \right) \right]^{-1}\ .
\end{equation}
Here, $\mu$ denotes the muon, $m_\mu$ its mass,
$\bar\nu_\mu$ the corresponding antineutrino, $G_F$ the
Fermi constant and $M_{ud}$ the pion mass. This combination was introduced in
\cite{DiCarlo:2019thl}. It is an observable defined in full QCD plus QED and
can be measured in experiment, assuming a value for
$|V_{ud}|$. We denote its physical value by $[F_{ud}]_\mathrm{qcd+qed}$ for which
we obtain
\begin{equation}
    \label{eq:fpiphys}
    [F_{ud}]_\mathrm{qcd+qed}= 131.711(45)~\mathrm{MeV}\ ,
\end{equation}
using results from \cite{ParticleDataGroup:2024cfk}, in particular
$V_{ud}=0.97367(32)$\footnote{There is some tension between the Particle Data
Group value $V_{ud}=0.97367(32)$ coming from nuclear $\beta$-decays and the
FLAG value $V_{ud}=0.97439(14)$~\cite{FlavourLatticeAveragingGroupFLAG:2024oxs,
FermilabLattice:2018zqv,Carrasco:2016kpy,ExtendedTwistedMass:2021qui,Miller:2020xhy,
Bazavov:2017lyh,Carrasco:2014poa,Dowdall:2013rya}
coming from the decay rate ratio of kaons and
pions, assuming the unitarity of the CKM matrix and using lattice computations
of the $f_K/f_\pi$ ratio with isospin-breaking correction taken from chiral
perturbation theory.  We choose to work with the PDG value here, since it
relies on fewer assumptions. We discuss the impact of the uncertainty of
$V_{ud}$ on our scale determination in Equation \eqref{eq:w0fromfpisys}.} and
$\Gamma= 3.8408(7)\times10^7/\mathrm{s}$. In pure QCD, this
observable corresponds to the usual pion decay constant, denoted here
$[F_{ud}]_\mathrm{qcd}$. Its value depends on the
scheme that defines the separation between the QCD and QED contributions in Equation \eqref{eq:Fud_def}, whereas $[F_{ud}]_\mathrm{qcd+qed}$ does not. In
any scheme, we can decompose an observable $O$ into QCD and QED parts, the latter
even into valence-valence, sea-valence, and sea-sea contributions:
\begin{equation*}
    [O]_\mathrm{qcd+qed}= [O]_\mathrm{qcd} + [O]_\mathrm{qed,vv} + [O]_\mathrm{qed,vs} + [O]_\mathrm{qed,ss}\ .
\end{equation*}

We adopt the valence-valence QED contribution from the lattice
computation of \cite{DiCarlo:2019thl}, which employs the scheme of Gasser,
Rusetsky and Scimemi (GRS) \cite{Gasser:2003hk}. This scheme is defined in such a
way that the strong coupling $\alpha_s$ and the quark masses
$m_u$, $m_d$ and $m_s$, renormalized at some fixed scale $\Lambda$, remain constant as the
electromagnetic coupling is turned on or off. In particular, the authors of
\cite{DiCarlo:2019thl} determine\footnote {Ref.~\cite{DiCarlo:2019thl} gives
$\delta R_\pi= 0.0153(19)$, which value includes a XPT estimate of sea effects
of $0.003$. Here, we need a value with the valence contribution only, therefore
we use the value quoted in the text.} that the valence-valence QED contribution
relative to the strong one is
\begin{equation}
    \label{eq:rpi}
    [F_{ud}]_\mathrm{qed,vv}/[F_{ud}]_\mathrm{qcd}= (1+\delta R_\pi)^{1/2} - 1
    \quad\text{with}\quad
    \delta R_\pi= 0.0150(18)\ .
\end{equation}
To reconstruct the full decay rate, the other QED
contributions must also be computed in the GRS scheme.  This can be
achieved without having to match renormalized quark masses in QCD and
in QCD plus QED at a fixed renormalization scale. It is sufficient to have
the decomposition of the pion and kaon masses in the GRS scheme.  The
authors of \cite{DiCarlo:2019thl} provide these. For the QCD
contributions they obtain \cite{Giusti:2017dmp}:
\begin{align}
    \label{eq:rm123a}
    [M_{ud}]_\mathrm{qcd++}&= 135.0(2)~\mathrm{MeV}\ ,\\
    \label{eq:rm123b}
    [\tfrac{1}{2}(M_{us}+M_{ds})]_\mathrm{qcd++}&= 494.6(1)~\mathrm{MeV}\ .
\end{align}
A subtle point arises here. This QCD++ component is obtained by subtracting
the valence-valence QED contribution from the physical QCD plus QED
values. Thus, the QCD component in Ref.~\cite{DiCarlo:2019thl} includes not only pure QCD but also sea-valence and sea-sea QED effects, as signaled by the label
QCD++. For a generic observable $O$ we have
\begin{equation*}
    [O]_\mathrm{qcd++}\equiv [O]_\mathrm{qcd} + [O]_\mathrm{qed,vs} + [O]_\mathrm{qed,ss}\ .
\end{equation*}
Using $\delta R_\pi$ and the full QCD plus QED value from above, the pion decay constant
in this scheme can be written
\begin{align}
    \label{eq:rm123c}
    [F_{ud}]_\mathrm{qcd++}&= [F_{ud}]_\mathrm{qcd+qed}/(1+\delta R_\pi)^{1/2}= 130.73(12)~\mathrm{MeV}\ .
\end{align}
As we will see soon, we specifically
require the QCD++ ---not only the QCD---component of these quantities.

We also note that, in Ref.~\cite{DiCarlo:2019thl}, the authors used
the QCD pion decay constant from Equation~\eqref{eq:rm123c} to set the
scale for their calculations of the QCD++ components of the masses in
Equations~\eqref{eq:rm123a} and~\eqref{eq:rm123b}. An alternative scale
choice---differing from the QCD pion decay constant by isospin-breaking
corrections---would only impact the results in
Equations~\eqref{eq:rm123a} and~\eqref{eq:rm123b} at the level of
neglected $O(e^4)$ terms. This is because the measurement specifically
targeted the valence QED component of these masses, which is an
$O(e^2)$ quantity.

We now proceed to determine the product $w_0F_{ud}$ in full QCD plus QED.  We
parameterize its quark-mass and electric-charge dependence as
\begin{align}
    \label{eq:w0fpi}
    w_0 F_{ud}=
    A &\ +
    B  \ F_{ud}^{-2}M_{ud}^2 +
    C  \ F_{ud}^{-2}(M_{us}^2+M_{ds}^2-M_{ud}^2)/2 +
    E  \ e_v^2 + F\ e_ve_s + G\ e_s^2\ ,
\end{align}
where $e_v$ and $e_s$ denote the unit electric charges associated with the
valence and sea quarks, and where we neglect all $O(e^4)$ and other NLO
isospin-breaking corrections, as we do throughout this section and paper. The
quantity $w_0$, as well as all masses and decay constants appearing in this
expression, are understood to be those obtained on the lattice in the full QCD
plus QED theory, for a variety of quark masses and electric charges.
Accordingly, these quantities are not labeled with the subscript
$\mathrm{qcd+qed}$, which we reserve for their physical values, e.g.  in
Equation \eqref{eq:fpiphys}. We note that $F_{ud}$ on the left- and right-hand
sides of this equation refer to the same quantity.

In Equation~\eqref{eq:w0fpi}, the quantity $w_0 F_{ud}$ is treated as a
function of the independent variables $F_{ud}^{-2}M_{ud}^2$,
$F_{ud}^{-2}(M_{us}^2+M_{ds}^2-M_{ud}^2)/2$, $e_v$ and $e_s$. For instance, $G$
is the second, partial derivative of $w_0 F_{ud}$ with respect to $e_s$, at
fixed values of the other variables. More generally, the parameters
$A,\cdots,G$, which are not those for $w_0 M_\Omega$ in Equation
\eqref{eq:w0om}, can be obtained by fitting Equation~\eqref{eq:w0fpi} to the
dependence of $w_0 F_{ud}$ on the independent variables for varying quark
masses and charges around their physical values. All dependence on the
independent variables is explicit in Equation~\eqref{eq:w0fpi}\footnote{An
additional term, like $e_s^2\partial_{02}(F_{ud}^{-2}M_{ud}^2)$, is not needed
in Equation~\eqref{eq:w0fpi}. In that Equation $F_{ud}^{-2}M_{ud}^2$ and
$e_s^2$ are treated as independent variables, which means that
$F_{ud}^{-2}M_{ud}^2$ can be adjusted independently of $e_s^2$. In the following
step we change the set of independent variables, replacing $F_{ud}^{-2}M_{ud}^2$ and
$F_{ud}^{-2}(M_{us}^2+M_{ds}^2-M_{ud}^2)/2$ by the bare light and strange quark masses,
and $F_{ud}^{-2}M_{ud}^2$ becomes a function of the bare parameters, ie. also of $e_s^2$.}.

The observables $F_{ud}^{-2}M_{ud}^2$ and
$F_{ud}^{-2}(M_{us}^2+M_{ds}^2-M_{ud}^2)/2$ are themselves functions of
QCD plus QED's bare parameters, as is $w_0 F_{ud}$. After appropriate
renormalization, they are finite. Because we are working to first order
in QED, the bare parameters $e_v$ and $e_s$ are equal to their
renormalized values, and are also finite. Thus,
Equation~\eqref{eq:w0fpi} relates independent, finite observables in
the full theory, implying that the parameters $A,\cdots,G$ are finite
and independent of any scheme used for separating these observables
into pure QCD and QED contributions. The remainder of this subsection
explains how we compute these parameters, which are needed to determine
the physical value of $w_0F_{ud}$.

We begin with $A$, $B$ and $C$. Since our target observable is
symmetric under $u\leftrightarrow d$ replacement, and we work to
leading order in isospin breaking, no term is required for the strong-isospin
breaking (the isospin-breaking effect from the up-down quark mass
difference). Now, by definition, QCD plus QED observables are equal to
their pure QCD counterparts, up to order-charge-squared corrections, in
any separation scheme. Thus, setting $e_v=e_s=0$,
Equation~\eqref{eq:w0fpi} reduces to
\begin{align}
    \label{eq:w0fpiiso}
    w_0 f_{ll}=
    A +
    B\ f_{ll}^{-2} M_{ll}^2 +
    C\ f_{ll}^{-2} (M_{ls}^2-\tfrac{1}{2}M_{ll}^2)\ ,
\end{align}
where $f_{ll}$ and $M_{ll}$ are the pion decay constant and mass,
respectively, computed in pure isospin-symmetric QCD.  This means that
$A$, $B$ and $C$ can be obtained with a pure isospin-symmetric QCD
calculation, by fitting the dependence of $w_0 f_{ll}$ on $f_{ll}^{-2}
M_{ll}^2$ and $f_{ll}^{-2} (M_{ls}^2-\tfrac{1}{2}M_{ll}^2)$ to that
equation.

We now turn to the electromagnetic coefficients $E$, $F$, and $G$. As
discussed above, the valence-valence coefficient $E$ has already been
determined in \cite{DiCarlo:2019thl}. To complete the calculation of
$[w_0F_{ud}]_\mathrm{qcd+qed}$, we must compute the sea-valence and
sea-sea electromagnetic coefficients $F$ and $G$. We begin with the
sea-sea electromagnetic coefficient $G$. We compute it via the bare,
second, partial derivative of Equation~\eqref{eq:w0fpi} with respect to
$e_s$, defined in Equation~\eqref{eq:o02}, and with the rule for second
derivatives of products of observables in Equation
\eqref{eq:ss-deriv-prod}. Here, the derivatives are taken with $e_v$
and the bare parameters of pure QCD held fixed.  Thus, $G$ is obtained
by fitting the sea-electric-charge derivative
\begin{align}
    \label{eq:w0fpiqed}
    G= \partial_{02} \left(
	w_0 F_{ud} -
	B\ F_{ud}^{-2} M_{ud}^2 -
	C\ F_{ud}^{-2}(M_{us}^2 + M_{ds}^2 - M_{ud}^2)/2
	\right)
\end{align}
to a constant, where the $B$ and $C$ parameters are taken from the
isospin-symmetric fit. Note that the terms proportional to $B$ and $C$ in
Equation \eqref{eq:w0fpiqed} effectively remove the divergences in the bare
derivative $\partial_{02} (w_0 F_{ud})$, which arise from the bare QED
corrections to the quark masses: $G$ is a physical, renormalized quantity\footnote{
The $B$ coefficient itself is a finite, renormalized quantity.
The divergence related to the QED correction on the light quark mass
appears in $\partial_{02}(F_{ud}^{-2} M_{ud}^2)$, that is multiplied by $B$. This
divergence together with the one coming from the $C$-term cancel the divergence
in $\partial_{02}(w_0 F_{ud})$ to give a finite $G$.}.
For this fit, we need the measurements of $f_{ll}$ and the
derivative of the quark determinant, as described in
Equation~\eqref{eq:o02}. The sea-valence term $F$ can be determined
from sea-valence derivatives. These are
flavor-$SU(3)$-suppressed and expected to be even smaller than
the $0.1\%$ contribution from sea-sea effects. 
Thus, for the level of precision required here, we can set $F=0$. We leave its computation for future work. 

We now have all the ingredients needed to determine the
physical value of $w_0 F_{ud}$. We decompose it as
\begin{equation}
    [w_0 F_{ud}]_\mathrm{qcd+qed} = [w_0 F_{ud}]_\mathrm{qcd++} + [w_0 F_{ud}]_\mathrm{qed,vv}\ .
\end{equation}
The QCD++ contribution is obtained via
\begin{equation}
    [w_0 F_{ud}]_\mathrm{qcd++}= A + B\ [F_{ud}^{-2}M_{ud}^2]_\mathrm{qcd++} + C  \ [F_{ud}^{-2}(M_{us}^2+M_{ds}^2-M_{ud}^2)/2]_\mathrm{qcd++}
    + F e^2 + G e^2\ ,
\end{equation}
where the $A,B,C,F,G$ parameters are determined as described above, and
the QCD++ values of the pion, kaon masses and decay constant are taken
from Equations \eqref{eq:rm123a}, \eqref{eq:rm123b} and
\eqref{eq:rm123c}. For the valence-valence QED part, we use Equation
\eqref{eq:rpi}. It can be rewritten as
\begin{equation}
    \frac{[w_0 F_{ud}]_\mathrm{qed,vv}}{[w_0 F_{ud}]_\mathrm{qcd}}=
    (1+\delta R_\pi)^{1/2} - 1\ ,
\end{equation}
because $w_0$ is defined in terms of
purely gluonic quantities and is therefore free of valence-valence QED
contributions.

Adding the QCD++ and valence-valence QED contributions, we get the physical value
\begin{equation}
    [w_0 F_{ud}]_\mathrm{qcd+qed}=
    [w_0 F_{ud}]_\mathrm{qcd++} + ( (1+\delta R_\pi)^{1/2} - 1 ) {[w_0 F_{ud}]_\mathrm{qcd}} 
    \approx [w_0 F_{ud}]_\mathrm{qcd++} (1+\delta R_\pi)^{1/2}\ ,
\end{equation}
where we drop terms that are higher order in $e^2$.  Finally, dividing by
the physical value of $F_{ud}$, and using Equation \eqref{eq:rm123c}, we arrive
at the physical value of $w_0$ itself
\begin{equation}
    \label{eq:w0fpimaster}
    [w_0]_\mathrm{qcd+qed}= [w_0 F_{ud}]_\mathrm{qcd++}/[F_{ud}]_\mathrm{qcd++}\ ,
\end{equation}
which is the main result of this subsection.

The parameterization in Equation \eqref{eq:w0fpi} can also be used to present
results in a different scheme, e.g.\ the Edinburgh consensus scheme, which has
now also been adopted by FLAG~\cite{FlavourLatticeAveragingGroupFLAG:2024oxs}. It is defined by
\begin{align*}
    [M_{ud}]_\mathrm{qcd,FLAG}&=135.0~\mathrm{MeV}\ ,\\
    [\tfrac{1}{2}(M_{us}+M_{ds})]_\mathrm{qcd,FLAG}&=494.6~\mathrm{MeV}\ ,\\
    [F_{ud}]_\mathrm{qcd,FLAG}&=130.5~\mathrm{MeV}\ .
\end{align*}
Substituting these values into Equation \eqref{eq:w0fpi} with $E,F$ and $G$ set
to zero and dividing by $[F_{ud}]_\mathrm{qcd,FLAG}$ gives us
$[w_0]_\mathrm{qcd,FLAG}$. The QED part in this scheme,
$[w_0]_\mathrm{qed,FLAG}$, can be computed by subtracting the QCD part from the
full physical value.

\subsection{Determination of $w_0$ using the pion decay rate - results}

\begin{table}[p]
    \centering
    \includegraphics{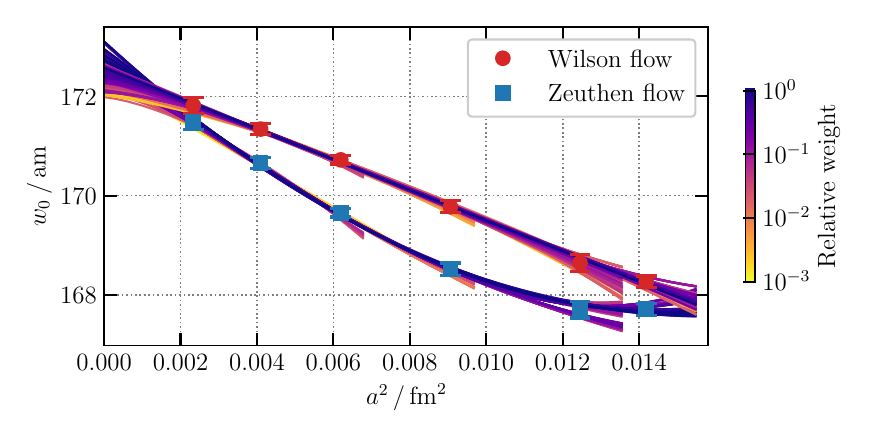}\\
    \includegraphics{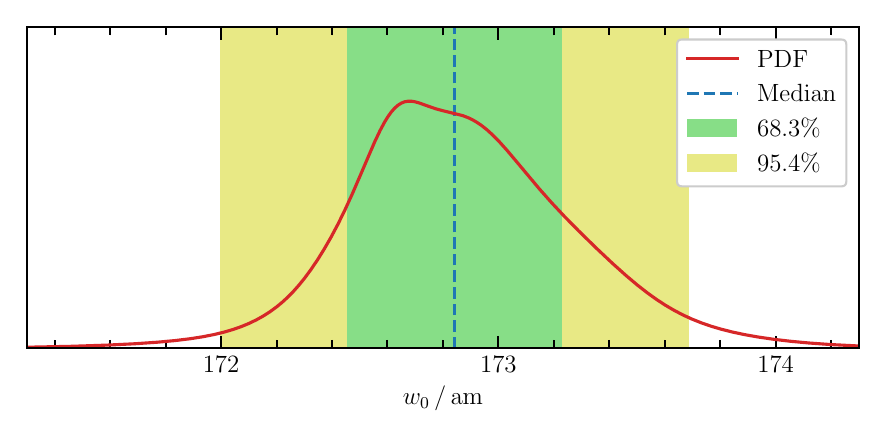}\\
    \input{figures/tables/w0fpi/tabular}
    \caption
    {
	\label{ta:res_w0fpi} Gradient-flow scale
	$w_0$ in attometers using the pion decay constant as input. Results are
	given in the isospin-symmetric point of the FLAG scheme and in the
	infinite volume limit.  The continuum extrapolations are shown for two
	definitions of the gradient flow, Wilson and Zeuthen.  More details on
	the plot conventions can be found in the first part of
	Section~\ref{se:si_win:table}. 
    }
\end{table}

\begin{figure}[t]
    \centering
    \includegraphics{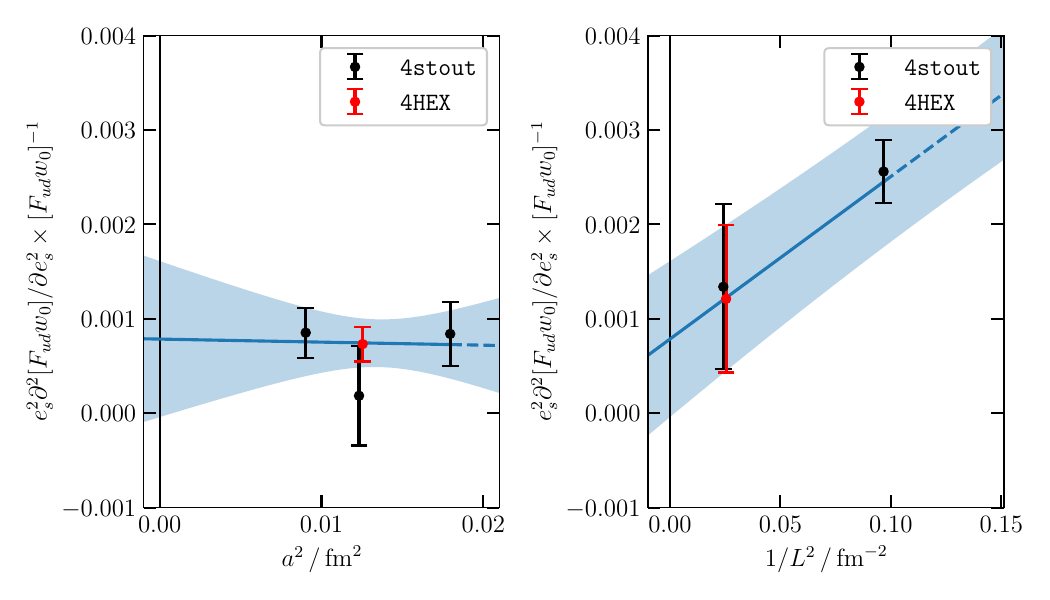}
    \caption
    {
	Renormalized second derivative of $[w_0 F_{ud}]$ with respect to the
	sea electromagnetic charge, which is called $G$ parameter in the text.
	We multiply by the physical charge squared and normalize by the isospin
	symmetric value of $[w_0 F_{ud}]$, such that we can interpret the
	quantity as a relative correction.  The left panel shows the continuum,
	the right panel the infinite volume extrapolation.
    }
    \label{fi:w0fpiqed}
\end{figure}

Here we discuss in detail the fit procedure, which gives
$[w_0]_\mathrm{qcd+qed}$. We start with the isospin-symmetric fits of
Equation \eqref{eq:w0fpiiso}, then we explain the fits to the sea-sea
contribution, Equation \eqref{eq:w0fpiqed}.

The isospin-symmetric fits determine the parameters $A,B$ and $C$. For the $A$
parameter we allow for polynomials of $a^2$ at most cubic order. Additionally,
we allow for modifying the $a^2$ term by an $\alpha_s^\gamma(a)$ factor, where
$\alpha_s(a)$ is the strong coupling at the scale of the lattice spacing and
the exponent $\gamma$ is taken from the set $\{0.0,0.5,1.0,1.5,2.0,2.5\}$.
With this variation, we address the uncertainties related to the {\em a priori}
unknown logarithmic corrections of the $a^2$ dependence (see
Reference~\cite{Husung:2019ytz} and the discussion in
Section~\ref{se:si_analysis_fit}). The $B$ and $C$ parameters correspond to
the mass dependences in the light and strange quarks. Here we allow for
a polynomial that is either constant or linear in $a^2$. We drop the
coarsest lattice spacing, $\beta=3.7000$, entirely from the analysis.
The fit results are shown in Table~\ref{ta:res_w0fpi}.

In these fits we use two different definitions for the gradient-flow: the first
is the original prescription of L\"uscher \cite{Luscher:2010iy}, for the second
we take the Zeuthen-flow \cite{Ramos:2015baa}, which is free from $O(a^2)$
artefacts. These effects are classical,
and therefore are not modified by logarithmic corrections\footnote{We thank D.~Nogradi and A.~Ramos for discussions
on lattice artefacts of gradient flows.}. In the continuum
extrapolation plots, we show results with both discretizations. The difference
between the results of these two flows becomes part of our systematic error.

To facilitate the comparison with other lattice collaborations, we present the
isospin-symmetric results in the FLAG scheme, as defined above.  The continuum-extrapolated result
at the isospin-symmetric point of FLAG is
\begin{equation}
    \label{eq:w0fromfpiqcd}
    [w_0]_\mathrm{qcd,FLAG}= \valuesWnullFromfpiQcd~\mathrm{fm}\ .
\end{equation}
We also show a histogram and error budget for these isospin-symmetric fits. The
largest source of systematic error comes from the variation in the
pseudoscalar fit range, closely followed by the variations in the type of the
gradient flow (Wilson vs Zeuthen) and in the functional form used for the continuum extrapolation.

We also investigate the impact of the new $a=0.048$~fm ensembles.  Without
these ensembles, we obtain $\valuesWnullFromfpiQcdNomon~\mathrm{fm}$, implying an
uncertainty reduction of $\valuesWnullFromfpiQcdEred$. This improvement comes mainly
from a better control of the uncertainties related to the fit form and from a
reduction of the difference between the results obtained from the Wilson and
Zeuthen flows: the finest lattice spacing improves the reliability of the
continuum extrapolation.

For the sea-sea contribution, we have to determine the $G$ parameter of Equation
\eqref{eq:w0fpiqed}. Here we assume a linear dependence in $a^2$ for the {\tt
4stout} ensembles; for the single {\tt 4hex} ensemble we either assume the same
artefacts as for the {\tt 4stout} or no artefacts at all. The difference of
these two cases becomes part of the systematic error. We also assume a volume
dependence, linear in $1/L^2$. Note, a $1/L$ behaviour is
excluded, since it must be structure independent. Some of these fits are shown
in Figure~\ref{fi:w0fpiqed}. The left panel shows the continuum extrapolation,
the right panel the infinite-volume extrapolation. We plot the $G$ parameter
multiplied by the physical value of the electric charge squared and divided by
the isospin-symmetric value of $w_0 F_{ud}$. We can read off from the plot
that the sea-sea contribution in the continuum is about $0.1$\% of the
isospin-symmetric value, with an error of $\pm0.1$\%. From the $SU(3)$ flavour
suppression, we estimate the sea-valence contribution to be about $20$\% of the
sea-sea contribution, i.e.\ about $0.02$\% of the isospin-symmetric value. This
is consistent with our partially-quenched XPT estimate, which is $-0.01$\%. It
is therefore reasonable to assume that the sea-valence contribution is
negligible at the current level of precision.

We can now put together the results of the above fits as required by Equation
\eqref{eq:w0fpimaster} to get
\begin{equation}
    \label{eq:w0fromfpi}
    [w_0]_\mathrm{qcd+qed}= \valuesWnullFromfpi~\mathrm{fm}\ ,
\end{equation}
which is our final result for the gradient flow scale in QCD plus QED. It has a
relative precision of $\valuesWnullFromfpiErr$ per mille. The first error is
statistical, the second is systematic, the third is these two added in
quadrature. The systematic error can be split up as
\begin{equation}
    \label{eq:w0fromfpisys}
    \valuesWnullFromfpiSys\ ,
\end{equation}
where the first error contains the systematic error of the isospin-symmetric
fits, the second arises from the uncertainty of the valence-valence
electromagnetic effects from the work of \cite{DiCarlo:2019thl}, the third is
the uncertainty of the electromagnetic sea effects, the fourth comes from the
uncertainty of $V_{ud}$ and the last from the uncertainty of the other
experimental inputs, see Equation \eqref{eq:fpiphys}. Finally, we can also give
the electromagnetic part of $w_0$ in the FLAG scheme, for which we subtract
Equation \eqref{eq:w0fromfpiqcd} from \eqref{eq:w0fromfpi} to get
\begin{equation}
    \label{eq:w0fromfpiqed}
    [w_0]_\mathrm{qed,FLAG}= \valuesWnullFromfpiQed~\mathrm{fm}\ .
\end{equation}
While the QCD and QED parts are scheme dependent, their sum can be directly
compared to our gradient flow scale determination using the Omega baryon as
input. This is done in Section~\ref{se:si_phys}.

%% file: figures/tables/w0fpi/tabular.tex
\begin{tabular}{l|r|r}
Number of fits & \multicolumn{2}{c}{12672} \\
Fits with $P>0.1$ & \multicolumn{2}{c}{74\%} \\
Median & \multicolumn{2}{c}{172.84} \\
\hline
Total error & 0.42 & 0.24\% \\
Statistical error & 0.34 & 0.20\% \\
Systematic error & 0.25 & 0.14\% \\
\hline
Pseudoscalar fits & 0.15 & 0.09\% \\
Finite volume XPT & 0.03 & 0.02\% \\
Type of flow (Ze/Wi) & 0.14 & 0.08\% \\
Lattice spacing cut & 0.04 & 0.02\% \\
Fit polynomial order & 0.11 & 0.06\% \\
Log corrections $\gamma$ & 0.07 & 0.04\% \\
\end{tabular}

%% file: si_phys.tex
\section{Physical point and isospin decomposition}
\label{se:si_phys}
\newcommand{\ISO}{\ensuremath{\mathrm{iso}}}

In our computations, we parameterize the quark-mass and electromagnetic-coupling
dependence of the observables using the gradient-flow-based $w_0$ scale and the
connected up, down, and strange pseudoscalar meson masses $M_{uu}$,
$M_{dd}$ and $M_{ss}$. These masses are computed by taking into account only
the quark-connected contributions to the corresponding two-point functions, as in
Ref.~\cite{BMW:2013mpk}, and are rigorously defined in a partially-quenched
theory. Concretely, the parameterization of an observable $Y$ is given as
\begin{equation}
    \label{eq:master}
    Y= A + B\ w_0^2 \hat{M}^2 + C\ w_0^2 M_{ss}^2 + D\ w_0^2 \Delta M^2 + E\ e_v^2 + F\ e_v e_s + G\ e_s^2\ ,
\end{equation}
where $e_v$ and $e_s$ are the electromagnetic couplings of the valence and sea
quarks. We use $\hat{M}$ to denote the average $\hat{M}^2 \equiv \tfrac{1}{2}
\left(M^2_{uu} + M^2_{dd}\right)$ and $\Delta M^2$ to denote the difference
$\Delta M^2\equiv M^2_{dd}-M^2_{uu}$, which is a measure of strong isospin
breaking. We called these kinds of parameterizations ``Type-II'' fits in our
2020 work. In this section, we first determine the physical point, then we
decompose the $Y$ observable into different isospin components.

\subsection{Physical point}

\begin{figure}[t]
    \centering
    \includegraphics{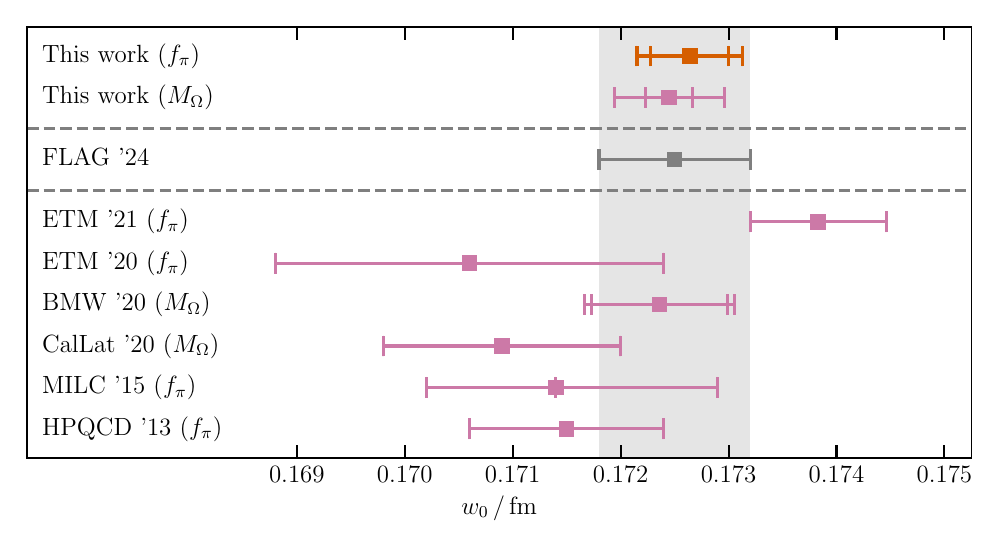}
    \caption
    {
	Comparison of recent determinations~\cite{Borsanyi:2020mff,
        ExtendedTwistedMass:2021qui,Miller:2020evg,MILC:2015tqx,Dowdall:2013rya,
	ExtendedTwistedMass:2020tvp} of the gradient-flow scale $w_0$.
	The upper two values correspond to our results from
	Sections~\ref{se:si_omega} and \ref{se:si_fpi} of this work, using the
	Omega-baryon mass and the pion leptonic-decay rate as input.
	The grey band shows the latest FLAG
	average~\cite{FlavourLatticeAveragingGroupFLAG:2024oxs}.
    }
    \label{fi:w0others}
\end{figure}

Our basis observables, $w_0$, $\hat{M}$, $M_{ss}$, $\Delta M$, cannot be
measured experimentally. Still, they have a well-defined continuum limit so that a
physical value can be associated with each of them. These values can be
computed in the full theory, i.e.\ QCD plus QED, where four experimentally
measured hadron masses are set to their physical values.

According to leading-order partially-quenched chiral perturbation theory
coupled to photons \cite{Bijnens:2006mk}, the physical value of $\hat{M}$
is equal to the neutral pion mass, i.e.\
\begin{equation}\label{eq:phys1}
    [\hat{M}]_\mathrm{qcd+qed}= 134.9768(5)~\mathrm{MeV}\ .
\end{equation}
This equality is valid up to next-to-leading order effects in
isospin breaking, which we neglect in this work. For the physical values of the remaining
meson masses, we use the analysis from 2020~\cite{Borsanyi:2020mff}. For completeness,
we give them here:
\begin{equation}\label{eq:phys2}
    \begin{aligned}
	[\Delta M^2]_\mathrm{qcd+qed} &=  13170(320)(270)[420]~\mathrm{MeV}^2\ ,\\
	[M_{ss}]_\mathrm{qcd+qed}     &= 689.89(28)(40)[49]~\mathrm{MeV}\ ,\\
    \end{aligned}
\end{equation}
where the first error is statistical, the second is systematic, and the third is
the total.

Finally, we turn to the gradient-flow scale $w_0$. Figure~\ref{fi:w0others}
shows determinations from various lattice collaborations. As input, either the
Omega baryon mass or the pion decay constant is used.  The band shows the latest
average from FLAG~\cite{FlavourLatticeAveragingGroupFLAG:2024oxs,Borsanyi:2020mff,
ExtendedTwistedMass:2021qui,Miller:2020evg,MILC:2015tqx,Dowdall:2013rya}.
In our 2020 work and also in a previous version of this
paper, we used the Omega baryon mass as input to compute the physical value
$[w_0]_\mathrm{qcd+qed}$.  Our latest determination using this approach
(presented in Section~\ref{se:si_omega}) is indicated by ``This work ($M_\Omega$)'' on the plot. As
explained in Section~\ref{se:si_omega}, all computations based on the Omega
baryon have a systematic error of unknown size, arising from multi-hadron
excited states in baryon propagators. As a consequence, we replace the Omega
baryon with the pion leptonic-decay rate as the input quantity of the scale setting
procedure, see Section~\ref{se:si_fpi}. The determination is the first that
takes both valence and sea quark electromagnetic effects into account. It gives
the physical value of the gradient flow scale of Equation \eqref{eq:w0fromfpi}.
This value is labelled by ``This work ($f_\pi$)'' in the comparison plot. It agrees nicely
with our determination using the mass of the Omega baryon as input. We use
this pion-decay-rate $w_0$ value in our computations of the hadronic vacuum polarization.
That is, on each ensemble the lattice spacing $a$ is defined as the physical
value of $[w_0]_\mathrm{qcd+qed}$ from Equation~\eqref{eq:w0fromfpi} divided by
the value of $[w_0/a]$ measured on the given ensemble.

To obtain the physical value of some observable $Y$, we implement the following procedure:
\begin{enumerate}
    \item The parameters $A,\dots,G$ in Equation \eqref{eq:master}
	are determined from the fitting procedure described in
	Section~\ref{se:si_analysis}.
    \item The physical values of the basis observables above are substituted into
	the equation to obtain the physical result.
\end{enumerate}
We incorporate the errors in the basis observables and also their correlations into
our analyses by a stochastic sampling of their respective distributions. The
procedure is described in detail in Section~\ref{se:si_analysis}. This approach
respects the statistical correlations, while most systematic ingredients are
treated as uncorrelated. We justify this choice by noting that our basis
observables have very different lattice artefacts compared to the ones with
which they are fitted.

\subsection{Isospin decomposition}

We can decompose an observable $Y$ into isospin-symmetric and isospin-breaking
contributions.  This decomposition is, of course, not unique and depends on
how the isospin-symmetric theory is defined. Thus, pure QCD results
have a scheme ambiguity, which one has to keep in mind when performing
comparisons. We have already seen several possible choices: the FLAG, GRS and
``Type-I'' schemes in Sections~\ref{se:si_omega} and \ref{se:si_fpi}.

In our 2020 work, we put forward a scheme that is based on the
parameterization in Equation \eqref{eq:master}. In that scheme, which we call
``BMW'' scheme, we define the physical values of the basis observables, $w_0$,
$M_{uu}$, $M_{dd}$ and $M_{ss}$, to be the same in QCD and in the full theory,
ie. QCD plus QED. In particular, the electromagnetic part of the gradient-flow
scale is zero by definition, $[w_0]_\mathrm{qed,BMW}=0$, as opposed to other schemes, where
it can attain a non-zero value--see Equations \eqref{eq:w0fromomqed} for
the ``Type-I'' or Equation \eqref{eq:w0fromfpiqed} for the FLAG scheme. The
physical value of the observable $Y$ can be decomposed into isospin-symmetric,
strong-isospin-breaking and electromagnetic parts as
\begin{equation}
    \label{eq:decomp1}
    [Y]_\mathrm{qcd+qed}= [Y]_\mathrm{iso,BMW} + [Y]_\mathrm{sib,BMW} + [Y]_\mathrm{qed,BMW}
\end{equation}
with
\begin{equation}
    \label{eq:decomp2}
    \begin{aligned}
	[Y]_\mathrm{iso,BMW}&= A + B\ [w_0^2 \hat{M}^2]_\mathrm{qcd+qed} + C\ [w_0^2 M_{ss}^2]_\mathrm{qcd+qed}\ ,\\
	[Y]_\mathrm{sib,BMW}&= D\ [w_0^2 \Delta M^2]_\mathrm{qcd+qed}\ ,\\
	[Y]_\mathrm{qed,BMW}&= E\ e^2 + F\ e^2 + G\ e^2\ .
    \end{aligned}
\end{equation}
The sum of the first two gives the QCD contribution $[Y]_\mathrm{qcd,BMW}$. In
this work, the isospin contributions are given in the ``BMW'' scheme, unless indicated otherwise.

\subsection{Kaon mass decomposition in different schemes}

\begin{figure}[btp]
    \centering
    \includegraphics{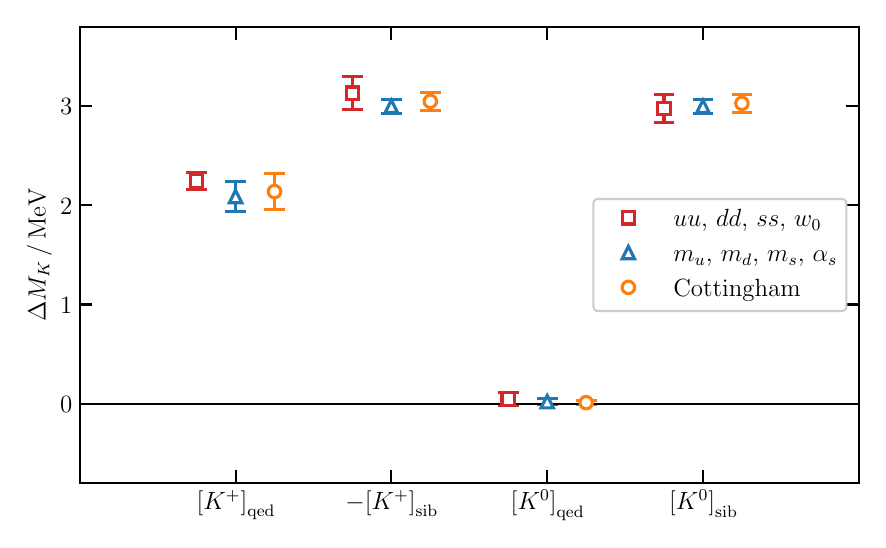}
    \caption
    {
	Decomposition of the neutral and charged kaon masses. Red squares stand
	for the BMW scheme, which is based on the observables
	$\{M_{uu},M_{dd},M_{ss},w_0\}$; blue triangles for the GRS scheme,
	based on quark masses and strong coupling constant taken from
	\cite{DiCarlo:2019thl}. Orange circles are obtained by computing the
	electromagnetic self-energies using the Cottingham-formula
	\cite{Stamen:2022uqh}.
    }
    \label{fig:kadecomp} 
\end{figure}

The isospin-symmetric point depends on the observables that define it, an
effect commonly referred to as scheme dependence. A similar scheme to ours was
put forward already in Ref.~\cite{Horsley:2015vla}, though the physical values
of the defining observables were not computed there.  A scheme based on the
Lagrangian parameters of the theory, which keeps renormalized quark masses and
the strong coupling fixed while turning on the electromagnetic interaction, was
proposed by Gasser, Rusetsky and Scimemi \cite{Gasser:2003hk}.  This scheme is
called the GRS scheme in the literature and was implemented on the lattice in
Refs.~\cite{Giusti:2017dmp,DiCarlo:2019thl}.

To compare our scheme with others, we decompose the neutral and charged kaon
masses into isospin components. For this purpose, we fit the kaon masses to
Equation~\eqref{eq:master}, and use Equations \eqref{eq:decomp1} and
\eqref{eq:decomp2} to compute the various components. The analysis includes
about ten thousand fits to estimate systematics related to, e.g.\ continuum
extrapolation and choice of the mass fit range. The decomposition we obtain is
\begin{equation}
    [M_{us}]_\mathrm{iso}=494.55[31]~\mathrm{MeV}\ ,
    \begin{aligned}
	\quad\quad [M_{ds}]_\mathrm{sib}=& +2.98[14]~\mathrm{MeV}\ , \quad\quad & [M_{ds}]_\mathrm{qed}&= 0.05[7]~\mathrm{MeV}\ ,\\
	\quad\quad [M_{us}]_\mathrm{sib}=& -3.13[17]~\mathrm{MeV}\ , \quad\quad & [M_{us}]_\mathrm{qed}&= 2.25[8]~\mathrm{MeV}\ ,\\
    \end{aligned}
\end{equation}
where the uncertainties given are the statistical and systematic errors added
in quadrature. Figure~\ref{fig:kadecomp} shows a comparison with the same
decomposition in the GRS scheme. The values are taken from
Ref.~\cite{Giusti:2017dmp}. Note that they are computed in the electro-quenched
approximation, whereas our results also include sea-valence electromagnetic
effects. Sea-sea electromagnetic effects are absent in mass isospin splittings.
The neglected sea-valence effects in Ref.~\cite{Giusti:2017dmp} should be
small, due to $SU(3)$ flavour suppression. It is possible to
compute the electromagnetic self-energies of hadrons using the Cottingham
formula~\cite{Cottingham:1963zz}. For kaons, this was done in
Ref.~\cite{Stamen:2022uqh}. We also show these results in the Figure. We find
good agreement between the different decompositions.

%% file: si_analysis.tex
\section{Analysis procedure}
\label{se:si_analysis}

\subsection{Fit functions}
\label{se:si_analysis_fit}

To obtain the physical isospin-symmetric values of different
hadronic vacuum polarization windows we perform global fits to the lattice
spacing and quark mass dependence of these observables.

Na{\"i}vely, we expect the leading discretization errors to scale like
$a^2$. In general, however, this is modified by the anomalous dimensions of
operators in the Symanzik effective theory, to become a sum of terms
proportional to $a^2\alpha_s^\gamma(a)$ with different values of the exponent
$\gamma$~\cite{Husung:2019ytz}, where $\alpha_s(a)$ is the strong coupling at
the scale of the lattice spacing. The set of values of $\gamma$ is not yet
known for our lattice action and observables. (For the case of unrooted
staggered fermions with different number of flavours results have been presented
in \cite{Husung:2025nsv}.) Also the prefactors appearing in these terms can be
very different. Getting a stable fit with multiple $a^2$ terms having different
$\gamma$'s is difficult with the currently available lattice spacings. Instead,
we perform fits with a single $a^2\alpha_s^\gamma(a)$ term at a time and vary
the $\gamma$ subsequently to address the uncertainty related to these
logarithmic corrections\footnote{We thank R.~Sommer for a discussion on
possible fit forms.}. In particular, we use the following function
\begin{equation}
    A(a)= A_0 + A_2\cdot (a/w_0)^2 \alpha_s^\gamma(a) + A_4\cdot (a/w_0)^4 + A_6\cdot (a/w_0)^6
\end{equation}
to parameterize the lattice artefacts, where we allow the variation of $\gamma$.
Here $A_4$ and $A_6$ describe higher order corrections in the lattice spacing.

For the quark mass dependence, the variables
\begin{equation}
    X_l= \hat{M}^2w_0^2  - {\big[\hat{M}^2w_0^2 \big]}_\mathrm{qcd+qed}
    \quad\text{and}\quad
    X_s= M_{ss}^2w_0^2   - {\big[M_{ss}^2w_0^2  \big]}_\mathrm{qcd+qed}
\end{equation}
describe the deviation from the physical light and strange-quark mass
respectively. Here ``qcd+qed'' denotes the physical values given in
Equations~\eqref{eq:phys1}, \eqref{eq:phys2} and \eqref{eq:w0fromfpi}. In the case of
isospin-symmetric fits $\hat{M}$ is given by $M_{ll}$. For $w_0/a$ we take
the Zeuthen version of the gradient flow. In our global fit we
consider terms linear in the $X_l$ and $X_s$ variables. No higher orders are needed, since we
work close to the physical point. We allow these terms to depend on the lattice
spacing, so their coefficients become polynomials of $(a/w_0)^2$, denoted by \(B(a)\) and \(C(a)\) in the following.

Putting all this together, the global fit function is
\begin{equation}
    Y = A(a) + B(a) X_l + C(a) X_s\ ,
\end{equation}
where $Y$ is one of our dimensionless target observables.  We consider
different polynomial orders for \(A\), \(B\), and \(C\), as well as omitting up
to three of the coarsest lattice spacings (out of a total of six) from the
fits. For \(A\) we consider linear, quadratic or cubic polynomials in
\(a^2\). In the linear term we allow for a modification by a factor of
$\alpha_s^\gamma(a)$, with $\gamma$ chosen from the set $\{0,0.5,1.0,1.5,2.0,2.5\}$.
For \(B\) and \(C\), we consider both constant and linear polynomials in
\(a^2\). For each combination of polynomial orders, we consider the lattice
spacing cuts that leave at least one more lattice spacing than the total number
of coefficients contained in \(A\) or \(A'\), as well as at least two more
lattice spacings than the maximum number of coefficients contained in \(B\) or
\(C\).

When presenting results for the observables, we will show continuum
extrapolation plots: results as a function of the $a^2$ for a representative
subset of the fits. On the plots the points are obtained by shifting the
observables on the ensembles to the physical values of the light and strange
masses, using the $B$ and $C$ coefficients from the fit. The fit curves
correspond to the fit functions with only the $A$ terms kept.

\subsection{Distribution of observables}

For a given observable, fit function, lattice spacing cut and other
systematic choices we assign a weight using the Akaike Information Criterion (AIC)
\cite{Akaike1973,Akaike:1974vps,Akaike1978c}
in a modified version as derived in Ref.~\cite{Borsanyi:2020mff}:
\begin{equation}
    w= \exp\left[-\frac{1}{2}\left(\chi^2 + 2n_\mathrm{par}-n_\mathrm{data}\right)\right]\label{eq:aic}\ ,
\end{equation}
computed from the \(\chi^2\) of the fit, the number of parameters
\(n_{\mathrm{par}}\), and the number of data points included in the fit
\(n_{\mathrm{data}}\). The first two terms in the exponent correspond to the
standard AIC, and the last term is introduced to weight fits with a different
number of ensembles, due to our cuts in the lattice spacing. From these
weighted fits we build a probability distribution from which the central value,
statistical and systematic errors for the observable $Y$ are constructed. The
technique is described in detail in \cite{Borsanyi:2020mff}, we briefly
summarize it below.

To estimate the statistical and systematic errors, the fit procedure is
carried out on each jackknife sample with many different choices of
the systematic ingredients. There are two different types of systematics: one
where the different possibilities enter with equal weight (flat-weighted) and
another where they enter with the AIC weight of their corresponding fit
qualities (AIC-weighted). We label collectively the flat-weighted systematics
with the indices $i,j$ and the AIC-weighted with the indices $a,b$. For each
analysis, given by a pair of indices $(i,a)$, we have an average value
$y_{ia}$, a jackknife error $\sigma_{ia}$ and an AIC weight $w_{ia}$ from Equation
\eqref{eq:aic}. From these inputs we construct a probability distribution
function (PDF) for the observable $Y$
\begin{equation}
    \label{eq:pdf}
    \mathrm{PDF}(Y) = \sum_{i,a}\frac{w_{ia}\cdot \mathcal{N}(y_{ia}, \sigma_{ia}; Y)}{\sum_{b} w_{ib} \cdot \sum_{j} 1}\ ,
\end{equation}
which includes both statistical and systematic variations. The statistical
variations are assumed to follow a normal distribution, i.e.\ \(\mathcal{N}(y,
\sigma; Y)\) is a normal PDF with mean $y$ and standard deviation $\sigma$.

The central value of $Y$ is defined by the median of the constructed PDF. The
lower and upper total errors are defined by the standard one-sigma quantiles of
the corresponding cumulative distribution function (CDF), close to
\qty{16}{\percent} and \qty{84}{\percent}.  A further estimate of the lower and
upper errors comes from halving the intervals obtained from the standard
two-sigma quantiles, close to \qty{2}{\percent} and \qty{98}{\percent}.
Any difference between the two estimates is related to deviations of our
distribution away from a normal distribution. To be conservative, for each
quantity we consider here, we take the larger of these two error estimates to
be our final error. The figures attached to this section show the PDF, the
median and corresponding error bands. This procedure gives the total
uncertainty of $Y$. It is also possible to decompose the total error into
statistical and systematic components, and the latter into each individual
systematic ingredient \cite{Borsanyi:2020mff}. 

In Ref.~\cite{Jay:2020jkz} the same PDF as in Equation \eqref{eq:pdf} is used,
however error estimates are constructed using variances of this distribution
instead from quantiles. Also the AIC criterion of \cite{Jay:2020jkz} differs
from ours in Equation \eqref{eq:aic} in the way the lattice spacing cuts are
implemented.
This criterion is derived by assuming that
removing a data point is equivalent to adding to the model a parameter that fits the data point exactly. 
Our AIC criterion is derived by directly computing the
dependence of the Kullback--Leibler divergence \cite{Kullback:1951zyt}
on the number of data points entering the analysis.
For the detailed error budgets given in the following tables we use the
variance approach of \cite{Jay:2020jkz} instead of the procedure used for the
total uncertainty, since
it gives very similar results for much less computational cost.

\subsection{Combining distributions via random sampling}
\label{se:si_analysis_comb}

We also need a technique to perform a stochastic sampling of the
previously described PDFs. This becomes useful when combining several such
PDFs, each with tens of thousands of analyses, where taking into account all
possible combinations would be unfeasible.  We perform an importance-weighted
stochastic sampling of the systematic ingredients. Let us consider two
observables, $Y$ and $Z$, which share some systematic ingredients with the
remaining ones considered independent. Then a random sample is constructed
in the following way:
\begin{enumerate}

    \item We make a common random selection for the systematic ingredients shared by
	$Y$ and $Z$. (In our case the shared ingredients are always
	flat-weighted, so we have a uniform distribution in this step.)

    \item We make a random selection for the remaining independent ingredients
	with a probability given by
	\begin{equation}
	    P(i,a) = \frac{w_{ia}}{\sum_{b} w_{ib} \sum_{j} 1}\ ,
        \end{equation}
	where the shared ingredients $i$ are fixed to the values
	selected in step 1. This is conveniently accomplished by choosing a
	uniform random number $r$ in the $[0,1]$ interval and picking the
	analysis where the CDF built from the $P(i,a)$ values first reaches the
	number $r$. This step has to be done for $Y$ and $Z$ independently.

    \item We build the combination observable from $Y$ and $Z$ using the shared
	and unshared ingredients from steps 1 and 2.

\end{enumerate}
To properly account for statistical correlations, we preserve the jackknife
samples during the whole construction. Repeating these steps $N_R$ times, we
obtain the desired distribution for the combination observable, which has a
single systematic ingredient labelled by the sample index, and each of them
comes with a flat weight of $1/N_R$.  The statistical, systematic and total
errors of the combination can then be obtained by the procedure
described above. A similar sampling technique was recently proposed in
Ref.~\cite{Boyle:2024grr}. 

To obtain the physical value of an observable $Y$ we need the physical
values of $w_0$ and $M_{ss}$ as inputs. The corresponding distributions are to
be taken from the analyses described in Section~\ref{se:si_phys} and we include
them as follows.  First we perform the analysis for $Y$ at two fixed values of
$w_0$ and two fixed values of $M_{ss}$, given by the edges of the central
one-sigma bands of their distributions. We label the outcome of these analyses
with $Y_k$ with $k=1,\dots4$.  Then we use the above importance-sampling to
select random samples, $w_{0}^r$ and $M_{ss}^r$. For a given sample, we perform a
bilinear interpolation of the fit values and their corresponding \(\chi^2\) values
from the $Y_{k}$ obtained at fixed values of $w_0$ and $M_{ss}$ to the sampled
values $w_{0}^r$ and $M_{ss}^r$. Finally, we compute the weights corresponding to the
interpolated \(\chi^2\) values and perform the above importance-sampling once
more to obtain the desired sample $Y^r$. After repeating $N_R$ times, we obtain the
corresponding distribution, which can be handled in the usual way. We find that with
$N_R=10^6$ the stochastic error is well below the uncertainties of our
observables, so we use this value in our combinations.

%% file: si_win.tex
\section{Window observables}
\label{se:si_win}

The lattice contribution of this work consists of computing the leading-order
hadronic contribution to the muon magnetic moment, $a_\mu^\mathrm{LO-HVP}$,
from zero Euclidean distance to $t_\mathrm{cut}=2.8~\mathrm{fm}$.  Both in our
2017 \cite{Borsanyi:2017zdw} and in our 2020 work \cite{Borsanyi:2020mff} we
used lattice results up to a given temporal distance, beyond which upper
and lower bounds were used to constrain the long-distance tail of the light and
disconnected correlators. We chose
$t_\mathrm{cut}=3~\mathrm{fm}$ and $4~\mathrm{fm}$ in 2017 and 2020,
respectively. In the present paper we improve on this procedure and beyond
$t_\mathrm{cut}=2.8~\mathrm{fm}$ we use state-of-the-art results from a
data-driven approach. The reasons for choosing this value of $t_\mathrm{cut}$
and details of the approach are described in Section~\ref{se:si_tail}.

Regarding $a_\mu^\mathrm{LO-HVP}$ and its various contributions we use the
definition and notations from our 2020 work. In particular, since we consider
only the LO-HVP contribution, we drop the superscript and multiply the result
by $10^{10}$, i.e.~$a_\mu$ stands for $a_{\mu}^\mathrm{LO-HVP} \times 10^{10}$
throughout the Supplementary Information. We will consider several different
window observables, where we restrict the integration in Euclidean time to a
region between $t_1$ and $t_2$ with the standard window function defined in
Ref.~\cite{RBC:2018dos}. We use the notation $15-19$ for a window between
$t_1=1.5~\mathrm{fm}$ and $t_2=1.9~\mathrm{fm}$ and accordingly for the other
windows. An important feature of the definition is that two adjacent windows
can be joined by simple addition, like $04-10$ and $10-28$ equals $04-28$.

A minor difference compared to our 2020 work concerns the splitting up of
$a_\mu$ into nonperturbative and perturbative parts. Back then we introduced a
momentum $Q_\mathrm{max}=\sqrt{3}~\mathrm{GeV}$ to separate the two regions:
below $Q_\mathrm{max}$ we used the lattice computation, above that perturbation
theory. Here we remove this separation by sending $Q_\mathrm{max}$ to infinity.
In practice this change only affects the $00-04$ window.

In this work we compute all major contributions to $a_\mu$: light, strange,
charm and disconnected\footnote{By disconnected contributions we refer to the
disconnected diagrams of light and strange quarks.}.  In our earlier work for
each individual flavour component we considered the contributions to $a_\mu$
obtained by integrating the correlator over all Euclidean times.  Here we cut
the integral via a one-sided window function that ends at $t=t_\mathrm{tail}$,
eliminating a noisy contribution with large finite-size and taste-violation
effects. We also split the time into several windows. This allows the fit
functions to be different in the different windows, giving us more control
over statistical and systematic errors.  Regarding finite-size effects we
perform the continuum extrapolations in a finite box, called the reference
box, with $L_\mathrm{ref}$ spatial and $T_\mathrm{ref}$ temporal extent. To
get the infinite-volume result we compute corrections in dedicated lattice
simulations, as described in Section~\ref{se:si_finvol}.

In the following subsections we begin by presenting our blinding procedure.
Then we show results for several windows that are available in the
literature.
In particular we consider the short-distance
window $00-04$, the intermediate-distance window $04-10$, and also a window at
longer distances $15-19$. For some of these we use a subset of the
configurations. On these observables we show our fit procedure in detail. The
results can then be also compared with those in the literature\footnote{We
perform the comparisons in the isospin-symmetric theory, where a scheme
ambiguity is present, see Section~\ref{se:si_phys}. This has been studied
numerically in the recent work of RBC \cite{RBC:2023pvn}. In case of the
$04-10$ window an ambiguity of $\delta a_{\mu,04-10}^\mathrm{light}=
0.10(24)(07)$ was found, much smaller than current uncertainties.}. Afterwards
we explain how we obtain the lattice result for the window $00-28$. Then we
describe how the rest of the contributions (strange, charm and isospin
breaking) were obtained. Finally, we combine them with the tail contribution to
get the all-flavour result for $a_\mu$. In this final section we also give
all-flavour results in the intermediate $04-10$ and long-distance $10-\infty$
windows.

\phantomsection\label{se:si_win:table} The results for the different windows
will be presented in Tables and here we give their general description. In the
first row, continuum extrapolations as a function of $a^2$ are shown.  For
readability, the points are projected to the physical quark masses using an
appropriate fit from the plot. The normalized weight of each fit is
indicated via the plotted colour scale. For even better readability we also make
the data points and fits appearing in the plots available under Reference~\cite{web:zenodo}. In the second row we
show with red the probability distribution function including both statistical
and systematic variations.  The median is given by the blue vertical line, the
1-sigma and 2-sigma bands, shown in green/yellow, contain 68.3\% and 95.4\% of
the distribution, centred at the median. Where results from other
collaborations are available in the literature, we show them in another figure
in the second row. The references to the other works are then given in the
text. If available, the statistical error is given as an inner error-bar.
The table in the third row contains the number of fits, the percentage of
fits having a P-value larger than $0.1$, the median of the observable, the
total, statistical and systematic uncertainties and an error budget.
All results correspond to the
reference box-size, $L_\mathrm{ref}=6.272~\mathrm{fm}$, except in the
comparison plot, where infinite-volume results are given.

\subsection{Blinding}

To eliminate human bias we performed our analysis in a blinded fashion with
independent crosschecks at each step.
When constructing the window observables from the current correlation functions
they are multiplied by a window dependent blinding factor. The full analysis
from fitting the window observables to performing the error analysis on the
global fits is implemented and performed independently by two or more groups,
who do not know the values of the blinding factors. The results of the analysis
are then crosschecked at each stage. In every case consensus was reached. Once
the analysis procedures were finalized with full crosschecks, the blinding
factors were revealed and divided out to obtain the final results.  A
special script was written, which carried out the unblinding, added the results
to the manuscript and produced the plots of the paper, without the need for
human intervention. Our procedure guarantees that no bias was involved in
obtaining the final results.

After unblinding, and during the peer review process, concerns were raised 1.
about consistency of window functions for the isospin-breaking corrections; 2.
about the possibility of an unknown excited-state systematic in the lattice
determination of the $\Omega$ mass, as described in Section~\ref{se:si_omega};
and 3. about the validity of continuum extrapolations with polynomials of
$\Delta_{KS}(a)$. Based on these criticisms, 1. the isospin-breaking
corrections were re-computed with the exact window functions; 2. the value of
$w_0$ scale was instead calculated in an alternate blinded analysis based upon
the experimental muonic pion decay rate, see Section~\ref{se:si_fpi}; 3.
the window observables were re-calculated in an alternate blinded analysis
replacing $\Delta_{KS}(a)$ by $a^2\alpha_s(a)^\gamma$, as described in
Section~\ref{se:si_analysis}. The values of the isospin-breaking corrections, the
$w_0$ and the window observables from these new determinations were
computed blind and not known at the time of making our decisions about the
final procedure, but the original isospin-breaking corrections, the $w_0$
and window values from the old analyses were.

In response to referee concerns, additional uncertainties were added to the
long-distance tail contributions discussed in Section~\ref{se:si_tail} after
unblinding.  These only increased the uncertainty of this small contribution without
changing the central value.

The net effect of these four changes on the final result
was to increase the central value by 1.0 (0.14\% of the final value), and increase the
uncertainty by 0.1 (3\% of the total uncertainty).  This differences are very small, and
much smaller than the uncertainty of our result.

\subsection{Short-distance window}

\begin{table}[p]
    \centering
    \includegraphics{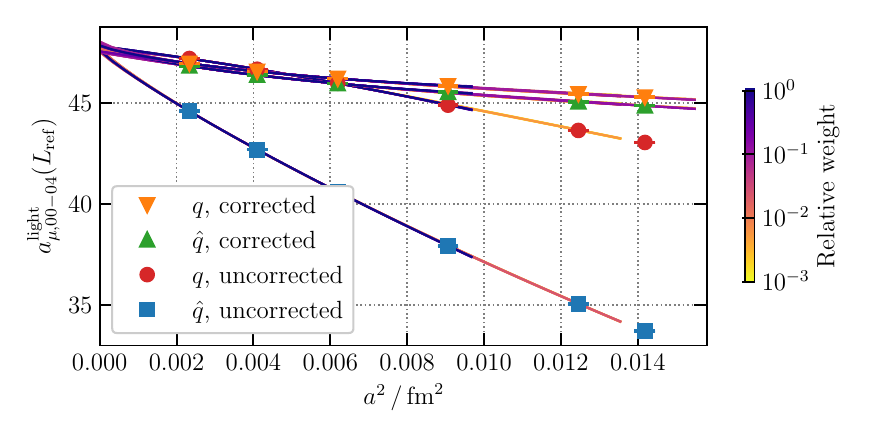}\\
    \includegraphics{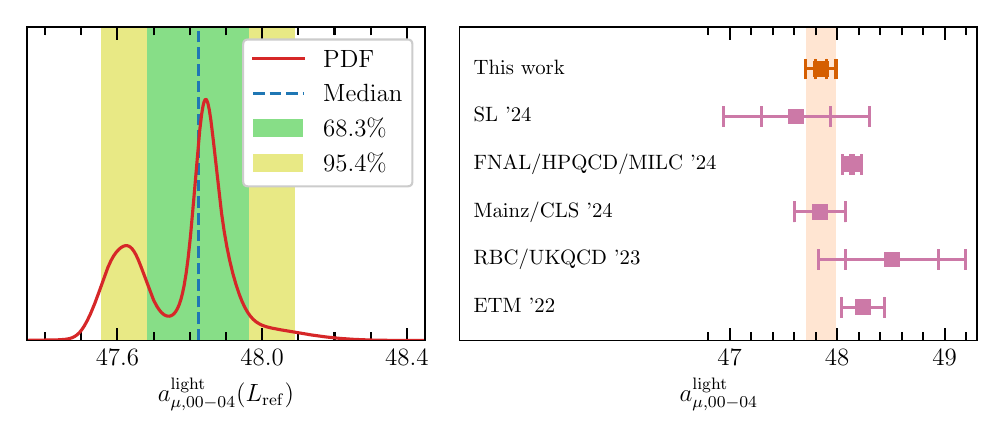}\\
    \input{figures/tables/li_00_04/tabular}
    \caption
    {
	\label{ta:res_li_00_04} Light-connected
	window observable $a_{\mu,00-04}^\mathrm{light}$. The plot conventions
	are described in the first part of Section~\ref{se:si_win:table}.
	The continuum extrapolations are shown with two different kernel
	functions, denoted by $q$ and $\hat{q}$. The probability distribution
	function displays two dominant peaks, corresponding to the variation
	between $q$ and $\hat{q}$ in the uncorrected case. We compare our
	result with others from the
	literature~\cite{ExtendedTwistedMass:2022jpw,RBC:2023pvn,Kuberski:2024bcj,MILC:2024ryz,Spiegel:2024dec}.
    }
\end{table}

\begin{table}[p]
    \centering
    \includegraphics{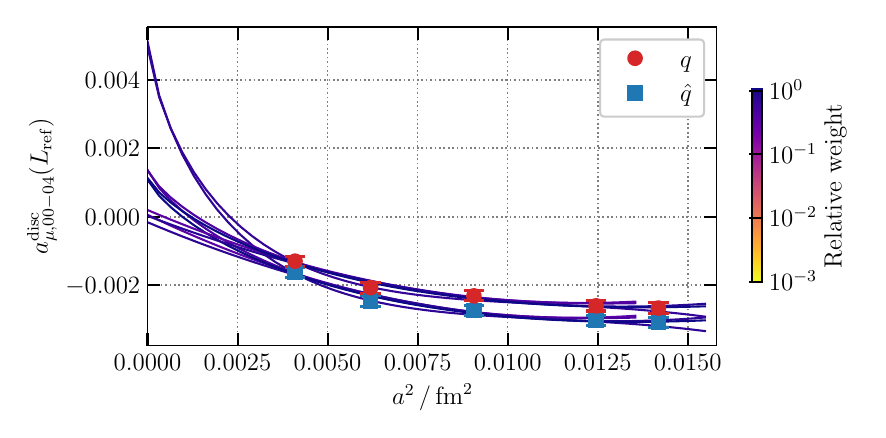}\\
    \includegraphics{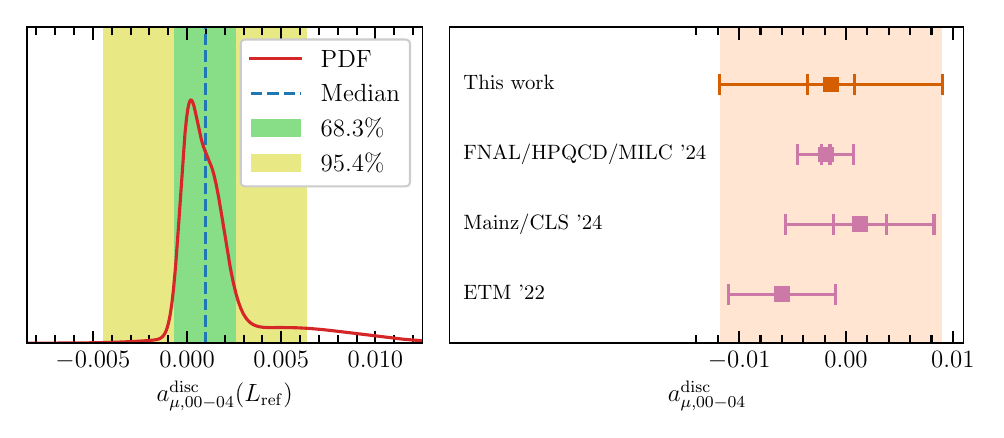}\\
    \input{figures/tables/di_00_04/tabular}
    \caption
    {
	\label{ta:res_di_00_04} Disconnected
	window observable $a_{\mu,00-04}^\mathrm{disc}$. The plot conventions
	are described in the first part of Section~\ref{se:si_win:table}.
	The continuum extrapolations are shown with two different kernel
	functions, denoted by $q$ and $\hat{q}$. We compare our result with
	others from the
	literature~\cite{ExtendedTwistedMass:2022jpw,Kuberski:2024bcj,
	MILC:2024ryz}.
    }
\end{table}

The short distance window is plagued by lattice artefacts that are
logarithmically enhanced compared to the usual case with on-shell observables
\cite{Ce:2021xgd}. These artefacts arise from arbitrary small time separations:
$a_{\mu,00-04}$ is not an on-shell quantity. We can improve the behaviour by
removing part of these discretization errors using lattice perturbation theory.
We define our tree-level improved observable by the transformation
\begin{equation}
    a^\mathrm{light}_{\mu,00-04} \to a^\mathrm{light}_{\mu,00-04} + a_{\mu,00-04}^\mathrm{tree}(0) - a_{\mu,00-04}^\mathrm{tree}(a)\ ,
\end{equation}
where $\mathrm{tree}$ stands for leading-order, infinite-volume, massless
staggered perturbation theory.  The positive effect of the improvement can be
seen in the top figure of Table~\ref{ta:res_li_00_04} for the light case. We
only apply this improvement to the light contribution, since the
disconnected component vanishes at this order in perturbation theory.

We also investigate the lattice artefacts by varying the kernel function the
current correlator is multiplied with.  In particular we change the square
bracket in Equation (65) of \cite{Borsanyi:2020mff}
\begin{equation}
    \left[ t^2 - \frac{4}{(aQ)^2}\sin^2\left(\frac{aQt}{2}\right)\right] \to
    \left[ t^2 - \frac{4}{(a\hat{Q})^2}\sin^2\left(\frac{aQt}{2}\right)\right]\ ,
\end{equation}
where $Q$ is the momentum of the hadronic vacuum polarization and $\hat{Q}$
denotes the lattice momentum defined by $\hat{Q}=2\sin(aQ/2)/a$. The
obtained data points are also shown in the top figure of
Table~\ref{ta:res_li_00_04} for the light and that of Table~\ref{ta:res_di_00_04} for
the disconnected contribution. Lattice artefacts are larger with $\hat{Q}$.
In tree-level perturbation theory one can show that the coefficient of the
logarithmically enhanced cutoff effect has a different sign, if one uses $Q$ or
$\hat{Q}$ in the kernel. The light results with $\hat{Q}$ approach the continuum
limit from below, whereas the ones with $Q$ show a characteristic turnover
behaviour. 

In our analyses we include data with and without tree-level improvement and
also using $Q$ and $\hat{Q}$.  Tree-level improvement may not remove
all logarithmic cutoff effects, so we also include fit functions with
logarithmic terms. The $a$-dependent part of our fit functions is as follows:
\begin{equation}
\label{eq:a2log}
    A_2 a^2+A_4 a^4\ ,\quad
    A_2 a^2+A_l a^2 \log(a^2/w_0^2)\ ,\quad
    A_2 a^2+A_4 a^4 + A_l a^2 \log(a^2/w_0^2)\ .
\end{equation}
Note that the true asymptotic form of the discretization errors is
unknown, since there is no theory analogous to Symanzik's for the
discretization errors of this short-distance quantity. Therefore, the choices
given in \eqref{eq:a2log} are phenomenological.

The results for in the light case can be found in Table~\ref{ta:res_li_00_04}.
To this we can add the finite-size correction from Table~\ref{ta:fv_res} to get
\begin{equation}
    a_{\mu,00-04}^\mathrm{light}= \valuesLightSd
\end{equation}
with statistical, systematic and total errors. This is our infinite-volume
result for the light contribution to the short-distance window, which is
compared in the Table to previous lattice results of ETM
\cite{ExtendedTwistedMass:2022jpw}, RBC/UKQCD \cite{RBC:2023pvn}, Mainz/CLS
\cite{Kuberski:2024bcj}, Fermilab Lattice/HPQCD/MILC \cite{MILC:2024ryz} and
Spiegel \& Lehner \cite{Spiegel:2024dec}. The results for the disconnected case
can be found in Table~\ref{ta:res_di_00_04}.  Note that the number of fits is
smaller than in the light case because we have measurements on fewer lattice
spacings. Again we perform the finite-size correction and obtain
\begin{equation}
    a_{\mu,00-04}^\mathrm{disc}= \valuesDiscoSd
\end{equation}
as the infinite-volume result. The relatively large systematic error comes from
the uncertainty of the $I=0$ finite-size effect. This value is compared to the
results of ETM \cite{ExtendedTwistedMass:2022jpw}, Mainz/CLS
\cite{Kuberski:2024bcj} and Fermilab Lattice/HPQCD/MILC \cite{MILC:2024ryz} in
a figure of the Table.

\begin{figure}[t]
    \centering
    \includegraphics{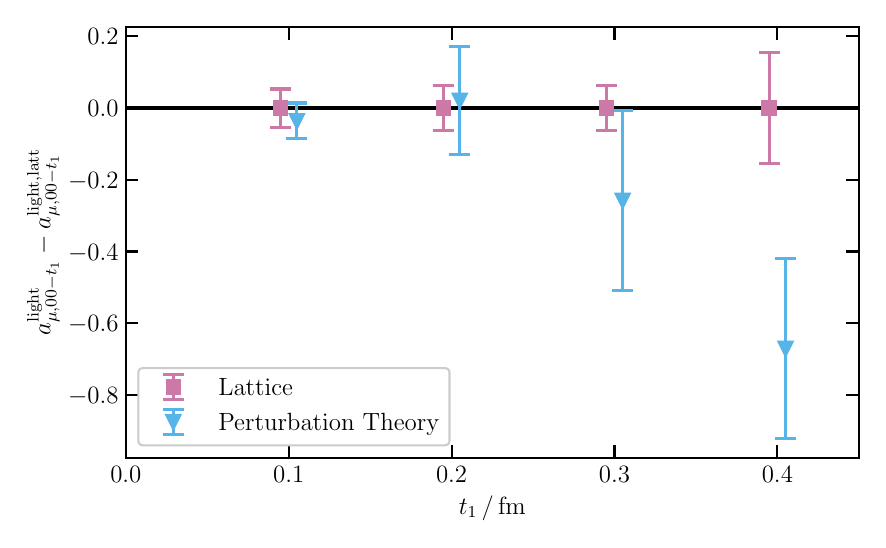}
    \caption
    {
	Comparison between the lattice results and those obtained from
	perturbation theory for the short distance window observables
	$a_{\mu,00-t_1}^\mathrm{light}$ as a function of $t_1$. The points have
	the central value of the lattice results subtracted off, and a slight
	horizontal offset for readability.
    }
    \label{fi:sd_vs_pert}
\end{figure}

In order to verify the validity of our continuum extrapolation procedure in
the short distance region, we compare our lattice results with the predictions
of perturbation theory. For this purpose we calculated light windows
$00-01$, $00-02$ and $00-03$ using the same procedure as was used for the
short distance window $00-04$. Additionally, we calculated the same
quantities, using $N_f=4$ massless perturbation theory
\cite{Chetyrkin:2010dx}. We change the scale parameter $\mu$ in the range
$[0.8, 2]\,\mathrm{GeV}$ and this variation enters as the systematic
uncertainty of the perturbative calculation. In Figure~\ref{fi:sd_vs_pert} we
plot the lattice and perturbative results for the light
windows $a_{\mu,00-t_1}^{\mathrm{light}}$ as the function of $t_1$. It
can be clearly seen that for shorter distances $00-01$, $00-02$ and $00-03$
perturbation theory agrees with the lattice data.

\subsection{Intermediate-distance window}

\begin{table}[p]
    \centering
    \includegraphics{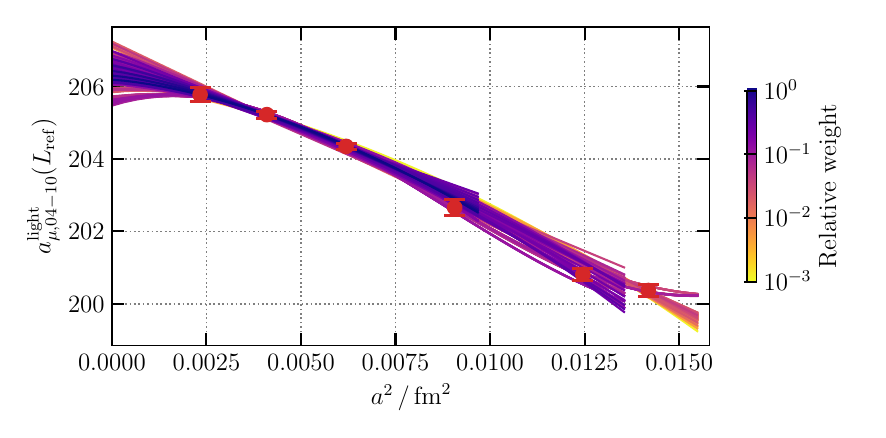}\\
    \includegraphics{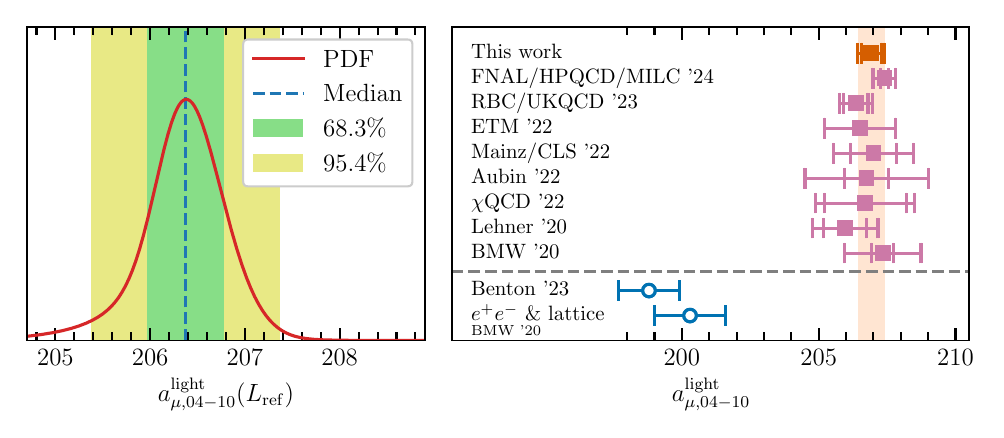}\\
    \input{figures/tables/li_04_10/tabular}
    \caption
    {
	\label{ta:res_li_04_10} Light-connected
	window observable $a_{\mu,04-10}^\mathrm{light}$. The plot conventions
	are described in the first part of Section~\ref{se:si_win:table}.  We
	compare our result with others from the literature, both
	lattice~\cite{Borsanyi:2020mff,Lehner:2020crt,
	Wang:2022lkq,Aubin:2022hgm,Ce:2022kxy,
	ExtendedTwistedMass:2022jpw,RBC:2023pvn,MILC:2024ryz} and
	data-driven~\cite{Borsanyi:2020mff,Benton:2023dci}.
    }
\end{table}

\begin{table}[p]
    \centering
    \includegraphics{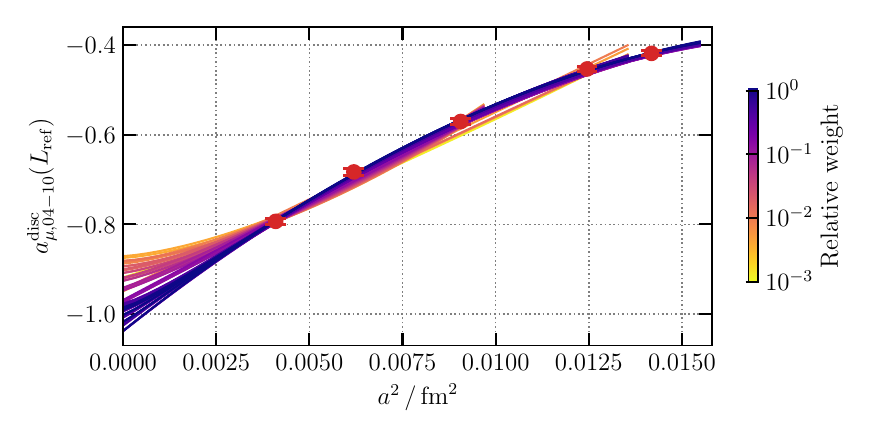}\\
    \includegraphics{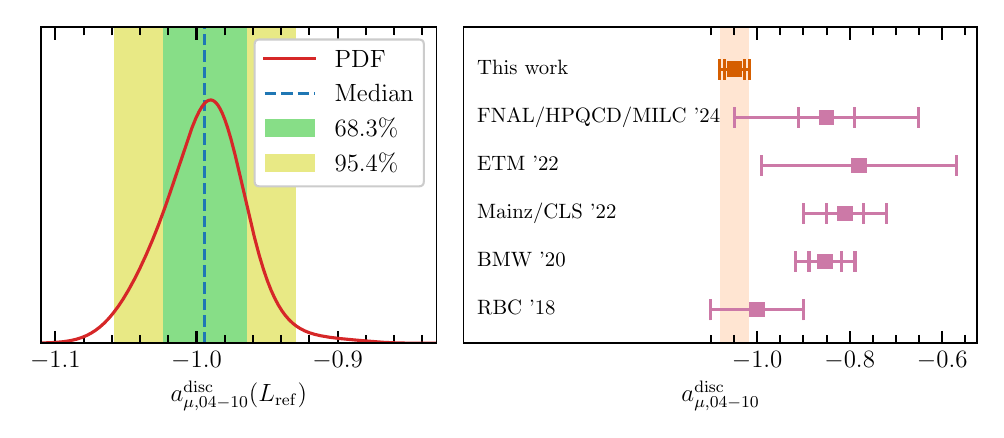}\\
    \input{figures/tables/di_04_10/tabular}
    \caption
    {
	\label{ta:res_di_04_10} Disconnected
	window observable $a_{\mu,04-10}^\mathrm{disc}$.  The plot conventions
	are described in the first part of Section~\ref{se:si_win:table}.  We
	compare our result with others from the literature~\cite{RBC:2018dos,
	Borsanyi:2020mff,Ce:2022kxy,ExtendedTwistedMass:2022jpw, MILC:2024ryz}.
    }
\end{table}

Let us consider first our results for the light-connected part of the so-called
intermediate-distance window, the window $04-10$ in our notation. This is a
quantity designed to compare different lattice calculations in a controlled
way, since its systematic errors are usually well under control. The results
are shown in Table~\ref{ta:res_li_04_10}. The dominant uncertainties are the
scale setting error and the one arising from the variation of the
polynomial order of the fit functions. After applying the finite-size
correction we get the infinite-volume result as
\begin{equation}
    \label{eq:lightid}
    a_{\mu,04-10}^\mathrm{light}= \valuesLightId\ .
\end{equation}
The comparison with the literature is shown in the figure of the second row of
Table~\ref{ta:res_li_04_10} where, besides our current and previous results,
those of Lehner and Meyer \cite{Lehner:2020crt}, $\chi$QCD \cite{Wang:2022lkq},
Aubin et al \cite{Aubin:2022hgm}, Mainz/CLS \cite{Ce:2022kxy}, ETM
\cite{ExtendedTwistedMass:2022jpw}, RBC/UKQCD \cite{RBC:2023pvn} and
Fermilab Lattice/HPQCD/MILC
\cite{MILC:2024ryz} are given. Also shown are two results from the
data-driven approach: the determination from our 2020 work, where we subtracted
the other-than-light lattice contributions from the total data-driven value,
and a recent pure data-driven computation from Benton et al
\cite{Benton:2023dci}.

In order to quantify the improvement on the uncertainty of
$a_{\mu,04-10}^\mathrm{light}$ due to the new $a=0.048\,\text{fm}$ ensemble, we
compare the results obtained with and without that ensemble. For consistency,
the latter is computed using the distribution of $w_0$ obtained also without
the new ensemble. We obtain $\valuesLightIdFvNomon$ without finite volume
corrections, to be compared with the median number and errors in
Table~\ref{ta:res_li_04_10}. Accordingly, the error reduction due to the finest
ensemble is $\valuesLightIdFvEred$.

In the disconnected case the results are given in Table~\ref{ta:res_di_04_10}.
The analysis gives
\begin{equation}
    \label{eq:discoid}
    a_{\mu,04-10}^\mathrm{disc}= \valuesDiscoId\ ,
\end{equation}
again after the application of the finite-size correction. We also provide a
comparison to the results of RBC/UKQCD \cite{RBC:2018dos}, Mainz/CLS
\cite{Ce:2022kxy}, ETM \cite{ExtendedTwistedMass:2022jpw} and Fermilab
Lattice/HPQCD/MILC \cite{MILC:2024ryz}.  The difference compared to the value
of our 2020 work comes from the addition of a new lattice spacing and that now
we also include quadratic fits in the analysis - previously we only used linear
continuum extrapolations.

\subsection{Long-distance window $15-19$}

\begin{table}[p]
    \centering
    \includegraphics{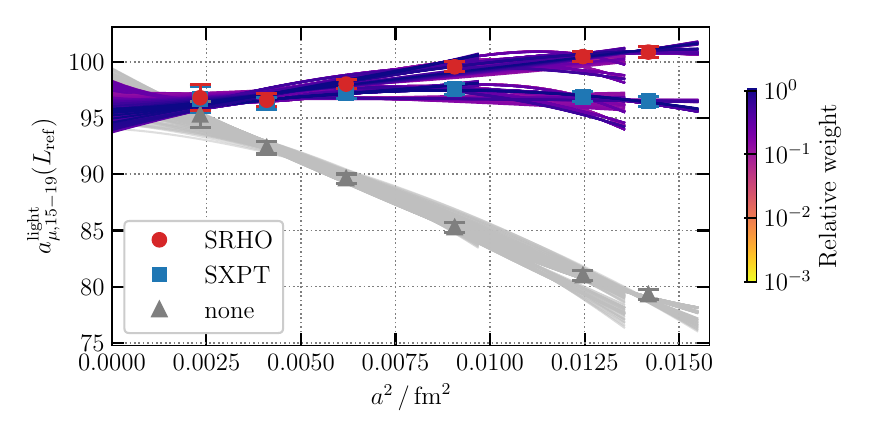}\\
    \includegraphics{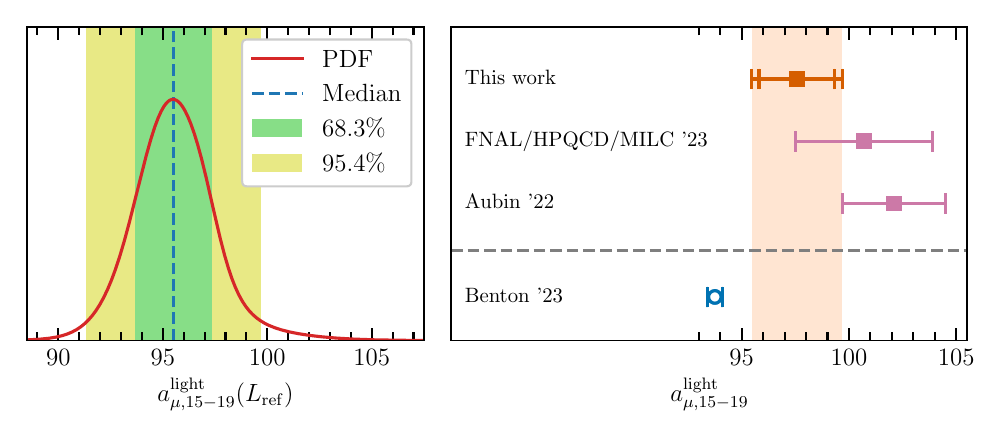}\\
    \input{figures/tables/li_15_19/tabular}
    \caption
    {
	\label{ta:res_li_15_19} Light-connected
	window observable $a_{\mu,15-19}^\mathrm{light}$. The plot conventions
	are described in the first part of Section~\ref{se:si_win:table}. The
	continuum extrapolations are shown with no, NNLO XPT and SRHO taste
	improvements. We compare our results with others from the literature,
	both lattice~\cite{Aubin:2022hgm,FermilabLatticeHPQCD:2023jof} and
	data-driven~\cite{Benton:2023dci}.
    }
\end{table}

A longer-distance window was proposed by Aubin et al \cite{Aubin:2022hgm}, for
the range $1.5-1.9~\mathrm{fm}$, which is called ``W2'' there and corresponds
to $15-19$ in our notation. A feature of this window is the strong
presence of taste-breaking effects.  Here we apply the same taste improvement
procedure that we had in our 2020 work.
For obtaining the central value of our result, we use the staggered version of
the rho-pion-gamma model (SRHO) of Jegerlehner and Szafron
\cite{Jegerlehner:2011ti}, originally proposed by Sakurai
\cite{Sakurai:1960ju}. In the context of removing taste-breaking effects in
$a_{\mu}$, it has already been used by the HPQCD collaboration
\cite{Chakraborty:2016mwy}.
In order to assign a systematic error associated with the taste improvement,
we also compute the taste-breaking effects using next-to-next-to-leading order
staggered chiral perturbation theory (NNLO SXPT)
\cite{Lee:1999zxa,Sharpe:2004is,Aubin:2006xv,Aubin:2019usy}.

The data with these improvements and also without any improvement are shown in
the top figures of Table~\ref{ta:res_li_15_19}. There is a strong non-linearity
in the unimproved data, that gets eliminated by the improvement. In our
analysis we use the SRHO model to get the central value of our result.  As a
systematic error associated with the taste improvement we assign the difference
of the result obtained with NNLO SXPT and SRHO.  We also assign a systematic to
the choice of the starting point of the taste improvement by performing the
analysis with two different values of this point, $1.2~\mathrm{fm}$ and
$1.4~\mathrm{fm}$.  Due to taste improvement the number of fits is higher than
it is in the $04-10$ window. We obtain
\begin{equation}
    a_{\mu,15-19}^\mathrm{light}= \valuesLightLdmilc
\end{equation}
in the infinite-volume limit. This value, together with those of Fermilab
Lattice/HPQCD/MILC \cite{FermilabLatticeHPQCD:2023jof} and of Aubin et al
\cite{Aubin:2022hgm}, is shown in the second row of
Table~\ref{ta:res_li_15_19}.

\subsection{Long-distance window $10-28$}

\begin{table}[p]
    \centering
    \includegraphics{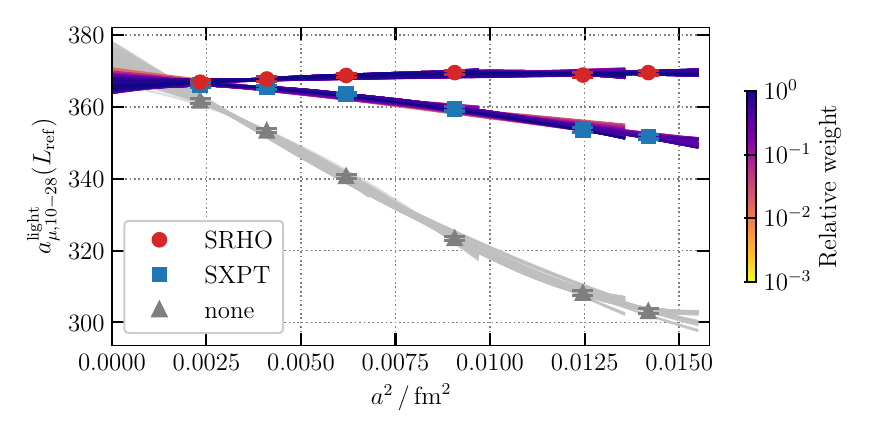}\\
    \includegraphics{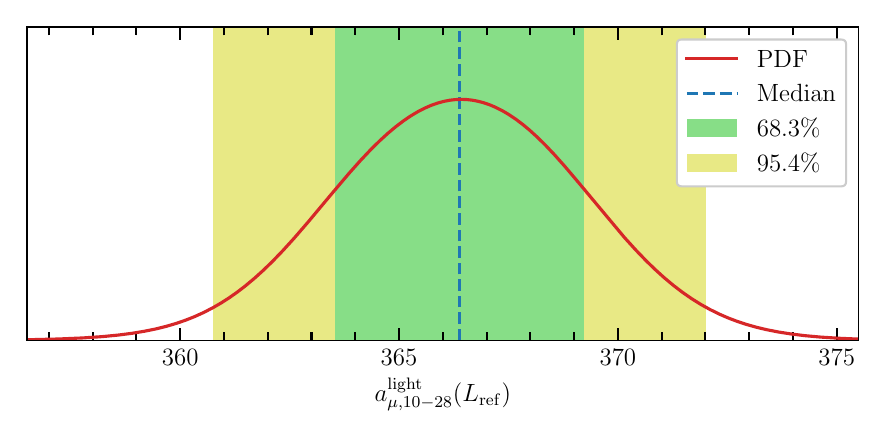}\\
    \input{figures/tables/li_10_28/tabular}
    \caption
    {
	\label{ta:res_li_10_28} Light-connected
	window observable $a_{\mu,10-28}^\mathrm{light}$. The plot conventions
	are described in the first part of Section~\ref{se:si_win:table}. The
	continuum extrapolations are shown with no, NNLO XPT and SRHO taste
	improvements.
    }
\end{table}

\begin{table}[p]
    \centering
    \includegraphics{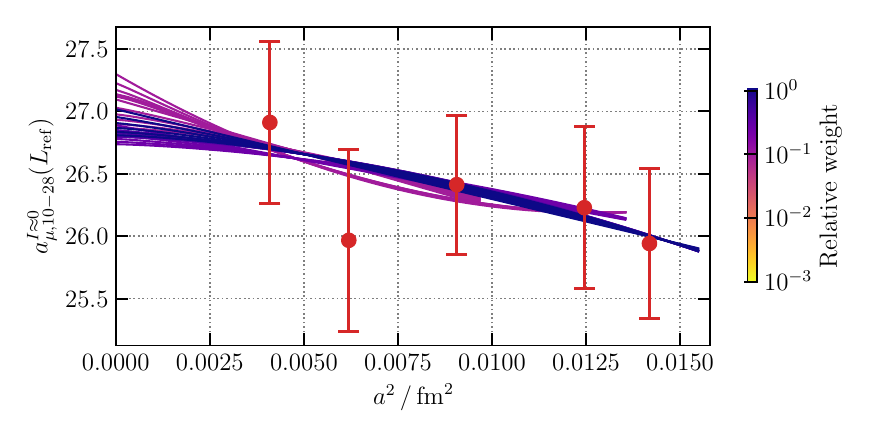}\\
    \includegraphics{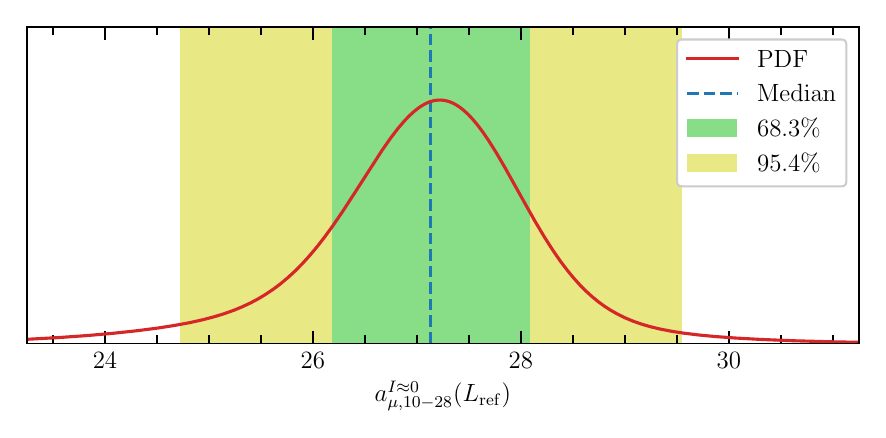}\\
    \input{figures/tables/i0_10_28/tabular}
    \caption
    {
	\label{ta:res_i0_10_28} $I\approx 0$ window
	observable $a_{\mu,10-28}^\mathit{I\approx 0}$. The plot conventions
	are described in the first part of Section~\ref{se:si_win:table}.
    }
\end{table}

We also compute the light-connected $10-28$ window, which, together
with contributions from other flavours and isospin-breaking, will be
complemented by a data-driven determination of the $28-\infty$ window to arrive at a
hybrid result of the all-flavour, $10-\infty$ window. The analysis of the
$10-28$ window proceeds the same way as above, and the results are presented in
Table~\ref{ta:res_li_10_28}. Regarding the disconnected contribution, we fit a
single window $10-28$ and use the following combination:
\begin{equation}
    \label{eq:iapprox0}
    a_{\mu,10-28}^\mathit{I\approx 0}\equiv \tfrac{1}{10}a_{\mu,10-28}^\mathrm{light} + a_{\mu,10-28}^\mathrm{disc}\ .
\end{equation}
This definition combines the light and disconnected contributions as they
appear in the $I=0$ channel. The total $I=0$ also contains the
contributions of the more massive $s,c,\dots$ quarks, which are not included in
$a_{\mu,10-28}^\mathit{I\approx 0}$.  The advantage of this observable is that
the large taste-breaking and finite-size effects of the $I=1$ channel are
cancelled.  The results are shown in Table~\ref{ta:res_i0_10_28}. The continuum
extrapolation is almost completely flat; the largest source of error here is
statistical.

Using the sampling procedure of Section~\ref{se:si_analysis_comb} we combine the
light-connected and the $I\approx0$ observable as
\begin{equation}
    a_{\mu,10-28}^\mathrm{light+disc}= \tfrac{9}{10}a_{\mu,10-28}^\mathrm{light} + a_{\mu,10-28}^\mathit{I\approx 0}
\end{equation}
also taking into account the distributions of our scale-setting quantities
$w_0$ and $M_{ss}$. We obtain
\begin{equation}
    \label{eq:amuw_10_28}
    a_{\mu,10-28}^\mathrm{light+disc}= \valuesLipdiLdbody
\end{equation}
in infinite volume, where we also added the finite-size corrections, both to the
$I=1$ and $I=0$ values, from Table~\ref{ta:fv_res}.

\subsection{All-distance window $00-28$}
\label{sec:all-distance}

Our procedure for the $00-28$ window introduces breakpoints at $0.4$~fm,
$2.0$~fm and $2.4$~fm time separations, the corresponding window
observables are denoted:
\begin{equation}
    a_{\mu,00-04}\ ,\quad
    a_{\mu,04-20}\ ,\quad
    a_{\mu,20-24}\ ,\quad
    a_{\mu,24-28}\ .
\end{equation}
We perform the analyses for these observables and sum them up to get the
$00-28$ contribution. The first one, $00-04$, is the short distance window,
which we have already presented. The other three windows are fitted together.
In the end, we combine the short distance window with the three others and also
with the distributions of $w_0$ and $M_{ss}$, as described in
Section~\ref{se:si_analysis_comb}.

\begin{figure}
    \centering
    \includegraphics[width=0.85\textwidth]{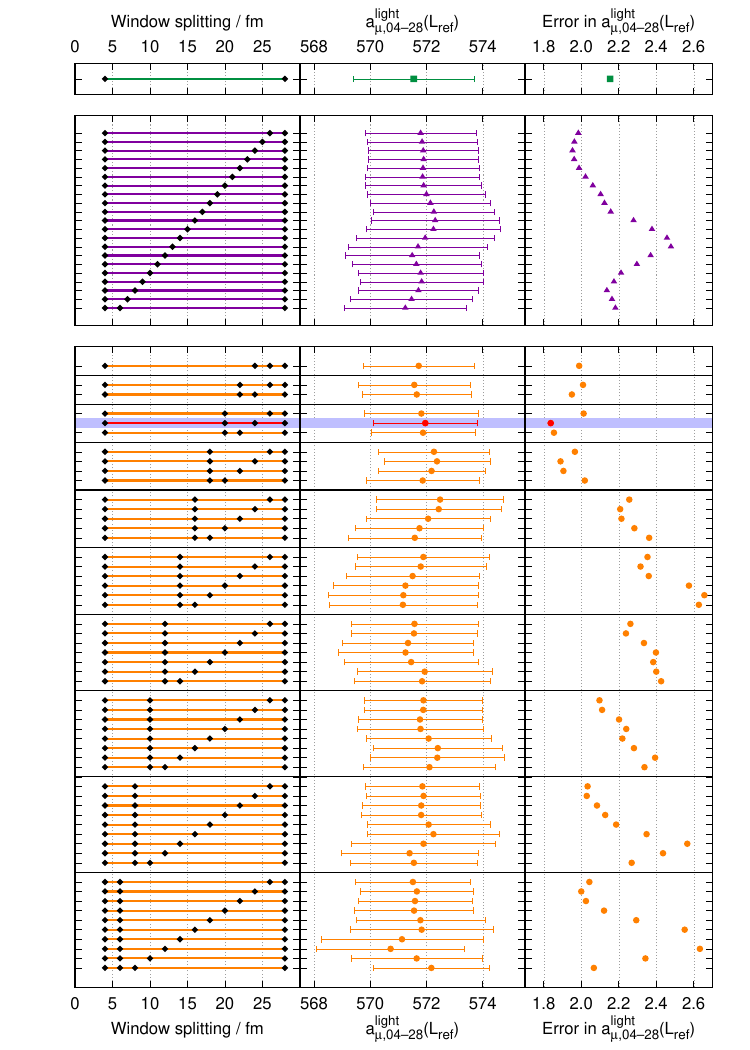}
    \caption
    {
	Value and error of the light-connected window observable
	$a_{\mu,04-28}^\mathrm{light}$ depending on the fitting procedure.
	Each line corresponds to a particular splitting of the window $04-28$.
	The green point (first panel) corresponds to fitting the window $04-28$
	without splitting. The purple points (second panel) show the values and
	errors when splitting into two windows was used. The orange points
	(third panel) show the case of splitting the window $04-28$ into three
	parts.  The first column demonstrates the type of splitting. The second
	column shows the value and error of the result of the particular
	fitting procedure. The third column shows the error once again for
	better readability. Our choice, the splitting into windows $04-20$,
	$20-24$ and $24-28$, is highlighted with a blue shaded band and
	coloured red.  Note that the shown errors do not include the
	uncertainties arising from the scale setting,
	which is actually the dominant source of systematics in
	this window.
    }
    \label{fi:win_select}
\end{figure}

In choosing the times at which the different windows begin and end, one
must make sure that they do not become too narrow, to avoid lattice
artefacts, nor too numerous, to avoid problems associated with strong
correlations. We allow the fit functions to be different in each window; the
$\chi^2$ of the fit includes the residuals and also their correlation matrix.
We then loop over all possible combinations of the fit functions and apply our
standard analysis procedure, as described in Section~\ref{se:si_analysis}.
There is a clear advantage of this procedure over fitting the total $04-28$
range at once.  Namely, we can have high-order fit functions at shorter
distances, where the data is precise, while we can use low-order fit functions
at longer distances, where the data is noisy. The favourable combinations are
selected automatically by the AIC weight.

We perform the multi-window analysis for different partitions of the $04-28$
interval, varying the number of windows and the times at which they are joined
(see Figure~\ref{fi:win_select}). We then monitor the continuum extrapolation
errors. For this study, the uncertainty coming from scale-setting is irrelevant
and therefore not included. However, we note that it still is the dominant
source of systematic uncertainty in the $04-28$ window. When using two
windows instead of one, we find the optimal split to be at $2.4$~fm, which decreases
the error by about 13\%. We also look at splitting the interval into three
windows and find a small improvement with respect to the two-window case. As a
result, we choose the three windows, $04-20$, $20-24$ and $24-28$, for our
final analysis.  The window-independent systematics like scale setting,
pseudoscalar fit ranges or taste-improvement procedure give us $256$
variations that are shared among the three windows. Then in each window we have
$198$ different fit functions, lattice spacing cuts,  and so on.  We use
all possible combinations of these, so we perform $256\times198\times
198\times 198$ fits altogether.  Representative continuum extrapolations,
histograms and error budgets are shown in Table~\ref{ta:res_li_04_28}.

\begin{table}[p]
    \centering
    \includegraphics{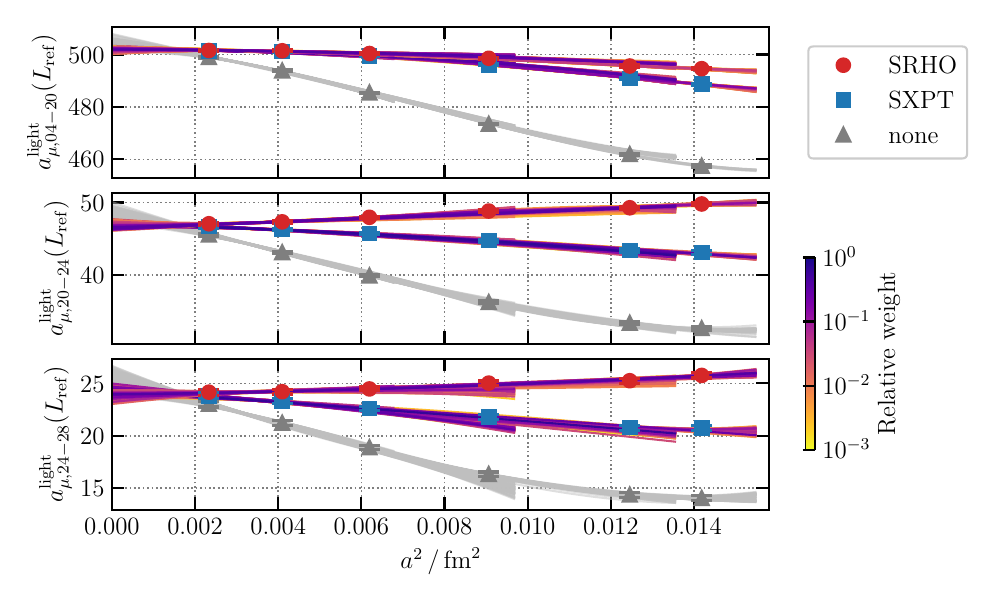}\\
    \includegraphics{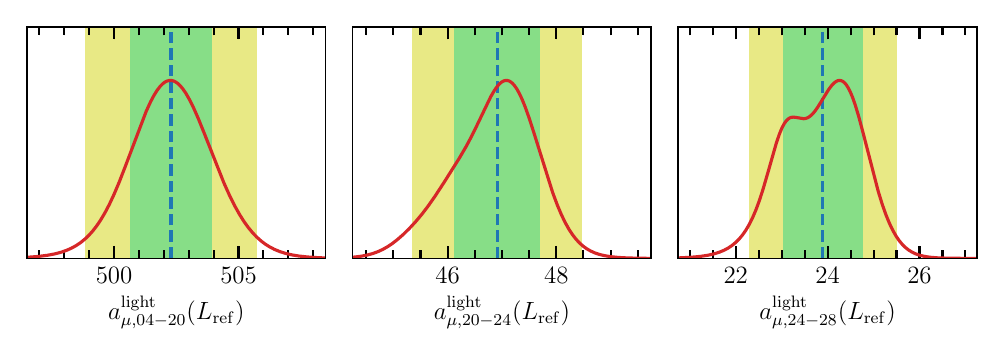}\\
    \input{figures/tables/li_04_28/tabular}
    \caption
    {
	\label{ta:res_li_04_28} Light-connected
	window observable $a_{\mu,04-28}^\mathrm{light}$ obtained as sum of
	three windows $04-20$, $20-24$ and $24-28$, fitted together.  The plot
	conventions are described in the first part of
	Section~\ref{se:si_win:table}. The continuum extrapolations are shown
	with no, NNLO XPT and SRHO taste improvements. The two-peak structure in the $24-28$
	histogram comes from the taste improvement variation.
    }
\end{table}

\begin{table}[p]
    \centering
    \includegraphics{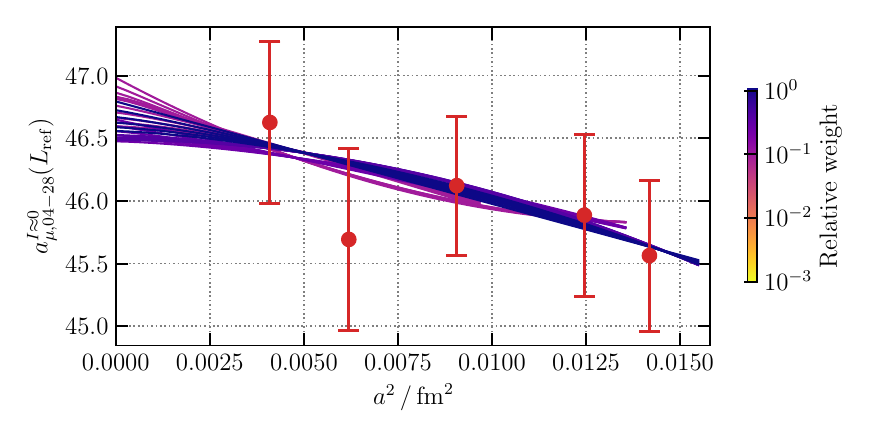}\\
    \includegraphics{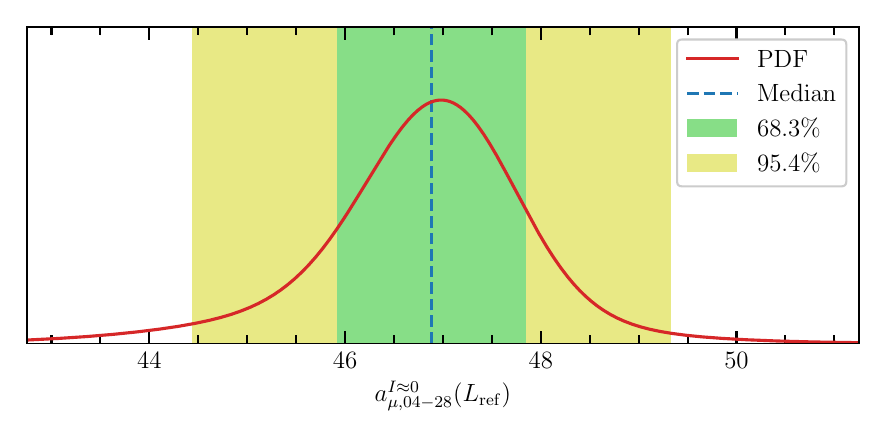}\\
    \input{figures/tables/i0_04_28/tabular}
    \caption
    {
	\label{ta:res_i0_04_28} $I\approx 0$
	window observable $a_{\mu,04-28}^\mathit{I\approx 0}$.  The plot
	conventions are described in the first part of
	Section~\ref{se:si_win:table}.
    }
\end{table}

Regarding the disconnected contribution, we fit a single window $04-28$ and use
the $I\approx0$ combination as in Equation \eqref{eq:iapprox0}. The results are
shown in Table~\ref{ta:res_i0_04_28}.

Finally, we combine the three-window fits of the light, the single-window fit
of the $I\approx0$ observable in the $04-28$ window and the light and disconnected
contributions in the $00-04$ window
\begin{equation}
    a_{\mu,00-28}^\mathrm{light+disc}= \tfrac{9}{10}a_{\mu,04-28}^\mathrm{light} + a_{\mu,04-28}^\mathit{I\approx 0} +
    a_{\mu,00-04}^\mathrm{light} + a_{\mu,00-04}^\mathrm{disc}
\end{equation}
also taking into account the distributions of our scale-setting quantities
$w_0$ and $M_{ss}$. We obtain
\begin{equation}
    \label{eq:amuw_00_28}
    a_{\mu,00-28}^\mathrm{light+disc}= \valuesLipdiBody
\end{equation}
in infinite volume, where we added the finite-size correction from
Table~\ref{ta:fv_res}.

\subsection{Long-distance window $10-\infty$ and comparison with 2020 result}
\label{sse:ldwin}

\begin{table}[p]
    \centering
    \includegraphics{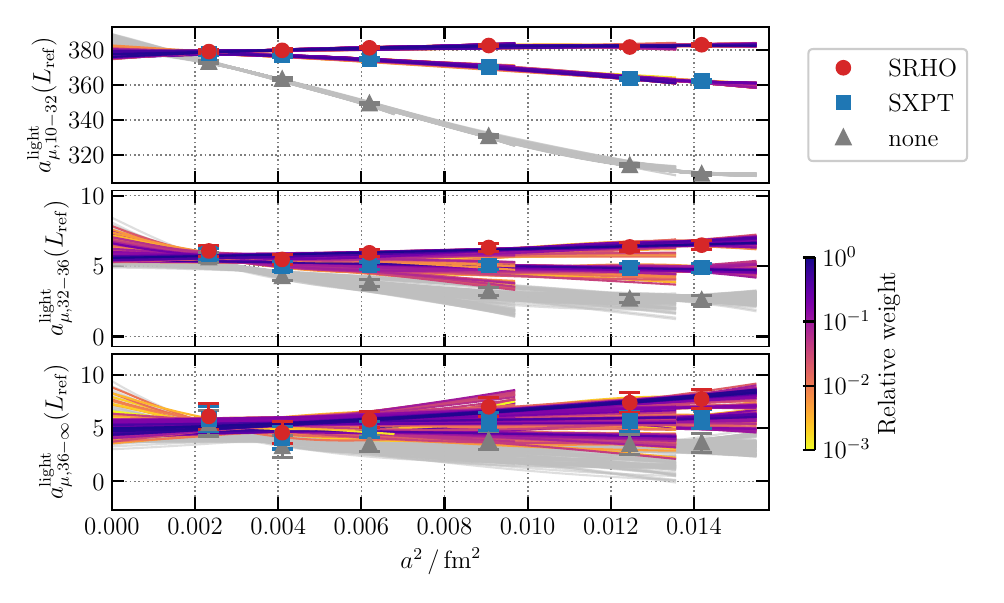}\\
    \includegraphics{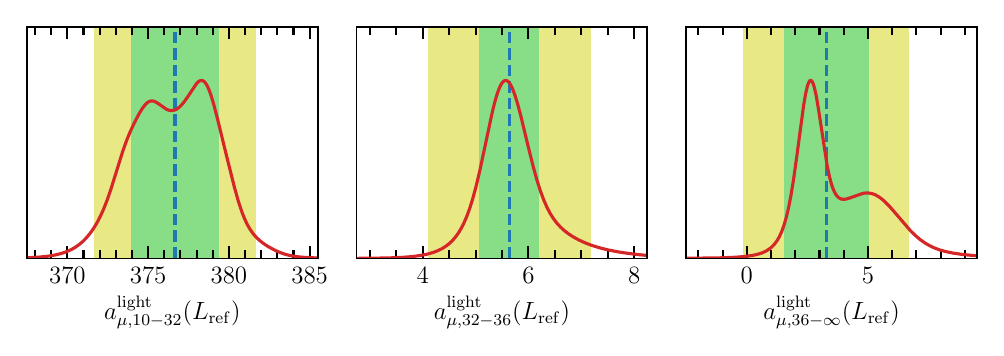}\\
    \input{figures/tables/li_10_inf/tabular}
    \caption
    {
	\label{ta:res_li_10_inf} Light-connected
	window observable $a_{\mu,10-\infty}^\mathrm{light}$ obtained as sum of
	three windows $10-32$, $32-36$ and $36-\infty$, fitted together. The
	plot conventions are described in the first part of
	Section~\ref{se:si_win:table}. The continuum extrapolations are shown
	with no, NNLO XPT and SRHO taste improvements, and for the case of the upper bound. The two-peak structure
	in the histogram of $10-32$ and $36-\infty$ comes from the taste improvement
	variation and the bounding, respectively.
    }
\end{table}

The standard long-distance window starts at Euclidean time $1.0$~fm and extends to
infinity. It is a central observable of the averaging procedure of the Theory
Initiative~\cite{Aliberti:2025beg}. Here we provide a pure lattice computation
of this quantity\footnote{We thank H.~Wittig for discussions on this
point.}. Results are given at the isospin symmetric point
of the WP25 scheme defined in Equation (3.9) of the White
Paper~\cite{Aliberti:2025beg}. According to this reference, results in the WP25 and BMW
schemes can be directly compared, without including matching factors, at the current
level of precision.

We apply the bounding method, originally proposed in
\cite{Lehner:2016,Borsanyi:2016lpl} and described in detail in
Section~13 of our 2020 work. Lattice data is used up to a Euclidean time of
$4.0$~fm, after which it is replaced either with zero (lower bound) or
with a two-pion correlator (upper bound). Here, to reduce
uncertainties, this bounding method is adapted to the situation in
which the $10-\infty$ window is obtained as the sum of smaller windows,
that is
\begin{equation}
    a_{\mu,10-\infty}^\mathrm{light} = a_{\mu,10-32}^\mathrm{light} 
    + a_{\mu,32-36}^\mathrm{light} 
    + a_{\mu,36-\infty}^\mathrm{light}\ .
\end{equation}
Thus, the lower bound is implemented via
\begin{equation}
    \label{eq:boundlo}
    a_{\mu,36-\infty}^\mathrm{light}|_\mathrm{lo}= a_{\mu,36-40}^\mathrm{light}\ ,
\end{equation}
whereas for the upper bound, we consider
\begin{equation}
    \label{eq:boundhi}
    a_{\mu,36-\infty}^\mathrm{light}|_\mathrm{up}= a_{\mu,36-40}^\mathrm{light} +
    \left(\frac{a_{\mu,39-40}^\mathrm{light}}{a_{\mu,39-40}^\mathrm{2-pion}}\right) a_{\mu,40-\infty}^\mathrm{2-pion}\ .
\end{equation}
The contributions to $a_\mu$ labelled ``2-pion'' are built
from a correlation function that falls off exponentially with a
rate given by the energy of two non-interacting pions, each with
one unit of momentum, i.e.\ the ground state in the $I=1$ channel.

In this approach, the $10-32$, $32-36$ and $36-\infty$ windows are
fitted together, like the three contributions to the $04-28$ window in
Section~\ref{sec:all-distance}. And again, in the interval from $1.0$
to $4.0$~fm only lattice data are used and above $4.0$~fm, these
are replaced by the bounds described above.

Because the 2-pion correlator, used in the upper bound of Equation \eqref{eq:boundhi}, increases more slowly than the
lattice correlator as Euclidean time is decreased, the ratio of the lattice to the
2-pion $39-40$ window, in parentheses in that equation, is larger
than $da_{\mu}^\mathrm{light}/dt$ at $t=4.0$~fm. In turn, this means that the
upper bound of Equation \eqref{eq:boundhi} is even weaker than the one
obtained by equating $da_{\mu}^\mathrm{2-pion}/dt$ to
$da_{\mu}^\mathrm{light}/dt$ at $t=4.0$~fm. We choose the former
because it smooths out possible fluctuations in the lattice correlator
around $t=4.0$~fm.

The lower and upper bounds for $a_{\mu,10-\infty}^\mathrm{light}$,
resulting from our procedure, are used as a two-point systematic
in our histograms. Thus, our central value for the $10-\infty$
window corresponds approximately to their average, and their
half-difference becomes a systematic uncertainty.

\begin{figure}[t]
    \centering
    \includegraphics{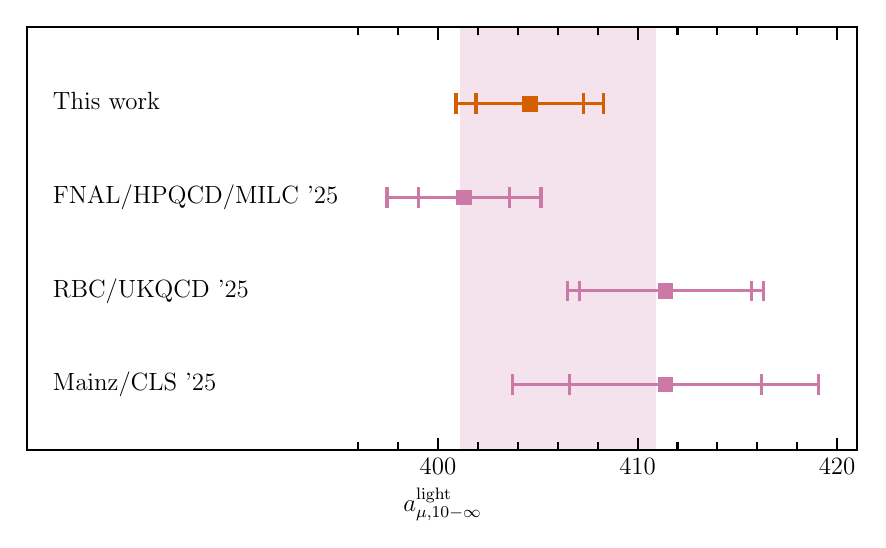}
    \caption
    {
	Comparison between different lattice results
	for the light connected long-distance window,
	$a_{\mu,10-\infty}^\mathrm{light}$. The results are given in the WP25
	scheme defined in Equation (3.9) of the White Paper
	\cite{Aliberti:2025beg}. The pink band is the average of all the lattice
	results excluding ours, as published in \cite{Aliberti:2025beg}.
    }
    \label{fi:amuld}
\end{figure}

Our results are presented in Table~\ref{ta:res_li_10_inf}.
In the long-distance-window results collected in the current White Paper,
finite-size corrections are obtained via phenomenology, not
from a lattice calculation. For a fair comparison, we apply the same choice
here. We use the data-driven determination
from Table~\ref{tab:fv_pheno_result_final} and for the much smaller finite-time correction, NNLO XPT.
For the latter, we obtain $\Delta
a_\mu^{I=1}(L_\mathrm{ref},T_\mathrm{ref}\to L_\mathrm{ref},\infty)=-0.47$.
Adding these corrections to the lattice results, we obtain in infinite volume,
\begin{equation}
    \label{eq:amuw_10_inf}
    a_{\mu,10-\infty}^\mathrm{light}= \valuesLightLd\ .
\end{equation}
This number can be compared to the other lattice results
\cite{RBC:2024fic,Djukanovic:2024cmq,FermilabLatticeHPQCD:2024ppc}, published
while the present paper was being refereed. Those results were
subsequently averaged by the Theory Initiative to provide a new recommended value for
$a_\mu$ in the 2025 White Paper \cite{Aliberti:2025beg}.  In Figure~\ref{fi:amuld},
we compare our result with those from the other groups, together with the White
Paper average.

In what follows, we address three questions relevant to our
analysis. First, we quantify the gain in precision achieved by
including the finest lattice spacing. Second, we assess the improvement
obtained when the lattice determination of the $28-\infty$ tail is
replaced by our data-driven approach. Third, we examine how our current
determination of the light-connected contribution differs from our
earlier work. For this purpose, it is sufficient to restrict attention
to results obtained in our reference volume. Our starting point is the
$10-\infty$ window presented in Table~\ref{ta:res_li_10_inf}, for which
we obtain:
\begin{equation}
    \label{eq:amuw_10_inf_fv}
    a_{\mu,10-\infty}^\mathrm{light}(L_\mathrm{ref})= \valuesLightLdFv\ .
\end{equation}

To answer the first question, we perform the analysis without the finest
ensembles, i.e.\ those with lattice spacing $a=0.048$~fm. We obtain
\begin{equation}
    \label{eq:amuw_10_inf_fv_nomon}
    a_{\mu,10-\infty}^\mathrm{light}(L_\mathrm{ref})= \valuesLightLdFvNomon\quad\quad
    \text{(without the finest lattice).}
\end{equation}
From this and the previous result, we conclude that the inclusion of the new lattice spacing
reduces the error by $\valuesLightLdFvEred$. Note that in our 2020 work, the error
on the light-connected contribution, Equation \eqref{eq:amu_light_old}, is
smaller than that of Equation \eqref{eq:amuw_10_inf_fv}, obtained by including our finest lattices. The reason
for this is that our current analysis is more conservative than the 2020 one.
Here, we include fit functions that are cubic in $a^2$, whereas they only went
up to quadratic order in our 2020 work. Moreover, the largest lattice spacing used here is only
$a=0.119$~fm, while previously it was $a=0.132$~fm.
There are also other differences in the analysis, e.g.\ we now
use a multi-window fit procedure, whereas earlier we fitted the whole time range at
once.

To answer the second question, we take the value of the $10-28$ window from Table~\ref{ta:res_li_10_28}:
\begin{equation}
    a_{\mu,10-28}^\mathrm{light}(L_\mathrm{ref})= \valuesLightLdbodyFv\ .
\end{equation}
Comparing its uncertainty with that of the result in
Equation~\eqref{eq:amuw_10_inf_fv}, our hybrid approach---where the lattice
determination of the $28-\infty$ tail is replaced by a data-driven
estimate whose uncertainty can be neglected for the purposes of this
exercise---achieves a $\valuesLightLdbodyEred$ improvement in precision
relative to a purely lattice-based computation employing the bounding
method.

For addressing the third question, we combine our pure lattice determination for the $10-\infty$ window
with the short- and intermediate-distance windows to get
\begin{equation}
    \label{eq:amu_light_new}
    a_{\mu}^\mathrm{light}(L_\mathrm{ref})= \valuesLightFv\ ,
\end{equation}
which is larger than our 2020 result
\begin{equation}
    \label{eq:amu_light_old}
    a_{\mu}^\mathrm{light}(L_\mathrm{ref})= \valuesLightFvNature\quad\quad
    \text{(our 2020 estimate)}
\end{equation}
by $\valuesLightFvThisvsnatDiff$ units. According to Equations
\eqref{eq:amuw_10_inf_fv} and \eqref{eq:amuw_10_inf_fv_nomon}, about
$\valuesLightLdFvDiff$ units come from the addition of a finer lattice
spacing. An additional contribution of approximately $4.0$ units arises from the fact that, in
the nomenclature of Ref.~\cite{Borsanyi:2020mff}, our 2020 result was
obtained using a Type-I fit, in which the lattice scale is set
directly using the mass of the $\Omega^-$ baryon, whereas the present result is
based on Type-II fits, which employ the intermediate scale $w_0$.
The size of this contribution is obtained directly from the difference
between the results for the light-connected contribution from the
Type-II and Type-I analyses, using the 2020 dataset.  Any residual
discrepancy can be attributed to the differences in the analysis
procedures described above.

We can also give an estimate of the error on the difference between Equations
\eqref{eq:amu_light_new} and \eqref{eq:amu_light_old}, taking correlations into
account. For the statistical correlation coefficient between the 2020 and
current results, we obtain $r=0.41$. To correlate the systematic errors, we use the
error budgets from Table~\ref{ta:res_li_10_inf} of the present paper and from
Table~14 of our previous work. The taste-correction uncertainty is treated as fully correlated.
However, there is still a large residual uncertainty on the difference due to the significant
reduction from $3.8$ in our 2020 work to $1.5$ in the current one.  We assigned no
uncertainty to the bounding method in 2020 (it was assumed to be covered by the
statistical error), so the uncertainty of $1.2$ that we assign this time is uncorrelated.
The remaining systematic was $1.8$ in 2020,
whereas it is $1.5$ now. As the two analyses are very
different, one could argue for taking zero correlation
between their remaining systematics. However, as another extreme, we also consider the case of full
correlation. Assuming zero or full correlation, we obtain for the
total uncertainty of the difference $\valuesLightFvThisvsnatErru$ or
$\valuesLightFvThisvsnatErrc$, respectively. This means that our new light
connected value is $\valuesLightFvThisvsnatSigu\sigma$ or
$\valuesLightFvThisvsnatSigc\sigma$ higher than the 2020 one, depending on the
assumption made about the correlations in their remaining systematics.

\subsection{Strange and charm contributions}

In addition to the light and disconnected parts of $a_{\mu}$, we also
performed an update of the strange and charm contributions.
The procedure for the strange quark is the same as
was used for the light. The main difference is that we split the strange
contribution $a_{\mu,00-28}^{\mathrm{strange}}$ into only two windows, the
short-distance $a_{\mu,00-04}^{\mathrm{strange}}$ and the remaining part
$a_{\mu,04-28}^{\mathrm{strange}}$, which are fitted separately. We have
$5632$ independent fits for the strange short-distance contribution, and for
both the remaining part and the intermediate window we have $2880$ fits each.
We obtain for the short-distance window
\begin{equation}
    a_{\mu,00-04}^\mathrm{strange}=\valuesStrangeSd\ ,
\end{equation}
for the intermediate-distance window
\begin{equation}
    \label{eq:strangeid}
    a_{\mu,04-10}^\mathrm{strange}=\valuesStrangeId\ ,
\end{equation}
and for a long-distance window
\begin{equation}
    \label{eq:strangeldbody}
    a_{\mu,10-28}^\mathrm{strange}=\valuesStrangeLdbody\ ,
\end{equation}
with the percentages of fits with $P$-value of at least $0.1$ being 88\%, 87\%
and 31\%, respectively.
Summing up the short-distance and the rest we get the contribution of
the strange quark to $a_{\mu}$:
\begin{equation}
    \label{eq:strangebody}
    a_{\mu,00-28}^\mathrm{strange}=\valuesStrangeBody\ .
\end{equation}

The behaviour of the charm quark contribution to $a_{\mu}$
is different from the light, disconnected or strange contributions,
and we modified the fit procedure accordingly. For the quark-mass
mistuning in the charm sector we use the variable
\begin{equation}
    X_c= M_{lc}^2w_0^2 - {\big[M_{lc}^2w_0^2]}_\mathrm{qcd}\ ,
\end{equation}
which describes the charm quark mass deviation of our ensembles from the
physical point. The physical value of $M_{lc}=M_D=1863.1(6)~\mathrm{MeV}$ in the
isospin symmetric limit is taken from Ref.\ \cite{Giusti:2017dmp}. The global
fit function then has the form
\begin{equation}
    Y = A(a^2) + D(a^2) X_c\ ,
\end{equation}
We used the same options for
$A(a^2)$ as for the light, short-distance function given in Equation (\ref{eq:a2log}).
For the short distance part we also included tree-level corrected result and
$Q/\hat{Q}$ variation in the fit procedure. In total we have $1408$ fits for the
charm short-distance contribution and $352$ fits for the remaining part,
with the percentages of fits with $P$-value above $0.1$ being 81\%, 47\%
and 27\%, respectively. The
final numbers for the individual windows together with the total result are
\begin{equation}
\begin{split}
    \label{eq:charm}
	a_{\mu,00-04}^\mathrm{charm}= \valuesCharmSd\ ,\\
	a_{\mu,04-10}^\mathrm{charm}= \valuesCharmId\ ,\\
	a_{\mu,10-28}^\mathrm{charm}= \valuesCharmLdbody\ ,\\
	a_{\mu,00-28}^\mathrm{charm}= \valuesCharmBody\ .
\end{split}
\end{equation}

\subsection{Other contributions, total and comparison with 2020 result}
\label{se:si_win_other}

\begin{table}[p]
    \centering
    \begin{tabular}{c|l|R|l}
	\hline
	$00-28$ & light and disconnected & \valuesLipdiBody   & this work, Equation \eqref{eq:amuw_00_28}\\
	$00-28$ & strange                & \valuesStrangeBody & this work, Equation \eqref{eq:strangebody}\\
	$00-28$ & charm                  & \valuesCharmBody   & this work, Equation \eqref{eq:charm}\\
	\hline
	$00-28$ & light qed & -1.86(53)(27)& this work\\
	$00-28$ & light sib & 7.06(64)(35)& this work\\
	$00-28$ & disconnected qed & -1.00(46)(28)& this work\\
	$00-28$ & disconnected sib & -4.39(58)(63)& this work\\
	$00-28$ & strange qed & -0.0136(85)(77)& this work\\
	\hline
	$00-\infty$ & disconnected charm & 0.0(1)& \cite{Borsanyi:2017zdw}, Section~4 in Supp. Mat.\\
	$00-\infty$ & charm qed & 0.0182(36)& \cite{Giusti:2019xct}\\
	$00-\infty$ & bottom & 0.271(37)& \cite{Colquhoun:2014ica}\\
	\hline
	$28-\infty$ & tail from data-driven & \valuesPhenoTailnew    & this work, Equation \eqref{eq:tail_final}\\
	\hline
	\hline
	$00-\infty$ & total & \valuesTotal &
    \end{tabular}
    \caption
    {
	List of all contributions to $a_\mu$.
    }
    \label{ta:amu}
\end{table}

\begin{table}[p]
    \centering
    \begin{tabular}{c|l|R|l}
	\hline
	$04-10$ & light        & \valuesLightId   & this work, Equation \eqref{eq:lightid}\\
	$04-10$ & disconnected & \valuesDiscoId   & this work, Equation \eqref{eq:discoid}\\
	$04-10$ & strange      & \valuesStrangeId & this work, Equation \eqref{eq:strangeid}\\
	$04-10$ & charm        & \valuesCharmId   & this work, Equation \eqref{eq:charm}\\
	\hline
	$04-10$ & light qed & 0.035(40)(44)& \cite{Borsanyi:2020mff}, Table~17\\
	$04-10$ & light sib & 0.753(40)(16)& \cite{Borsanyi:2020mff}, Table~17\\
	$04-10$ & disconnected qed & -0.117(17)(6)& \cite{Borsanyi:2020mff}, Table~17\\
	$04-10$ & disconnected sib & -0.237(9)(6)& \cite{Borsanyi:2020mff}, Table~17\\
	$04-10$ & strange qed & -0.0050(35)(37)& \cite{Borsanyi:2020mff}, Table~17\\
	\hline
	\hline
	$04-10$ & total & \valuesTotalId &
    \end{tabular}
    \caption
    {
	List of all contributions to the intermediate-window $a_{\mu,04-10}$.
    }
    \label{ta:amuwid}
\end{table}

\begin{table}[p]
    \centering
    \begin{tabular}{c|l|R|l}
	\hline
	$10-28$ & light and disconnected & \valuesLipdiLdbody   & this work, Equation \eqref{eq:amuw_10_28}\\
	$10-28$ & strange                & \valuesStrangeLdbody & this work, Equation \eqref{eq:strangeldbody}\\
	$10-28$ & charm                  & \valuesCharmLdbody   & this work, Equation \eqref{eq:charm}\\
	\hline
	$10-28$ & light qed & -1.79(52)(29)& this work\\
	$10-28$ & light sib & 6.24(64)(36)& this work\\
	$10-28$ & disconnected qed & -0.89(45)(29)& this work\\
	$10-28$ & disconnected sib & -4.13(57)(60)& this work\\
	$10-28$ & strange qed & -0.0094(55)(42)& this work\\
	\hline
	$28-\infty$ & tail from data-driven & \valuesPhenoTailnew    & this work, Equation \eqref{eq:tail_final}\\
	\hline
	\hline
	$10-\infty$ & total & \valuesTotalLd &
    \end{tabular}
    \caption
    {
	List of all contributions to the long-distance-window
	$a_{\mu,10-\infty}$.
    }
    \label{ta:amuwld}
\end{table}

Until now we discussed the light, strange, charm and disconnected contributions
to $a_{\mu,00-28}$, in the isospin-symmetric limit. The remaining ones are
listed in the second part of Table~\ref{ta:amu}, most of them are
isospin-breaking corrections. Here we give details on how we have computed
them.

In our 2020 work we computed isospin-breaking contributions for the total
$a_\mu$. In a previous version of this paper we took over the isospin-breaking
contributions from our 2020 work as they are, instead of computing them in the
$00-28$ window. We now remove this shortcoming. For this purpose we repeat the
very same analysis as in our 2020 work for isospin breaking in the
$a_{\mu,00-28}$ window, in the cases of the light, disconnected and strange
flavours. These are called ``Type-II'' fits and are detailed in Section~24 of
the 2020 paper \cite{Borsanyi:2020mff}. The results are reported in
Table~\ref{ta:amu}. It turns out, that the numbers obtained in 2020 for the total
range and the numbers obtained now for the $00-28$ window are in good agreement
with each other. However, we consider the errors of the $00-28$ window
quantities more reliable, since the long distance part can be very challenging
to estimate.

Now we add up the light, disconnected, strange, charm and isospin-breaking
contributions in the $00-28$ window determined on the lattice in this work,
plus the remaining contributions from previous lattice computations. These are
the first, second and third sections of Table~\ref{ta:amu}. We are not
correlating the isospin-breaking, the strange, and charm computations with each
other or with the light plus disconnected part. We justify this choice by
noting that these observables describe very different physics, each with its
own kind of lattice artefact and in case of the isospin breaking sometimes they
are measured on different sets of ensembles than the isospin symmetric ones.
Finally we add the $28-\infty$ window from the data-driven approach discussed
in Section~\ref{se:si_tail}.  Altogether we get for the leading-order hadronic
vacuum polarization contribution to the muon magnetic moment:
\begin{equation}
    a_\mu= \valuesTotal
\end{equation}
with statistical, systematic and total errors. This is the main result of this work.

This result differs from our 2020 determination by
$\valuesHvpthisvsnatDiff$ units (see also the discussion in
Subsection~\ref{sse:ldwin}). We next address whether this difference is
statistically significant.  We begin by outlining, in general terms,
the various sources of correlation between our current and previous
analyses, before quantifying their impact below.

Statistical correlations between our 2020 analysis and the present work
can be computed using jackknife resampling, and we find them to be small.
This dilution arises both from the inclusion
of a new lattice spacing and from our use of the $a_{\mu,00\text{–}28}$
window rather than $a_\mu$. 

In addition, the dominant source of systematic uncertainty differs
markedly between the two studies. In 2020, the leading systematic error
originated from the difference between the SRHO and NNLO
taste-improvement procedures. That uncertainty is now significantly
smaller, owing to the availability of finer lattices and to our
decomposition of the analysis into multiple windows.
This uncertainty is fully correlated, but due to its reduction in
the present work, there is still a large residual uncertainty on
the difference.

In the present work, the dominant systematic uncertainty instead stems
from the scale-setting quantity $w_0$, whereas the 2020 analysis relied
on $M_\Omega$ for scale setting. Because neither the uncertainties
associated with $M_\Omega$ nor those in $a_\mu$ are expected to
correlate with fluctuations in $w_0$, it is reasonable to treat these
sources of systematic as uncorrelated.

Regarding finite-size corrections, we assume full correlation between
the 2020 and present analyses, since these corrections were obtained
from the same {\tt 4hex} configurations and were applied to $a_\mu$ in
2020 and to $a_{\mu,00\text{–}28}$ here.

By contrast, the continuum-extrapolation uncertainties in the 2020 and
2025 analyses are expected to be largely uncorrelated, given the
substantial methodological differences in how each is obtained. In
2020, the continuum limit was obtained from a single fit of $a_\mu$,
whereas in the present work $a_{\mu,00\text{–}28}$ is decomposed into
four windows, each with its own fit function and lattice-spacing cuts.
This methodological difference provides a reasonable basis for treating
the associated continuum-extrapolation uncertainties as uncorrelated.
Nevertheless, for completeness, we also consider the opposite extreme
of full correlation. The same treatment is applied to the remaining
systematic uncertainties, such as those associated with the fit ranges
for the pseudoscalar masses, etc.

Accounting for the correlations discussed above, we estimate the
uncertainty on the difference between the 2020 and present results. The
statistical correlation coefficient is $r = 0.16$. The
taste-improvement systematic was $3.8$ in 2020 and $0.9$ in the current
analysis; assuming full correlation, the uncertainty on the difference
is $2.8$. Continuum-extrapolation and remaining systematic uncertainties
remain essentially unchanged at $1.9$.

For finite-size corrections, treated as fully correlated and using the
{\tt 4hex} simulations, the uncertainty on the difference is also $1.9$,
arising from the finite volume corrections beyond $2.8~\mathrm{fm}$
that do not enter into the hybrid result.

Combining all contributions in quadrature, and considering both zero
and full correlation for the continuum-extrapolation and remaining
systematics, we obtain total uncertainties of $\valuesHvpthisvsnatErru$
and $\valuesHvpthisvsnatErrc$, respectively. Thus, the new result
exceeds the 2020 value by $\valuesHvpthisvsnatSigu\sigma$ or
$\valuesHvpthisvsnatSigc\sigma$, in agreement with the difference
observed for the light-connected contribution in
Section~\ref{sse:ldwin}.

We can also give a complete result for the intermediate-window
observable. The light and disconnected contributions are given in Equations
\eqref{eq:lightid} and \eqref{eq:discoid}. The others were determined in our
previous work, which we give here for completeness in Table~\ref{ta:amuwid}.
Adding up all these contributions we obtain the value
\begin{equation}
    a_{\mu,04-10}= \valuesTotalId\ ,
\end{equation}
which is compared to other determinations, lattice and data-driven, in
Figure~2 of the main article.

We further provide a complete result for the long-distance window observable, where
the Euclidean time range goes from $1.0$~fm to infinity. The largest
contribution is given by the light flavour in the $10-28$ window, whose analysis
details can be found in Table~\ref{ta:res_li_10_28}. Adding the
disconnected contribution gives the result in Equation \eqref{eq:amuw_10_28}.
We performed analyses for all the other contributions as
in the case of the $00-28$ window, with the results listed in
Table~\ref{ta:amuwld}. The $28-\infty$ window is taken from the data-driven approach
again. Adding all of those contributions up, we obtain
\begin{equation}
    a_{\mu,10-\infty}= \valuesTotalLd\ .
\end{equation}

%% file: figures/tables/li_00_04/tabular.tex
\begin{tabular}{l|r|r}
Number of fits & \multicolumn{2}{c}{8192} \\
Fits with $P>0.1$ & \multicolumn{2}{c}{54\%} \\
Median & \multicolumn{2}{c}{47.82} \\
\hline
Total error & 0.14 & 0.29\% \\
Statistical error & 0.05 & 0.10\% \\
Systematic error & 0.13 & 0.28\% \\
\hline
Pseudoscalar fits & 0.00 & 0.00\% \\
$M_{ss}$ value & 0.00 & 0.00\% \\
$w_0$ value & 0.01 & 0.02\% \\
Tree correction and/or $\hat{q}$ & 0.09 & 0.19\% \\
Lattice spacing cut & 0.05 & 0.11\% \\
Fit polynomial order & 0.06 & 0.12\% \\
\end{tabular}

%% file: figures/tables/di_00_04/tabular.tex
\begin{tabular}{l|r|r}
Number of fits & \multicolumn{2}{c}{2560} \\
Fits with $P>0.1$ & \multicolumn{2}{c}{100\%} \\
Median & \multicolumn{2}{c}{0.0010} \\
\hline
Total error & 0.0027 & 284\% \\
Statistical error & 0.0022 & 226\% \\
Systematic error & 0.0016 & 172\% \\
\hline
Pseudoscalar fits & 0.0000 & 1\% \\
$M_{ss}$ value & 0.0000 & 1\% \\
$w_0$ value & 0.0002 & 16\% \\
Tree correction and/or $\hat{q}$ & 0.0001 & 6\% \\
Lattice spacing cut & 0.0005 & 47\% \\
Fit polynomial order & 0.0018 & 192\% \\
\end{tabular}

%% file: figures/tables/li_04_10/tabular.tex
\begin{tabular}{l|r|r}
Number of fits & \multicolumn{2}{c}{50688} \\
Fits with $P>0.1$ & \multicolumn{2}{c}{80\%} \\
Median & \multicolumn{2}{c}{206.37} \\
\hline
Total error & 0.50 & 0.24\% \\
Statistical error & 0.37 & 0.18\% \\
Systematic error & 0.33 & 0.16\% \\
\hline
Pseudoscalar fits & 0.01 & 0.00\% \\
$M_{ss}$ value & 0.01 & 0.00\% \\
$w_0$ value & 0.18 & 0.09\% \\
Lattice spacing cut & 0.07 & 0.04\% \\
Fit polynomial order & 0.18 & 0.09\% \\
Log corrections $\gamma$ & 0.10 & 0.05\% \\
\end{tabular}

%% file: figures/tables/di_04_10/tabular.tex
\begin{tabular}{l|r|r}
Number of fits & \multicolumn{2}{c}{30720} \\
Fits with $P>0.1$ & \multicolumn{2}{c}{48\%} \\
Median & \multicolumn{2}{c}{-0.994} \\
\hline
Total error & 0.032 & -3.2\% \\
Statistical error & 0.021 & -2.1\% \\
Systematic error & 0.024 & -2.4\% \\
\hline
Pseudoscalar fits & 0.000 & -0.0\% \\
$M_{ss}$ value & 0.000 & -0.0\% \\
$w_0$ value & 0.008 & -0.8\% \\
Lattice spacing cut & 0.013 & -1.3\% \\
Fit polynomial order & 0.012 & -1.2\% \\
Log corrections $\gamma$ & 0.015 & -1.5\% \\
\end{tabular}

%% file: figures/tables/li_15_19/tabular.tex
\begin{tabular}{l|r|r}
Number of fits & \multicolumn{2}{c}{50688} \\
Fits with $P>0.1$ & \multicolumn{2}{c}{100\%} \\
Median & \multicolumn{2}{c}{95.51} \\
\hline
Total error & 2.10 & 2.20\% \\
Statistical error & 1.75 & 1.84\% \\
Systematic error & 1.16 & 1.22\% \\
\hline
Pseudoscalar fits & 0.02 & 0.02\% \\
$M_{ss}$ value & 0.02 & 0.02\% \\
$w_0$ value & 0.62 & 0.65\% \\
Taste correction & 0.28 & 0.30\% \\
Lattice spacing cut & 0.22 & 0.23\% \\
Fit polynomial order & 0.62 & 0.64\% \\
Log corrections $\gamma$ & 0.20 & 0.21\% \\
\end{tabular}

%% file: figures/tables/li_10_28/tabular.tex
\begin{tabular}{l|r|r}
Number of fits & \multicolumn{2}{c}{50688} \\
Fits with $P>0.1$ & \multicolumn{2}{c}{99\%} \\
Median & \multicolumn{2}{c}{366.38} \\
\hline
Total error & 2.84 & 0.78\% \\
Statistical error & 1.76 & 0.48\% \\
Systematic error & 2.23 & 0.61\% \\
\hline
Pseudoscalar fits & 0.04 & 0.01\% \\
$M_{ss}$ value & 0.03 & 0.01\% \\
$w_0$ value & 1.64 & 0.45\% \\
Taste correction & 0.90 & 0.25\% \\
Lattice spacing cut & 0.40 & 0.11\% \\
Fit polynomial order & 0.33 & 0.09\% \\
Log corrections $\gamma$ & 0.36 & 0.10\% \\
\end{tabular}

%% file: figures/tables/i0_10_28/tabular.tex
\begin{tabular}{l|r|r}
Number of fits & \multicolumn{2}{c}{7680} \\
Fits with $P>0.1$ & \multicolumn{2}{c}{100\%} \\
Median & \multicolumn{2}{c}{27.13} \\
\hline
Total error & 1.21 & 4.45\% \\
Statistical error & 1.18 & 4.36\% \\
Systematic error & 0.24 & 0.88\% \\
\hline
Pseudoscalar fits & 0.01 & 0.03\% \\
$M_{ss}$ value & 0.01 & 0.02\% \\
$w_0$ value & 0.14 & 0.53\% \\
Lattice spacing cut & 0.03 & 0.12\% \\
Fit polynomial order & 0.22 & 0.83\% \\
Log corrections $\gamma$ & 0.00 & 0.01\% \\
\end{tabular}

%% file: figures/tables/li_04_28/tabular.tex
\begin{tabular}{l|r|r|r|r|r|r|r|r}
Number of fits & \multicolumn{8}{c}{248396544} \\
Fits with $P>0.1$ & \multicolumn{8}{c}{80\%} \\
Median & \multicolumn{2}{c|}{572.99} & \multicolumn{2}{c|}{502.28} & \multicolumn{2}{c|}{46.91} & \multicolumn{2}{c}{23.90} \\
\hline
Total error & 2.75 & 0.48\% & 1.72 & 0.34\% & 0.79 & 1.68\% & 0.88 & 3.67\% \\
Statistical error & 1.84 & 0.32\% & 1.24 & 0.25\% & 0.41 & 0.88\% & 0.41 & 1.70\% \\
Systematic error & 2.04 & 0.36\% & 1.19 & 0.24\% & 0.67 & 1.43\% & 0.78 & 3.25\% \\
\hline
Pseudoscalar fits & 0.05 & 0.01\% & 0.03 & 0.01\% & 0.02 & 0.03\% & 0.01 & 0.04\% \\
$M_{ss}$ value & 0.03 & 0.01\% & 0.02 & 0.00\% & 0.01 & 0.02\% & 0.00 & 0.02\% \\
$w_0$ value & 1.72 & 0.30\% & 1.20 & 0.24\% & 0.31 & 0.66\% & 0.21 & 0.86\% \\
Taste correction & 0.86 & 0.15\% & 0.31 & 0.06\% & 0.32 & 0.69\% & 0.45 & 1.89\% \\
Lattice spacing cut & 0.40 & 0.07\% & 0.16 & 0.03\% & 0.20 & 0.43\% & 0.15 & 0.62\% \\
Fit polynomial order & 0.40 & 0.07\% & 0.23 & 0.05\% & 0.20 & 0.42\% & 0.27 & 1.11\% \\
Log corrections $\gamma$ & 0.23 & 0.04\% & 0.06 & 0.01\% & 0.08 & 0.17\% & 0.10 & 0.42\% \\
\hline
& \multicolumn{2}{c|}{$04-28$}& \multicolumn{2}{c|}{$04-20$}& \multicolumn{2}{c|}{$20-24$}& \multicolumn{2}{c }{$24-28$} \\
\end{tabular}

%% file: figures/tables/i0_04_28/tabular.tex
\begin{tabular}{l|r|r}
Number of fits & \multicolumn{2}{c}{7680} \\
Fits with $P>0.1$ & \multicolumn{2}{c}{100\%} \\
Median & \multicolumn{2}{c}{46.89} \\
\hline
Total error & 1.22 & 2.61\% \\
Statistical error & 1.19 & 2.54\% \\
Systematic error & 0.28 & 0.59\% \\
\hline
Pseudoscalar fits & 0.01 & 0.02\% \\
$M_{ss}$ value & 0.01 & 0.01\% \\
$w_0$ value & 0.17 & 0.36\% \\
Lattice spacing cut & 0.04 & 0.08\% \\
Fit polynomial order & 0.25 & 0.54\% \\
Log corrections $\gamma$ & 0.01 & 0.02\% \\
\end{tabular}

%% file: figures/tables/li_10_inf/tabular.tex
\begin{tabular}{l|r|r|r|r|r|r|r|r}
Number of fits & \multicolumn{8}{c}{496793088} \\
Fits with $P>0.1$ & \multicolumn{8}{c}{100\%} \\
Median & \multicolumn{2}{c|}{386.06} & \multicolumn{2}{c|}{376.66} & \multicolumn{2}{c|}{5.64} & \multicolumn{2}{c}{3.28} \\
\hline
Total error & 3.64 & 0.94\% & 2.70 & 0.72\% & 0.77 & 13.66\% & 1.75 & 53.22\% \\
Statistical error & 2.68 & 0.70\% & 1.47 & 0.39\% & 0.63 & 11.16\% & 0.97 & 29.51\% \\
Systematic error & 2.46 & 0.64\% & 2.27 & 0.60\% & 0.44 & 7.89\% & 1.45 & 44.29\% \\
\hline
Pseudoscalar fits & 0.06 & 0.02\% & 0.05 & 0.01\% & 0.02 & 0.28\% & 0.02 & 0.55\% \\
Taste correction & 1.49 & 0.39\% & 1.40 & 0.37\% & 0.13 & 2.33\% & 0.04 & 1.34\% \\
Bounding & 1.22 & 0.32\% & 0.00 & 0.00\% & 0.01 & 0.20\% & 1.21 & 36.93\% \\
Lattice spacing cut & 0.54 & 0.14\% & 0.53 & 0.14\% & 0.07 & 1.25\% & 0.11 & 3.38\% \\
Fit polynomial order & 0.98 & 0.25\% & 0.57 & 0.15\% & 0.31 & 5.54\% & 0.30 & 9.24\% \\
Log corrections $\gamma$ & 0.64 & 0.16\% & 0.62 & 0.16\% & 0.03 & 0.61\% & 0.06 & 1.70\% \\
\hline
& \multicolumn{2}{c|}{$10-\infty$}& \multicolumn{2}{c|}{$10-32$}& \multicolumn{2}{c|}{$32-36$}& \multicolumn{2}{c }{$36-\infty$} \\
\end{tabular}

%% file: si_finvol.tex
\section{Finite-size effects}
\label{se:si_finvol}

Finite spatial and temporal box sizes cause a significant distortion of the
target observables. We remove this by the same procedure as
applied in our 2020 work \cite{Borsanyi:2020mff}.
The largest contribution comes from the
finite-size effects in the isospin-symmetric part in the $I=1$ channel.
The finite-size effects in the $I=0$ channel are much smaller and even smaller are
those in the isospin-breaking part. Our procedure defines two box-sizes,
the reference box size $L_\mathrm{ref}=6.272$~fm and a large box size
$L_\mathrm{big}=10.752$~fm.

\begin{table}[p]
    \centering
    \begin{tabular}{C|C|R|R|R}
	\text{window}& L\times T & M_\pi=104\text{ MeV} & M_\pi=121\text{ MeV} & M_\pi=110\text{ MeV}\\
	\hline
	00-28&56\times84&656.77(72)& 641.77(60)& 651.35(49)\\
	     &96\times96&669.39(55)& 649.02(48)& 662.03(43)\\
	\hline
	00-04&56\times84&42.84(00)& 42.85(00)& 42.84(00)\\
	     &96\times96&42.87(00)& 42.87(00)& 42.87(00)\\
	\hline
	04-10&56\times84&207.53(03)& 206.54(04)& 207.17(02)\\
	     &96\times96&208.19(02)& 206.96(02)& 207.75(01)\\
	\hline
	10-28&56\times84&406.40(72)& 392.38(59)& 401.33(48)\\
	     &96\times96&418.33(55)& 399.19(47)& 411.41(43)\\
	\hline
	10-\infty&56\times84&436.22(2.87)& 419.15(2.16)& 430.05(2.04)\\
	     &96\times96&459.72(1.94)& 434.22(1.73)& 450.50(1.40)\\
	\hline
	15-19&56\times84&107.83(15)& 103.79(14)& 106.37(11)\\
	     &96\times96&110.32(10)& 105.27(11)& 108.50(08)\\
	\hline
	28-35&56\times84&20.31(79)& 18.60(52)& 19.69(54)\\
	     &96\times96&25.31(33)& 21.50(42)& 23.93(28)\\
    \end{tabular}
    \caption
    {
	\label{ta:fv_4hex} Results for $a_\mu^\mathrm{light}$ in different
	windows for our {\tt 4hex} simulations. These are performed with
	$M_\pi=104$~MeV and $121$~MeV Goldstone-pion masses. The last column
	contains values interpolated to the Goldstone-pion mass of
	$M_\pi=110$~MeV, where the taste-averaged pion mass takes on the $\pi^0$'s
        physical, mass value.
    }
\end{table}

To compute the $I=1$ finite-volume effects in our 2020 work we prepared a
dedicated set of simulations, the {\tt 4hex} data-set. We designed it to be
sensitive to finite-size effects even in the presence of staggered taste
violations. For this purpose we utilized a new staggered action, the action
details can be found in our 2020 publication. We lowered the Goldstone-pion
masses to $M_\pi=110$~MeV, so that the average of the masses of the pion taste
multiplet equals the physical pion mass.  We performed simulations with two
different lattice geometries, one on a $56\times 84$ and another on a large
$96\times 96$, which have spatial sizes $L_\mathrm{ref}$ and $L_\mathrm{big}$.
For each case the simulations were performed using two different Goldstone-pion
masses, $M_\pi=104$~MeV and $121$~MeV. These were used to interpolate to the
physical point defined above. The simulations and measurements were already
available in our 2020 work, in Table~\ref{ta:fv_4hex} we give results for
$a_\mu^\mathrm{light}$ in the different windows that are used in this paper.

\begin{table}[p]
    \centering
    \begin{tabular}{C||C|C||C|C}
	&
	\multicolumn{2}{C||}{\Delta a_\mu^{I=1}(L_\mathrm{ref}\to \infty)}&
	\multicolumn{2}{C}{\Delta a_\mu^{I=1}(L_\mathrm{big}\to \infty)}
	\\
	\text{window}&\text{NLO}&\text{NNLO}&\text{NLO}&\text{NNLO}\\
	\hline
	00-28 & 6.90 & 8.97 & 0.39 & 0.43 \\
	00-04 & 0.01 & 0.02 & 0.00 & 0.00 \\
	04-10 & 0.33 & 0.40 & 0.01 & 0.01 \\
	10-28 & 6.56 & 8.55 & 0.37 & 0.42 \\
	10-\infty & 12.02 & 16.33 & 1.18 & 1.40 \\
	15-19 & 1.48 & 1.87 & 0.07 & 0.07 \\
	28-35 & 2.46 & 3.45 & 0.25 & 0.29
    \end{tabular}
    \caption
    {
	\label{ta:fv_xpt} Finite-volume effects in the $I=1$ channel computed
	in chiral perturbation theory. Results are shown for two different box
	sizes, $L_\mathrm{ref}$ and $L_\mathrm{big}$, and for two different
	orders, NLO and NNLO.
    }
\end{table}

In addition, we also utilized chiral perturbation theory (XPT), and its
staggered version (SXPT), to compute finite-size effects
\cite{Aubin:2019usy,Borsanyi:2020mff}. Values obtained for several different
windows are given for $L_\mathrm{ref}$ and $L_\mathrm{big}$ box sizes, both
with next-to-leading-order (NLO) and next-to-next-to-leading order (NNLO)
chiral perturbation theory, in Table~\ref{ta:fv_xpt}.

\begin{table}[p]
    \centering
    \begin{tabular}{C||C|C|C||C}
	\text{window} &
	\Delta a_\mu^{I=1}(L_\mathrm{ref}\to L_\mathrm{big})&
	\Delta a_\mu^{I=1}(L_\mathrm{big}\to \infty)&
	\Delta a_\mu^{I=1}(L_\mathrm{ref}\to \infty)&
	\Delta a_\mu^{I=0}(L_\mathrm{ref}\to \infty)\\
	& \text{from {\tt 4hex}} & \text{from NNLO} & \text{from {\tt 4hex}+NNLO} & \text{from N4LO est.} \\
	\hline
	00-28 & 8.94(57)(67) & +0.43-0.06 & 9.31(88) & 0.00(19)\\
	00-04 & 0.02(00)(00) & +0.00-0.00 & 0.02(00) & 0.00(01)\\
	04-10 & 0.48(02)(04) & +0.01-0.00 & 0.49(04) & 0.00(00)\\
	10-28 & 8.44(57)(64) & +0.42-0.06 & 8.80(85) & 0.00(18)\\
	10-\infty & 17.12(2.08)(1.29) & +1.40-0.84 & 17.68(2.45) & 0.00(55)\\
	15-19 & 1.78(12)(13) & +0.07-0.00 & 1.85(18) & 0.00(03)\\
	28-35 & 3.55(50)(27) & +0.29-0.11 & 3.73(57) & 0.00(16)\\
    \end{tabular}
    \caption
    {
	\label{ta:fv_res} Finite-size effects on $a_\mu$ in different windows.
	The second column gives the difference in $a_\mu^{I=1}$ between
	boxes $\mathrm{big}$ and $\mathrm{ref}$ computed with the {\tt 4hex}
	simulations. The third column gives our NNLO XPT computations for the
	finite-size effects of the $\mathrm{big}$ box: the first/second number
	corresponds to the finite-volume/finite-time correction. The fourth is
	the sum of the second and third columns with errors added in
	quadrature. The last column is an estimate of the finite-size effect in
	the $I=0$ channel from N4LO XPT.
    }
\end{table}

The {\tt 4hex} lattice data and
chiral perturbation theory are combined in the following three step procedure
to obtain our result in infinite volume:

\begin{enumerate}

    \item On each {\tt 4stout} ensemble we apply a shift on the
	observables which brings them from the simulation box size to
	$L_\mathrm{ref}$. These shifts are computed with NNLO SXPT using the
	taste-violation of the ensemble as input. We then perform a continuum
	extrapolation, which then corresponds to a box size of
	$L_\mathrm{ref}$.
	
    \item We add the difference in the observables between box sizes
	$L_\mathrm{big}$ and $L_\mathrm{ref}$ computed with the {\tt 4hex}
	simulation.

    \item We add the difference between the infinite and $\mathrm{big}$ box
	sizes computed in NNLO XPT, also including corrections for the
	finite-time extent.
\end{enumerate}

The {\tt 4hex} finite-size corrections in step 2 are obtained from the results
in Table~\ref{ta:fv_4hex}. We take into account lattice artefacts in the same
way as in our 2020 work: we reduce the measured finite-size effect by 7\%, and
assign a 7\% uncertainty to this correction step. The size of this correction
step was taken from the deviation of the total $a_\mu$ light from its continuum
value. The results obtained are shown in the second column of
Table~\ref{ta:fv_res}. They are computed from the $a_\mu^\mathrm{light}$ numbers from
Table~\ref{ta:fv_4hex} including a multiplication by the
$\left(\frac{9}{10}\right)$ charge factor. The first error is statistical and
the second is an estimate of the cutoff effect. We can also compare the
$L_\mathrm{ref}\to L_\mathrm{big}$ finite-size effect from the {\tt 4hex}
simulations with the prediction of XPT, the corresponding numbers can be found
in Table~\ref{ta:fv_xpt}, both for NLO and NNLO XPT. We find a good agreement
between the values of the {\tt 4hex} simulations and those of NNLO XPT.

Regarding step 3, which is the very small, residual finite-volume effect of the
large box, we employ NNLO XPT. Here we can use the numbers from the last column
of Table~\ref{ta:fv_xpt}. We can estimate the size of the neglected N3LO
contribution\footnote{We use N3LO for next-to-next-to-next-to-leading order and
the N4LO for the subsequent one.}, using the the NLO and NNLO values in the
Table. It turns out to be an order of magnitude smaller than the total error in
our {\tt 4hex} data set, so the NNLO approximation used here is sufficient. We
also performed these XPT computations taking into account a finite temporal
extent. The relevant formulae can be found in our 2020 paper.  These
finite-time effects are even smaller than the residual finite-volume effects.
Both the finite-volume and the finite-time effects are listed in the third
column of Table~\ref{ta:fv_res}.

There is also a finite-size effect in the $I=0$ channel. We expect it to be
N4LO, as discussed in our 2020 work. In the fifth column of
Table~\ref{ta:fv_res}, we give our estimates for the N4LO contribution in the
different windows. Those numbers are obtained from the NLO and NNLO values in
Table~\ref{ta:fv_xpt}. The estimates are taken to be an additional
systematic error in the final result. There are also finite-size effects in the
isospin-breaking contributions, for which our 2020 estimate was $\pm 0.10$ in
the total $a_\mu$. In the windows these effects have to be proportionally
smaller and can be safely neglected at the current level of precision.

To summarize, the finite-size effect in the $L_\mathrm{ref}$ box size can be
obtained by adding the $I=1$ and $I=0$ contributions from the last two columns
of Table~\ref{ta:fv_res}. The light and disconnected cases can be corrected
from the values in the Table according to
\begin{equation}
    \Delta a_\mu^\mathrm{light}=\tfrac{10}{9} \Delta a_\mu^{I=1}
    \quad\text{and}\quad
    \Delta a_\mu^\mathrm{disc}=-\tfrac{1}{9} \Delta a_\mu^{I=1} + \Delta a_\mu^{I=0}\ .
\end{equation}
These numbers have to be added to the continuum-extrapolated results of the
{\tt 4stout} data-set, which correspond to $L_\mathrm{ref}$ box size.

%% file: si_finvol_pheno.tex
\section{Data-driven checks of finite-volume corrections}
\label{se:finvol_pheno}

\begin{figure}[t]
    \centering
    \includegraphics{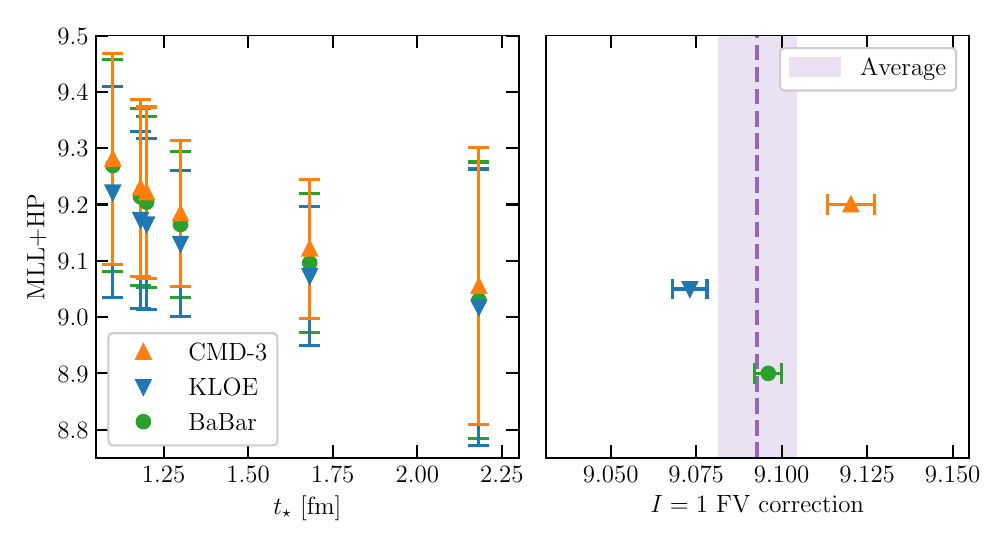}
    \caption
    {
	Results for a data-driven determination of finite-volume effects in
	the $00-28$ window. The left panel shows, on the y-axis, the sum of
	the MLL and HP (1-pion and regular) contributions to the $I=1$
	finite-volume correction, obtained using fits to the different
	experimental data sets, as a function of the time $t_{\star}$ at which
	we switch from HP to MLL.  In addition to the small error due to the
	fit parameters, the uncertainty reported here accounts for the
	truncation of either series and for the one associated with the
	regular part of HP, as described in the main text. The right panel
	contains again $I=1$ finite-volume corrections from individual data
	sets, but the uncertainty comes solely from the fit parameters that
	determine the phase shift.  On the same plot the weighted average is
	shown as a purple dashed vertical line, the purple band corresponds to
	its uncertainty multiplied by a factor of $\sqrt{\chidof}$, as
	described in the text.
    }
    \label{fig:pheno_FV_expfit_and_tstar}
\end{figure}

\begin{figure}[t]
    \centering
    \includegraphics{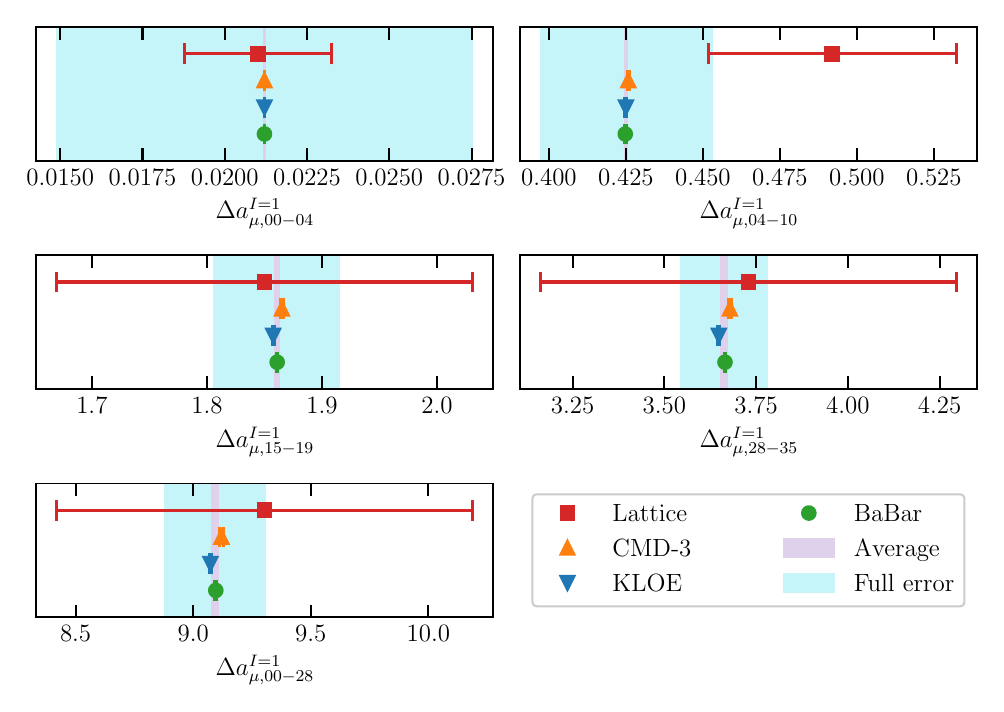}
    \caption
    {
	Comparison of results for the $I=1$ finite-volume correction in
	various windows. The red points correspond to the {\tt 4hex} lattice
	results. The other points correspond to the data-driven determinations
	from individual data sets.  Their averages are also displayed (inner
	purple band), as described in
	Figure~\ref{fig:pheno_FV_expfit_and_tstar}.  Finally, the outer cyan
	band displays the total uncertainty on our data-driven determination,
	accounting for the systematic effects described in this section.
    }
    \label{fig:pheno_FV_and_4hex_sharedplot}
\end{figure}

Using a data-driven approach, in this section we provide an alternative
evaluation of the finite-volume corrections to the various isovector window
observables considered in this paper. This serves as a cross-check for the
corresponding lattice calculation based on the {\tt 4hex} ensembles, and
despite being an interesting result on its own, it is not used in our final
evaluation. We generically denote these finite-volume corrections as $\Delta
a_\mu^{I=1}(L_{\rm ref} \rightarrow \infty)$.

We employ a combination of the Meyer-Lellouch-Lüscher~\cite{Meyer:2011um,
Luscher:1990ux, Lellouch:2000pv} (MLL) and the
Hansen-Patella~\cite{Hansen:2019rbh, Hansen:2020whp} (HP) methods.  They allow
the calculation of finite-volume corrections from the infinite-volume $\pi
\pi$ elastic phase shift $\delta_{11}$ in the $I^G(J^{PC})=1^+(1^{{--}})$
channel.  The methods additionally require the $I=1$ component of the pion
electromagnetic form factor $F_\pi^{I=1}(s)$, evaluated in the timelike domain
$s>0$ for MLL and in the spacelike region $s\leq 0$ for HP.

The MLL formalism has been widely used with the Gounaris-Sakurai
parametrization~\cite{Gounaris:1968mw} of the form factor and the phase shift.  We adopt instead a
data-driven approach based on the parametrization of
Ref.~\cite{DeTroconiz:2001rip, deTroconiz:2004yzs, Davier:2019can}. The
corresponding parameters are obtained via new individual fits to three
measurements of the $e^+e^-\to\pi^+\pi^-$ spectra by
BaBar~\cite{BaBar:2009wpw,BaBar:2012bdw}, CMD-3~\cite{CMD-3:2023alj} and
KLOE~\cite{KLOE:2008fmq,KLOE:2010qei,KLOE:2012anl,KLOE-2:2017fda}, according to
the procedure described in Ref.~\cite{Davier:2019can}.  For each data set we
then obtain the phase shift $\delta_{11}(s)$ and the form factor,
\begin{equation}\label{eq:phenoFV_R_times_Omega}
	F^{I=1}_\pi(s) = J(s) \Omega(s) \; ,
\end{equation}
where $\Omega(s)$ is the Omn\`es
function~\cite{Omnes:1958hv},
\begin{equation}\label{eq:omnes_representation}
    \Omega(s) = \exp \frac{s}{\pi}\int_{s_{\rm th}}^\infty ds' \, \frac{\delta_{11}(s')}{s'(s'-s)} \, ,
\end{equation}
and $J(s)= 1 + \alpha_V s$, where $\alpha_V$ is one of the fit parameters.
Equation~\eqref{eq:phenoFV_R_times_Omega} corresponds to the $I=1$ component
of the pion form factor, up to inelastic contributions modelled by $J(s)$.

Following Ref.~\cite{Borsanyi:2020mff}, for MLL eight finite-volume states are
used to construct the finite-volume correlator $G(t,L)$, an approximation that
is increasingly more accurate at larger Euclidean times $t$.  The last, eighth term
is taken as an estimate of the uncertainty due to the truncation of the sum
over states.  It is added to the finite-volume correlator with a 100\%
uncertainty.  To keep this uncertainty small and its estimate reliable, we use
MLL only at times larger than $t_{\star} = (M_\pi L / 4)^2 / M_\pi$, as
suggested in Ref.~\cite{Francis:2013fzp}. 

Below $t_{\star}$, where MLL poorly represents the finite-volume correlator,
the HP series of Ref.~\cite{Hansen:2020whp} is used, including the three terms
that have been computed, thus neglecting contributions of order $\exp \left( -
M_\pi L  \sqrt{2+\sqrt{3}} \right)$ and higher.  The last, third term is used
as an estimate of the uncertainty due to the truncation of the series.  The
calculation of these terms requires knowledge of the Compton scattering
amplitude of a pion off a spacelike photon. As shown in
Ref.~\cite{Hansen:2020whp} this is dominated by the exchange of a single pion.
This is in turn related to the spacelike pion form factor, for which we use
again Equation~\eqref{eq:omnes_representation}.  The result for the HP single
pion term is then added to the MLL contribution.  Their uncertainties are
summed assuming 100\% correlation, since they derive from the same phase shift.

The rest of the Compton amplitude was computed in NLO XPT in
Ref.~\cite{Hansen:2020whp}. We adopt the same expression, which amounts to a
small contribution to the result, called ``regular part'', to which we assign a
100\% uncertainty that is combined in quadrature with the other uncertainties.
Results for the combination of MLL and HP are listed in
Table~\ref{tab:fv_pheno_HP_and_MLL} for the windows of interest in this paper.

The choice of $t_{\star}$ should not be critical as long as one does not use
MLL at short distances, or HP at long distances, where the methods are not
reliable. We have checked the dependence of our result on different values of
$t_{\star}$. We find compatibility within the uncertainties, provided we do not
consider large or small values of $t_{\star}$, as shown in the left panel of
Figure~\ref{fig:pheno_FV_expfit_and_tstar} where we scan around
$t_\star=1.682\,\fm$, using the $00-28$ window as an example.

Another potential source of systematic uncertainty is the inelastic
contribution to $F_\pi^{I=1}(s)$ due to four-pion states. Given that
inelasticities become sizeable only at energies around $M_\omega +
M_\pi$~\cite{Colangelo:2018mtw}, one would expect them not to
propagate significantly into the finite-volume correction to $a_\mu$.
We note that XPT predicts that they begin at order $s^3$. In the
chiral regime, $s^3$ is two orders higher than the linear $s$ term in
$J(s)$. Thus we conservatively take the effect of that linear term as
an estimate of the possible size of such higher-order
contributions. This is achieved by repeating our calculation of the
finite-volume corrections without that linear term in
$J(s)$. This results in a small shift in the finite-volume
corrections, which we take as an additional systematic.

While we use the phase shift in the isospin limit, where we identify the pion
mass with $M_{\pi^0}$, our parametrization of the phase shift is fitted to
experimental data, where the pions are charged.  We regard this as an
isospin-breaking effect in the finite-volume corrections.  In order to estimate
this effect, we adopt the two-loop inverse-amplitude-method (IAM) expression of
$\delta_{11} (s)$ from Ref.~\cite{Niehus:2020gmf}, which has the advantage of
displaying dependence on the pion mass.  By using the inputs for the IAM provided
in Ref.~\cite{Colangelo:2018mtw,Colangelo:2021moe}, we perform our calculation
with the phase shift at the mass of neutral and of charged pions.  The
resulting shifts in the finite-volume corrections are taken as estimates of
this additional source of uncertainty.  We consider this effect to be
independent of the other systematics and we add it to them in quadrature.

Finally, we need to propagate the uncertainties from the fits to BaBar, KLOE
and CMD-3.  To this end we generate Gaussian samples for the fit parameters
associated with each experimental data set.  While the parameters associated
with different experiments are not correlated, they are correlated for a given
data set.  We account for those correlations and compute the standard
deviations on the resulting finite-volume corrections. Since the data sets are
independent, we compute an average of the corresponding finite-volume
corrections, each weighted by its uncertainty. The result is reported in the
right panel of Figure~\ref{fig:pheno_FV_expfit_and_tstar}, where the
uncertainty has been rescaled by a factor of $\sqrt{\chidof}$. With this
procedure, before including the various other systematic uncertainties
described above, the finite-volume correction determined from CMD-3 data is
not compatible with the one coming from fits to BaBar and KLOE data. This is
because the phase shift (the input to this calculation) is obtained by global
fits that extend up to $\sim 1$~GeV, thus including the region dominated by
the $\rho$ peak, where discrepancies between experiments are observed. As we
will see later, this effect is much smaller than the other systematics of the
calculation. The same applies for including experiments, other than the above
three: they would not significantly impact the final result and error for the
finite-volume effect.

The various sources of uncertainties on the finite-volume corrections studied
here are listed in Table~\ref{tab:fv_pheno_results_systematics}: truncation,
regular part of HP, inelastic, tension between experiments and isospin breaking.
They are assumed to be independent and are summed in quadrature.  Truncation,
inelastic and isospin uncertainties are computed for each data set and we only
keep the largest value for each source of uncertainty. The result of this
combination is shown for each window in
Table~\ref{tab:fv_pheno_results_systematics}, which contains our complete
data-driven determinations of the finite-volume corrections.  We find them to
be in agreement with the corresponding values computed using the {\tt 4hex}
simulations.  In Figure~\ref{fig:pheno_FV_and_4hex_sharedplot} we update the
right panel of Figure~\ref{fig:pheno_FV_expfit_and_tstar} to include the full
uncertainty (cyan band) and a comparison with the corresponding {\tt 4hex} value, and we
display results for all windows.

In Table~\ref{tab:fv_pheno_result_final} we compare finite-volume effects as
computed in the data-driven approach in this section to those computed in NNLO
XPT and in the {\tt 4hex} simulations, as described in the previous section. We
find a good agreement between the various approaches. In our final results the
bulk of the finite-size correction is taken from the {\tt 4hex} simulations,
and for a residual finite-size effect we use NNLO XPT.

\begin{table}[p]
    \centering
    \begin{tabular}{C|C|C|C|C}
	\text{window} & \text{MLL} & \text{HP(1-pion)} &  \text{HP(regular)} & \text{sum} \\
	\hline
	00-28 & 6.60(0.02) & 2.43(0.09) & 0.06(0.06) & 9.09(0.12) \\
	00-04 & 0.00(0.00) & 0.01(0.00) & 0.01(0.01) & 0.02(0.01) \\
	04-10 & 0.00(0.00) & 0.40(0.01) & 0.02(0.02) & 0.43(0.03) \\
	10-28 & 6.60(0.02) & 2.02(0.08) & 0.03(0.03) & 8.65(0.11) \\
	10-\infty & 15.11(0.02) & 2.02(0.08) & 0.03(0.03) & 17.16(0.11)\\
	15-19 & 1.13(0.01) & 0.72(0.03) & 0.01(0.01) & 1.86(0.04) \\
	28-35 & 3.66(0.00) & 0.00(0.00) & 0.00(0.00) & 3.66(0.00) \\
    \end{tabular}
    \caption
    {
	Results for the HP and MLL contributions in the windows of interest in
	this paper. HP is used to compute the finite-volume corrections to
	$a_\mu^{I=1}$, integrating the correlator from $t=0$ up to
	$t_\star=1.682\,\fm$. It is divided into the single pion contribution,
	reported in the third column, and the regular part, in the fourth
	column. From $t_\star$ to infinity, MLL is used. The uncertainty on the
	sum, reported in the last column, assumes 100\% correlation between the
	uncertainties on MLL and HP (1-pion). The uncertainty attached to the
	regular part of HP is added in quadrature.
    }
    \label{tab:fv_pheno_HP_and_MLL}
\end{table}

\begin{table}[p]
    \centering
    \begin{tabular}{C|C|C|C|C|C|C}
	\text{window} & \text{trunc. (MLL+HP)} & \text{HP(regular)} & \text{inel.} & \text{exp.} & \text{iso.} & \text{quad. sum} \\
	\hline
	00-28 & 0.11 & 0.06 & 0.14 & 0.01 & 0.10 & 0.21 \\
	00-04 & 0.00 & 0.01 & 0.00 & 0.00 & 0.00 & 0.01 \\
	04-10 & 0.01 & 0.02 & 0.00 & 0.00 & 0.01 & 0.03 \\
	10-28 & 0.10 & 0.03 & 0.14 & 0.01 & 0.09 & 0.20 \\
	10-\infty & 0.10 & 0.03 & 0.42 & 0.04 & 0.26 & 0.51\\
	15-19 & 0.04 & 0.01 & 0.03 & 0.00 & 0.02 & 0.05 \\
	28-35 & 0.00 & 0.00 & 0.10 & 0.01 & 0.06 & 0.12 \\
    \end{tabular}
    \caption
    {
	Uncertainties in the finite-volume corrections to $a_\mu^{I=1}$
	in the windows of interest in this paper. The numbers reported
	correspond to the data set (BaBar, KLOE or CMD-3) that has the largest
	uncertainty for the given entry. The second column is the sum of the
	truncation uncertainties of HP and MLL, which are assumed to be 100\%
	correlated.  The third column is due to the 100\% uncertainty on the
	leftover regular part of HP. The fourth column is the difference
	between the finite-volume correction result, with and without the
	inelastic contribution to the pion form factor. The fifth column shows
	the error associated with
	the weighted average over experiments, inflated by $\sqrt{\chidof}$. In
	the sixth column, we report an estimate of isospin-breaking effects.
	The last column contains the sum in quadrature of all these
	uncertainties.
    }
    \label{tab:fv_pheno_results_systematics}
\end{table}

\begin{table}[p]
    \centering
    \begin{tabular}{C|C|C|C}
	\text{window} & \text{data-driven} & \text{NNLO XPT} & \text{{\tt 4hex} + NNLO XPT}\\
	\hline
	00-28 & 9.09(0.21) & 8.97 & 9.37(88) \\
	00-04 & 0.02(0.01) & 0.02 & 0.02(00) \\
	04-10 & 0.43(0.03) & 0.40 & 0.49(04) \\
	10-28 & 8.65(0.20) & 8.55 & 8.86(85) \\
	10-\infty & 17.16(0.51) & 16.33 & 18.52(2.45) \\
	15-19 & 1.86(0.05) & 1.87 & 1.85(18) \\
	28-35 & 3.66(0.12) & 3.45 & 3.84(57) \\
    \end{tabular}
    \caption
    {
	Results for the finite-volume correction to $a_\mu^{I=1}$ in the
	windows of interest in this paper, using three different approaches:
	the data-driven approach, the pure NNLO XPT approach and {\tt 4hex}
	direct measurements supplemented by NNLO XPT for residual effects. The
	figures of the last two are taken over from Tables~\ref{ta:fv_xpt} and
	\ref{ta:fv_res} and they correspond to an infinite-time extent.
    }
    \label{tab:fv_pheno_result_final}
\end{table}

%% file: si_ibcheck.tex
\section{Verification of isospin-breaking contributions}
\label{se:si_ibcheck}

As already mentioned in Section~\ref{se:si_win}, we obtain the
isospin-breaking (IB) contributions to $a_{\mu,00-28}$ by repeating the same
analysis as in our 2020 work \cite{Borsanyi:2020mff}, now in the $00-28$
window. In order to further validate our previous procedures, we have
performed several additional checks.

\subsection{Isospin breaking contributions to $a_{\mu}^\mathrm{light}$}

In the present subsection we revisit the computation of the
strong-isospin-breaking (SIB) and the valence QED contributions to
$a_{\mu}^\mathrm{light}$, obtained with the derivatives
\begin{gather}
    [a_{\mu}^\mathrm{light}]'_m \equiv \left. m_l \, \frac{\partial
    [a_{\mu}^\mathrm{light}]}{\partial\, \delta
    m} \right|_{\delta m=0}
    \quad\text{and}\quad
    [a_{\mu}^\mathrm{light}]''_{20} \equiv \left. \frac{1}{2} \frac{\partial^2
    [a_{\mu}^\mathrm{light}]}{\partial e_v^2} \right|_{e_v=0}\ .
\end{gather}
Here $\delta m \equiv m_d - m_u$ denotes the difference in the
down and up quark masses, $m_l \equiv \frac{1}{2} \left( m_u + m_d \right)$
denotes their average, and $e_v$ denotes the valence electric charge.

In our work \cite{Borsanyi:2020mff}, these contributions were obtained using a
chiral extrapolation: we performed computations with valence quark masses that
were multiples $\kappa$ of the sea quark mass, $m_l$, then we used the
measurements at the values $\kappa = 3, 5, 7$ to perform an extrapolation to
the physical point of $\kappa = 1$ gauge-configuration by gauge-configuration.
Here we check the validity of this procedure by eliminating this extrapolation
and by performing measurements directly at $\kappa=1$. 

\begin{figure}
\centering
\includegraphics{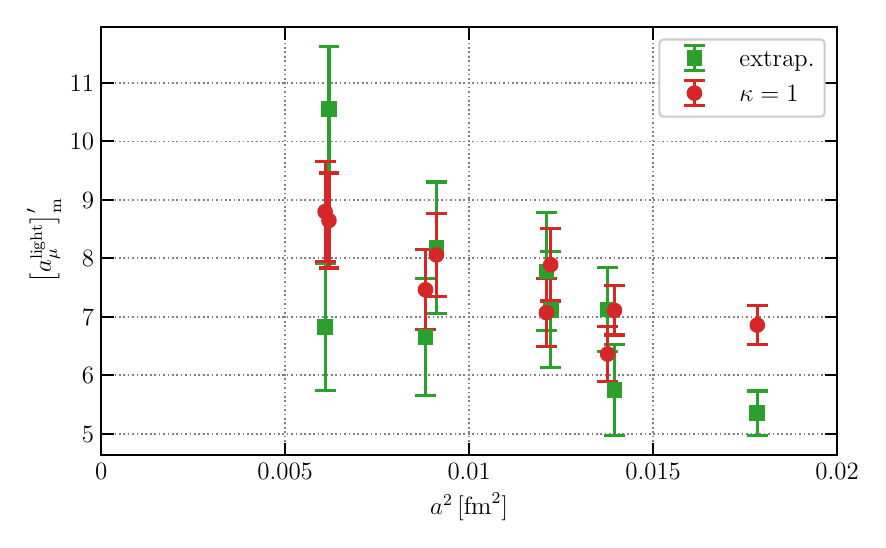}
\caption{
        Strong-isospin-breaking (SIB) contribution to
        $a_{\mu}^{\text{light}}$. The green squares show the results from
        \cite{Borsanyi:2020mff}, obtained via a chiral
        extrapolation from measurements performed at valence quark masses
        which are $\kappa = 3, 5, 7$ times larger than the
        physical light-quark mass $m_l$. The red circles correspond to the
        measurements performed directly at $\kappa=1$, using an
        LMA-based technique.
}
\label{fig:ibcheck_l_sib}
\end{figure}

In order to compute the SIB contribution $[a_{\mu}^\mathrm{light}]'_m$, we
have extended the Low Mode Averaging (LMA) technique \cite{Neff:2001zr,Giusti:2004yp,DeGrand:2004qw,Li:2010pw}, which we were already
using to reduce the noise in the isospin-symmetric contribution. We have
performed measurements at $\kappa=1$ on the same ensembles that were
previously used for the $\kappa$ extrapolation procedure. The new computation
confirms the results of \cite{Borsanyi:2020mff} within error-bars
(cf.~Figure~\ref{fig:ibcheck_l_sib}).

\begin{figure}
    \centering
    \includegraphics{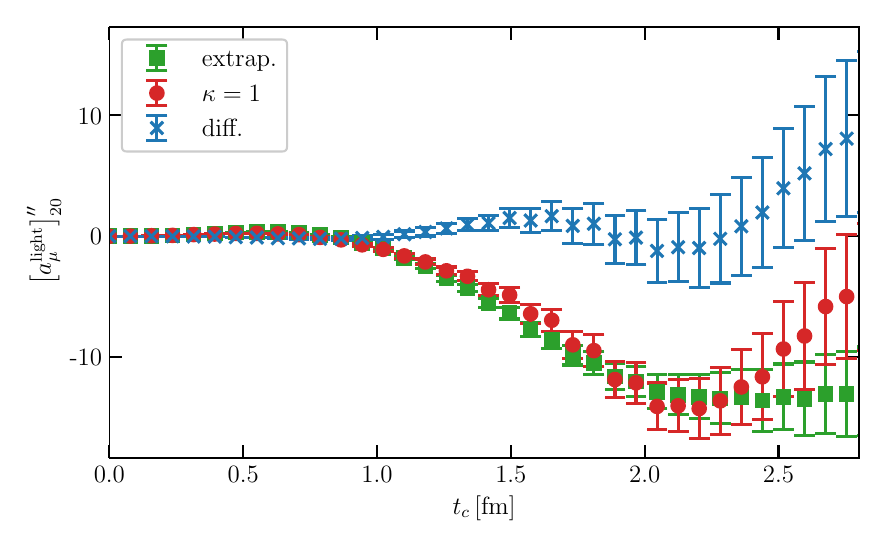}
    \caption
    {
	The valence QED contribution $[a_{\mu}^\mathrm{light}]''_{20}$ on a
	selected $a= 0.0787\,\fm$ configuration, as a function of the upper
	limit of the time integration $t_c$. The green squares show the results
	obtained through a chiral extrapolation from measurements performed at
	valence quark masses which are $\kappa = 3, 5, 7$ times larger than the
	physical light quark mass $m_l$. The red circles correspond to the
	measurements performed directly at $\kappa=1$, and the blue crosses
	show the difference of the two methods.
    }
    \label{fig:ibcheck_l_q20}
\end{figure}

For the valence QED contribution $[a_{\mu}^\mathrm{light}]''_{20}$, we
have checked the gauge-configuration by gauge-configuration extrapolation
procedure on one selected configuration. We have randomly chosen one QCD and
QED configuration from one of our $a= 0.0787\,\fm$ ensembles, and measured
the valence QED contribution using the LMA technique. For the exact eigen part
we projected out 4128 eigenvectors, and used 36000 random source vectors for
the $\kappa=3,5,7$ measurements, and 72000 random source vectors for the
$\kappa=1$ measurement. Figure~\ref{fig:ibcheck_l_q20} shows
$[a_{\mu}^\mathrm{light}]''_{20}$ as a function of the time cut $t_c$, the
upper limit of the integration of the Euclidean time current-current
correlator.

\subsection{Strong-isospin-breaking contribution to $a_{\mu}^\mathrm{disc}$}

\begin{figure}
\centering
\includegraphics{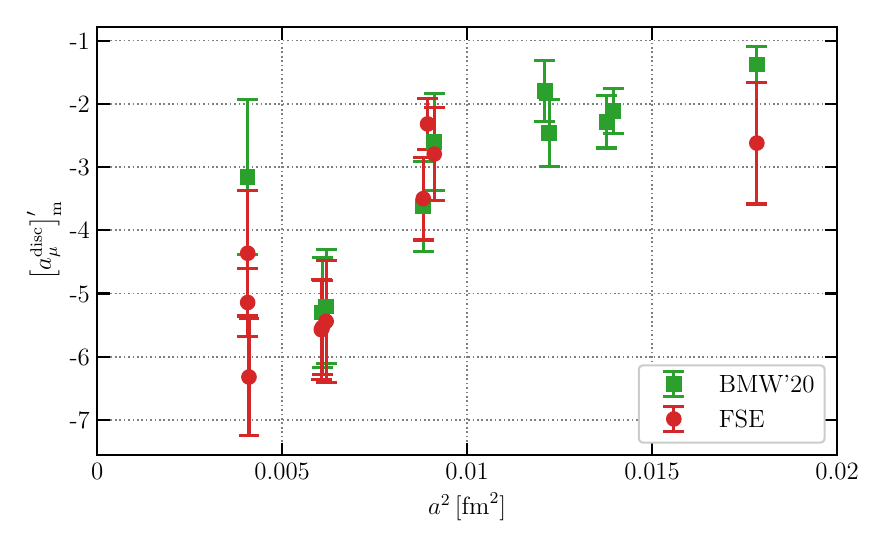}
\caption{
        Strong-isospin-breaking (SIB) contribution to
        $a_{\mu}^{\text{disc}}$. The green squares show the data points
        obtained using the finite difference-method of
        Ref.~\cite{Borsanyi:2020mff}, with 3000 random source vectors. The red
        circles are obtained using the FSE technique of Giusti
        et.~al.~\cite{Giusti:2019kff}, with 128 random-source vectors.
}
\label{fig:ibcheck_di_sib}
\end{figure}

In our previous work \cite{Borsanyi:2020mff} the SIB contribution to
$a_{\mu}^\mathrm{disc}$ was computed by performing a numerical derivative: the
mass derivative was approximated by the difference of the measurements at the
light quark masses $1.0\cdot m_l$ and $0.9\cdot m_l$.  Using the recently
developed frequency-splitting estimator (FSE) technique of
Ref.~\cite{Giusti:2019kff}, we have remeasured on many of our ensembles both
the disconnected contribution to $a_{\mu}$, and its SIB correction as an
exact mass derivative. By using the FSE method, the number of required random
source vectors drops by one to two orders of magnitude. A comparison of the new
FSE measurements of $[a_{\mu}^\mathrm{disc}]'_m$ and the results from
\cite{Borsanyi:2020mff} is shown in Figure~\ref{fig:ibcheck_di_sib}.

\subsection{Corrections}

We discovered two mistakes in our 2020 work, which affect the isopsin-breaking
contributions. None of them was numerically important for $a_\mu$, in the sense
that the associated changes are far less than our final uncertainty.

First we set too aggressive time-cuts for the light-quark sea-QED
contribution, $[a_\mu^\mathrm{light}]^{''}_{02}$. We used $1.0~\mathrm{fm}$ and
$1.5~\mathrm{fm}$, at which point the corresponding correlator is not yet
consistent with zero. An appropriate
choice, $2.2~\mathrm{fm}$ and $2.7~\mathrm{fm}$, decreases the contribution
by one standard deviation and the values in Figure~1 of the main text and in Table~15 of
the Supplemental Information in \cite{Borsanyi:2020mff} have to be changed as
follows:
\begin{gather}
    \label{eq:light_qed}
    \text{light qed:}\quad -0.93(35)(47) \to -1.57(42)(35)\ ,\\
    \text{light qed-ss:}\quad 0.37(21)(24)\to -0.36(27)(15)\ .
\end{gather}
Since in the present work we recompute the IB corrections to the $00-28$
window without further time-cuts, the numbers in Table~\ref{ta:amu} are
unaffected.

In addition, we mistakenly subtracted a term related to one-photon-reducible
contributions, $a_{\mu}^{1\gamma R}$, that we computed by means of infinite-volume
QED. For consistency with our lattice setup, this term should also be
computed in QED$_L$. Due to the zero spatial-momentum projection of the
correlation function the spatial momentum of the photon vanishes. However, in
the QED$_L$ prescription these particular modes are removed from the theory
such that this subtraction is identically zero.  Therefore in Table~16 of
\cite{Borsanyi:2020mff} the following change
\begin{gather}
    \text{one-photon-reducible subtraction:}\quad -0.321(11) \to 0
\end{gather}
has to be applied.

Regarding the total $a_\mu$, we keep using the originally published number as
our 2020 result, when we refer to it in this paper.  As we mentioned, fixing
the two mistakes would change $a_\mu$ only by a small fraction of its total
error. In the new numbers of this work, these mistakes are of course corrected.

%% file: si_tail.tex
\section{\ Long-distance contributions}
\label{se:si_tail}

\def\amutailhead{a_{\mu,28-35}}
\def\amutail{a_{\mu,28-\infty}}
\def\amuID{a_{\mu,04-10}}
\renewcommand{\amulohvp}{a_\mu}

As emphasized in the main text, our goal is to compute the leading-order
hadronic vacuum polarization contribution to the anomalous magnetic moment of
the muon, with fully controlled uncertainties that are significantly smaller
than available today.

The vast majority of our final result (over $95\%$) comes directly from lattice
simulations.  However, there is a contribution for which a data-driven approach
yields an order of magnitude smaller uncertainty than a lattice calculation.
It is the one corresponding to the large Euclidean-time ``tail'' of the current
correlator.  We choose it to be the contribution from a window~\cite{RBC:2018dos} that
extends from $2.8\,\fm$ to $\infty$, for reasons explained below.  We denote it
$\amutail$. Another set of contributions for which a data-driven approach can
lead to uncertainty reduction is the one corresponding to finite-volume
corrections, as discussed in Section~\ref{se:finvol_pheno}. However, in
the present work dedicated lattice simulations are used to compute those
corrections, and the corresponding data-driven determinations are only used
as cross-checks.

Before going into the details of how this tail contribution is obtained, it is
important to clarify a number of issues. As is well known, the 2020 ``Muon
$g-2$ Theory Initiative'' White Paper (WP~'20) data-driven determination of
the full LO-HVP contribution~\cite{Aoyama:2020ynm} leads to significant
tensions between the standard-model prediction and the experimentally-measured
value of $(g_\mu-2)$~\cite{Muong-2:2023cdq,Muong-2:2025xyk}. In addition, our
previous lattice determination of $\amulohvp$ is $2.1\sigma$ larger than the
one given in WP~20, a tension that rises to over $4\sigma$ when one
considers the intermediate-distance window contribution $\amuID$~\cite{
    MILC:2024ryz,RBC:2023pvn,ExtendedTwistedMass:2022jpw,Ce:2022kxy,Aubin:2022hgm,Wang:2022lkq,Lehner:2020crt,Davier:2023cyp,Aliberti:2025beg}.
To make the situation even more confusing, the recent measurement of the
$\epemtopippim$ spectrum by the CMD-3
collaboration~\cite{CMD-3:2023rfe,CMD-3:2023alj} leads to predictions for
$\amulohvp$ and for $\amuID$ that are in good agreement with our 2020 lattice
ones~\cite{Davier:2023fpl}. 

In that last reference~\cite{Davier:2023fpl}, some of us reappraised the
data-driven determination of $\amulohvp$, showing that results, based on
measurements of the $\pi^+\pi^-$ spectrum by
BaBar~\cite{BaBar:2009wpw,BaBar:2012bdw}, CMD-3~\cite{CMD-3:2023alj},
KLOE\cite{KLOE:2008fmq,KLOE:2010qei,KLOE:2012anl,KLOE-2:2017fda} and via
hadronic $\tau$ decays~\cite{Davier:2009ag,Davier:2013sfa}, display significant
tensions.  This reappraisal was prompted by a detailed study of radiative
corrections in initial-state radiation measurements.

\begin{figure}
    \centering  
    \includegraphics[]{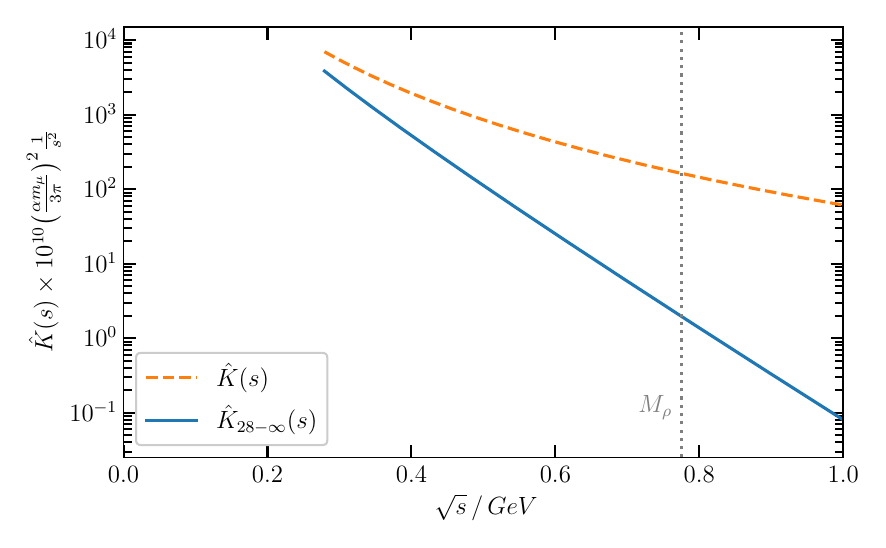}
    \caption
    {
	Kernels $\hat{K}(s)$ and $\hat{K}_{28-\infty}(s)$ plotted as a function
	of centre-of-mass energy $\sqrt{s}$ from the two-pion threshold to
	$1\,\gev$. These kernels multiply the R-ratio $R(s)$ in the integrals
	over $s$ which yield $\amulohvp$ and its tail.
    }
    \label{fi:kernels}
\end{figure}

\begin{figure}
    \centering  
    \includegraphics[width=0.45\textwidth]{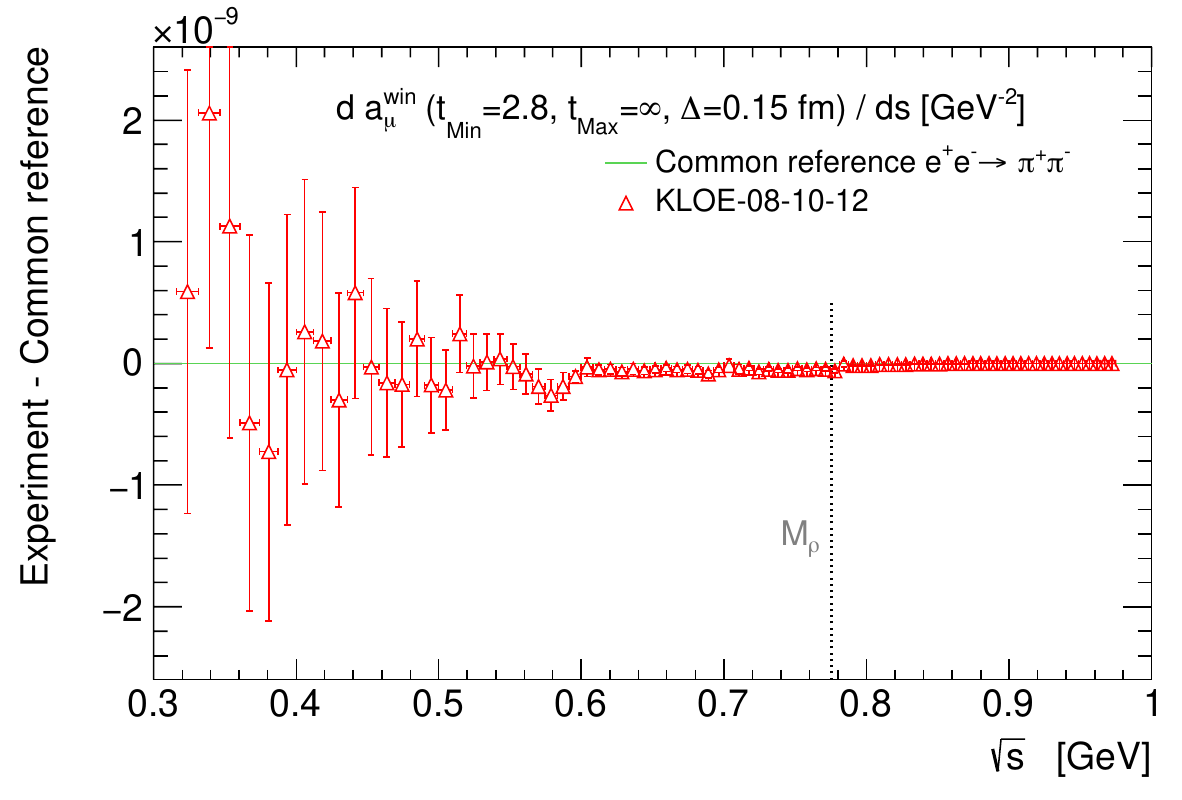}
    \includegraphics[width=0.45\textwidth]{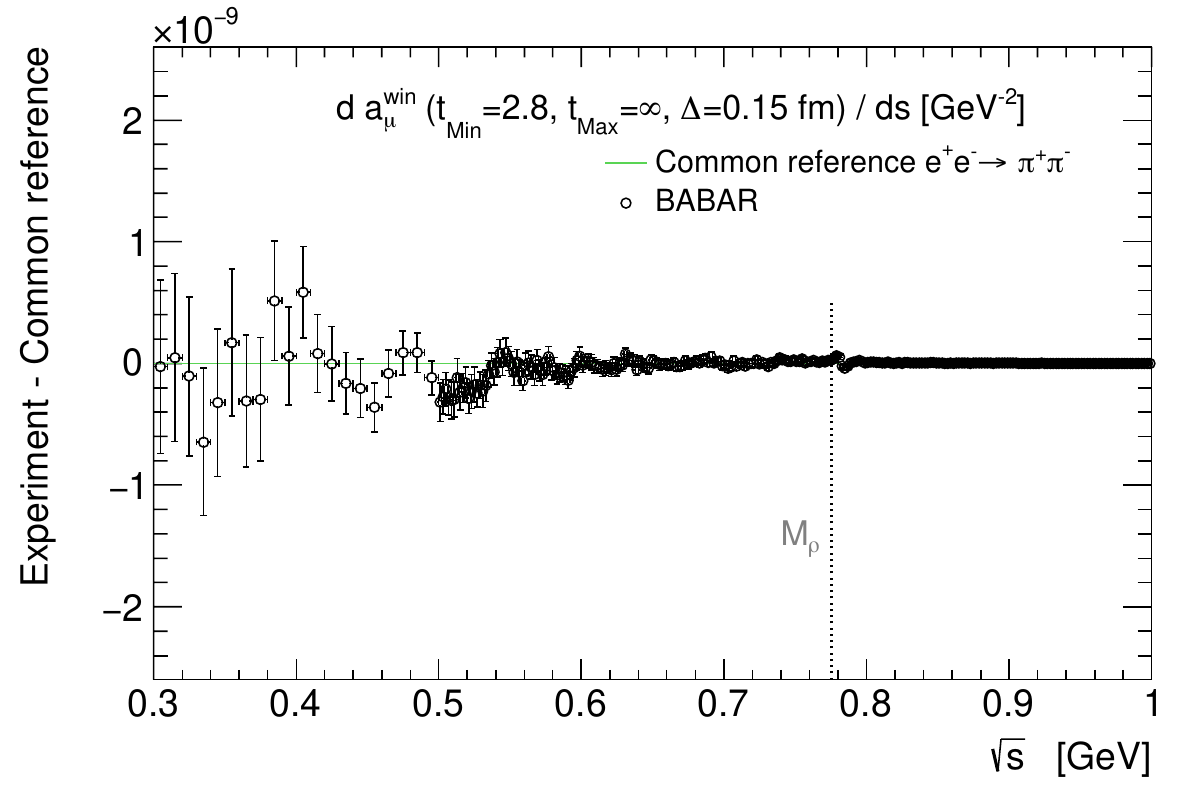}\\
     \includegraphics[width=0.45\textwidth]{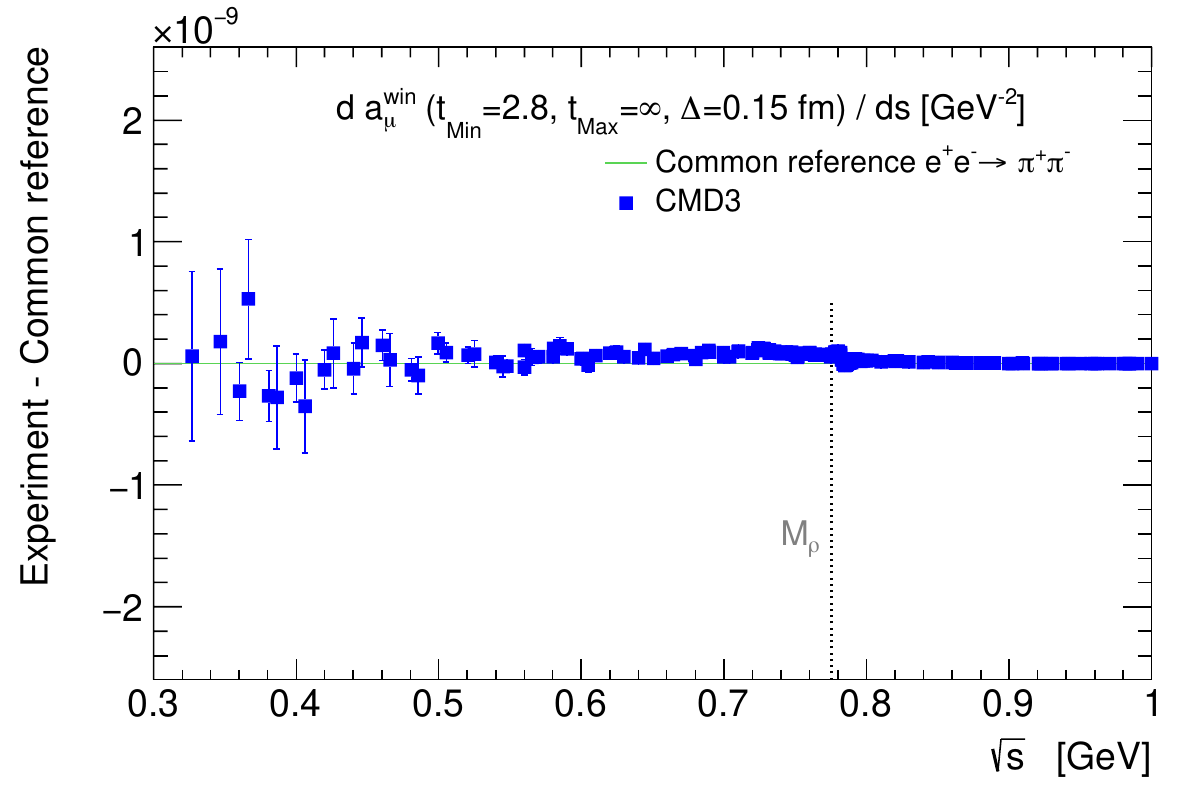}
    \includegraphics[width=0.45\textwidth]{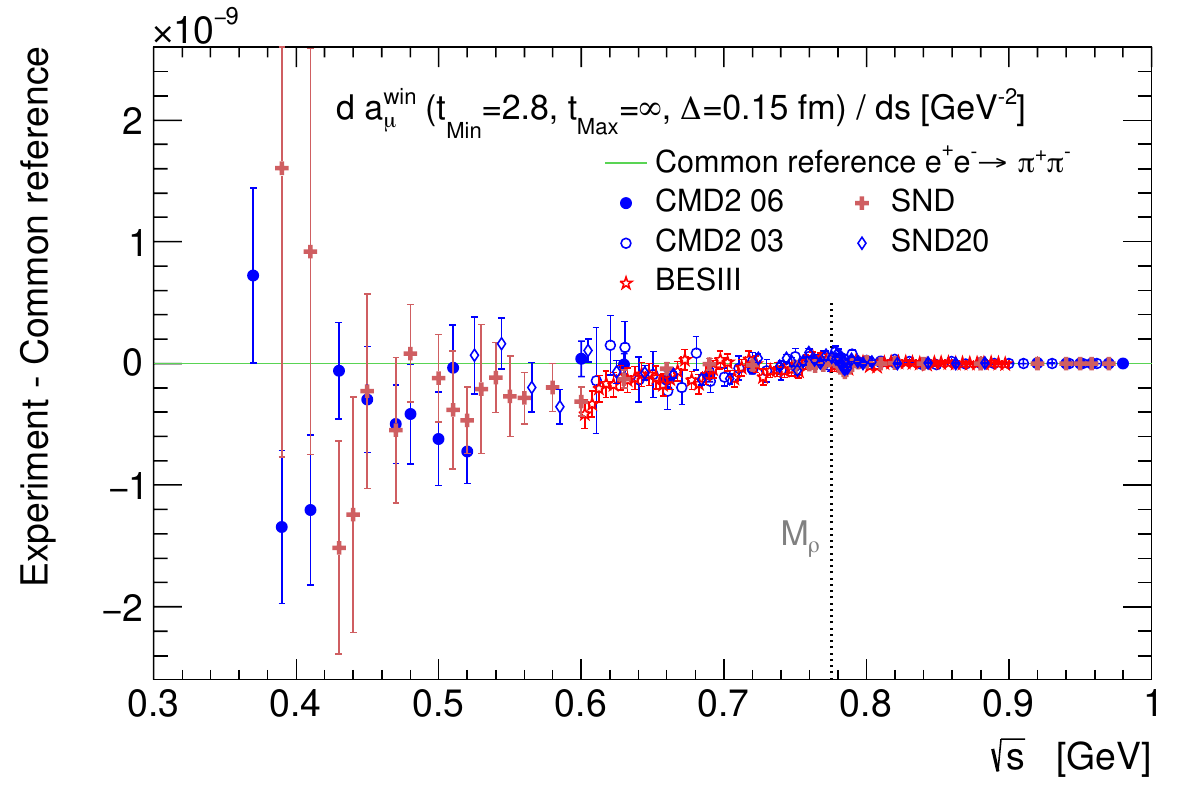}\\
    \includegraphics[width=0.45\textwidth]{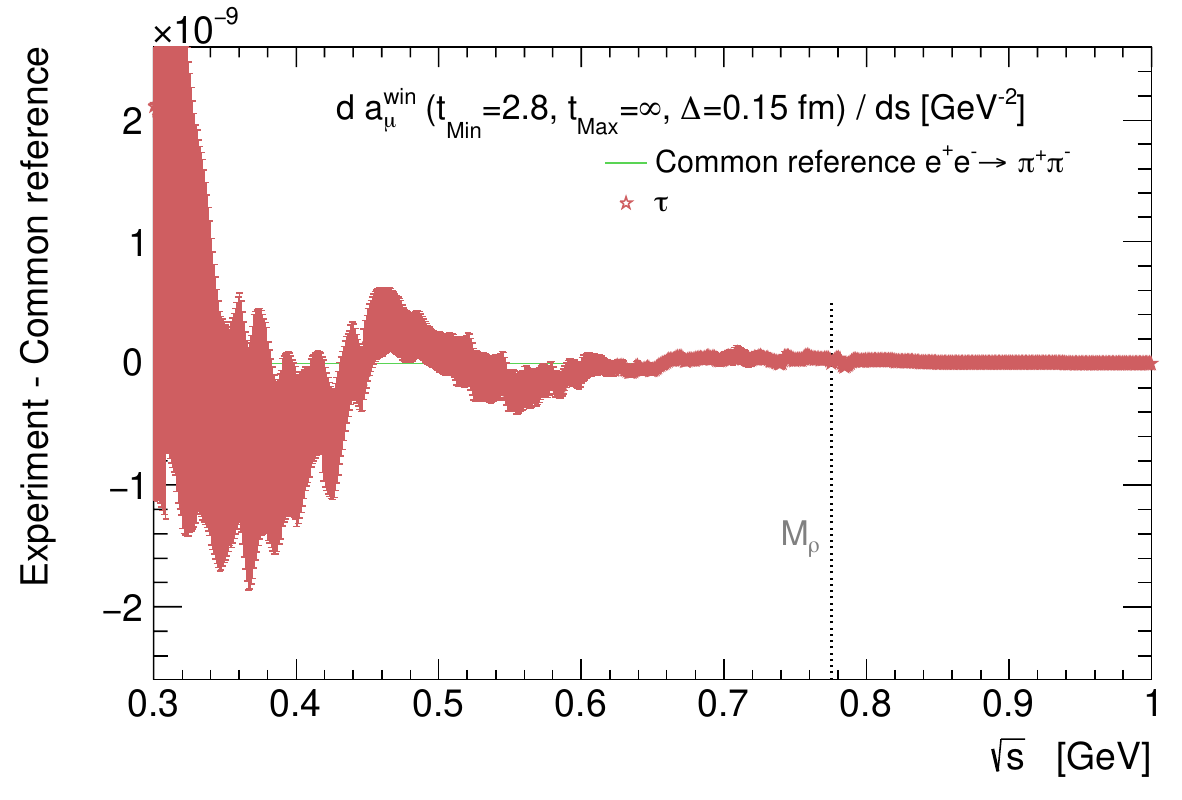}
    \caption
    {
	Difference with a common baseline of integrands for the tail contribution
    to $\amulohvp$ from threshold to $1\,\text{GeV}$.
    The data points are obtained by multiplying the normalized two-pion cross-section
    measurements from different experiments with the tail kernel, $\hat{K}_{28-\infty}(s)$, and by subtracting from them
    the common baseline plotted as the green line at $0$.
    Results are shown for KLOE~(top left), BaBar~(top right), CMD-3~(middle left),
    a set of measurements from BESIII, CMD-2 and SND~(middle right), and from $\tau$
    decays~(bottom).
    All experiments which have data for $\sqrt{s}\leq 0.55\,\text{GeV}$ agree
    well within their uncertainties. 
    The
    higher-mass contributions are highly suppressed by the kernel for the tail. 
    The difference in the areas caught between two datasets and the common reference (with an
    appropriate rescaling on the x-axis from $\sqrt{s}$ to $s$) measures the difference in
    their corresponding predictions for the two-pion contribution to $\amutail$.
    Those differences are visibly small compared to the full size of that contribution,
    whose value is more than $26\times 10^{-10}$ but also to the large uncertainties present at low mass.
    This is made more quantitative in Figure~\ref{fi:DD_tail} where we plot the individual
    results for $\amutail$, obtained using the data of each of the four measurements of
    the $\epemtopippim$ cross sections by BaBar~\cite{BaBar:2009wpw,BaBar:2012bdw},
    CMD-3~\cite{CMD-3:2023alj}, KLOE\cite{KLOE:2008fmq,KLOE:2010qei,KLOE:2012anl,KLOE-2:2017fda}
    and of the rate for $\tau^-\to\pi^-\pi^0\nu_\tau$ decays~\cite{Davier:2009ag,Davier:2013sfa}.
    }
    \label{fi:diff_damutail_ds_vs_sqrt_s}
\end{figure}

\begin{figure}
    \centering  
    \includegraphics{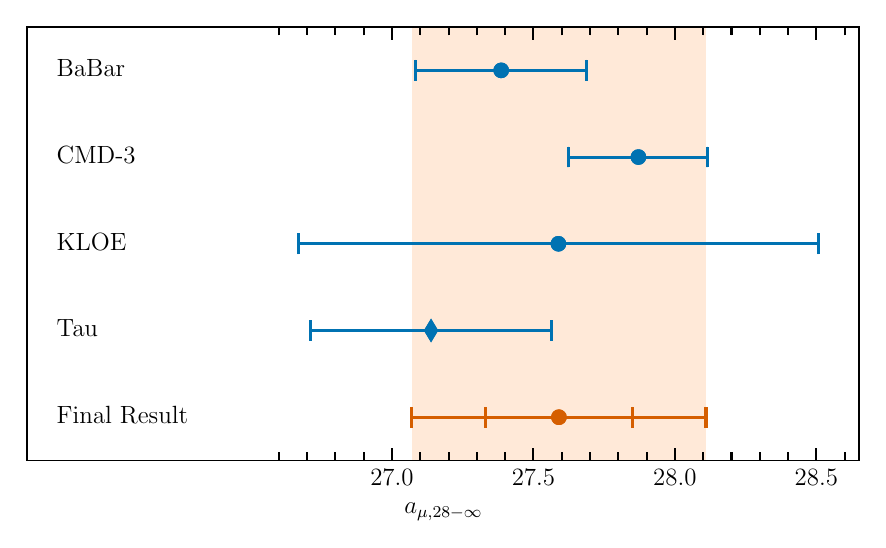}
    \caption
    {
	Results for $\amutail$ obtained using the $\pi^+\pi^-$ spectra measured
	by BaBar, KLOE, CMD-3 and in $\tau$ decays. 
	The orange circle at the bottom shows
	the weighted average. The outer error bars, and the corresponding shaded band
	include the additional uncertainty estimates that we
	conservatively include, obtained as described in the text.
    }
    \label{fi:DD_tail}
\end{figure}

\begin{figure}
    \centering  
    \includegraphics{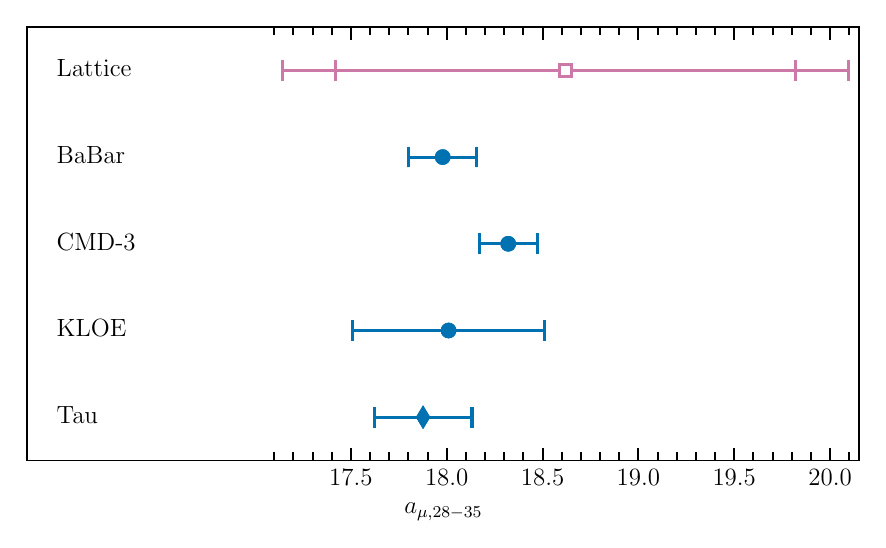}
    \caption
    {
    Results for $\amutailhead$. Figure description is the same as in
    Figure~\ref{fi:DD_tail}. Additionally, the pink square on the top
	represents our lattice result, which is in excellent agreement with the
	data-driven determinations.
    }
    \label{fi:DD-lat_2835_cmp}
\end{figure}

In light of those challenges, one may wonder whether it is justified to
consider a data-driven determination of $\amutail$. To understand why it is, we
emphasize a few important facts:
\begin{itemize}

    \item The kernel, $\hat{K}_{28-\infty}(s)$, which multiplies the
	R-ratio in the integral that yields $\amutail$ in the data-driven
	approach falls off much more quickly with increasing energy than the
	equivalent kernel for $\amulohvp$, $\hat{K}(s)$. This difference is
	illustrated in Figure~\ref{fi:kernels}. Thus, $\amutail$
	is far less sensitive than the full $\amulohvp$ to the $\rho$ peak,
	where the observed discrepancies between different experiments occur%
	~\cite{Davier:2023fpl,Aliberti:2025beg}.

    \item To visualize the effect of this exponential suppression on the
	data-driven determination of the tail we plot in
	Figure~\ref{fi:diff_damutail_ds_vs_sqrt_s} the difference, with respect to a
	common reference, of the data for the two-pion contribution to
	$d\amutail/ds$, coming from all major experiments.  The reference curve
	is used to give a common baseline to all the measurements. All
	experiments which have data for $\sqrt{s}\leq 0.55\,\text{GeV}$ agree
	well within their uncertainties.

    \item As shown in Figure~\ref{fi:DD_tail}, using the $\pi^+\pi^-$ spectrum
	from each experiment that fully covers the energy range that is critical for
	this calculation, all four determinations of this tail contribution are
	entirely consistent. The $\chidof$ associated with the
	weighted average of those four determinations is $1.0$.  The reason is
	that the tail contribution is dominated by the low-mass part of the
	spectrum, below the $\rho$ peak, where all four measurements are in
    good agreement, as discussed in the previous points.

    \item We have checked that the data-driven contribution to $\amulohvp$ from
	$t\ge2.8\,\fm$ is entirely compatible with our lattice calculation.
	This is shown in Figure~\ref{fi:DD-lat_2835_cmp} where we compare
	results for the window $28-35$ that accounts for close to $70\%$ of
	the full tail, obtained from the four measurements and in lattice QCD.
	Here again the data-driven determinations are entirely compatible among
	themselves, with a $\chidof$ of $1.1$, and also with our lattice
	result. The lattice calculation shown here validates the
	data-driven approach to the $8$\% level. This somewhat disadvantageous
	$8$\% uncertainty is the reason why we used the data-driven method for
	the tail observable.

    \item The total uncertainty on our average of the data-driven
    $\amutail$, including the additional conservative uncertainties we
    add below, is $0.52$ in our $10^{-10}$ units, a number that must
    be compared to our total uncertainty of $\valuesTotalerr$ on
    $\amulohvp$. Even if the uncertainty on the tail were arbitrarily doubled, the effect on the
    total uncertainty would be insignificant.

\end{itemize}

Now, the reasons for choosing the tail to start at $t_\mathrm{cut}=
2.8~\mathrm{fm}$ are the following:
\begin{itemize}

\item To ensure the result for $\amulohvp$ presented in this paper is
  dominated by the lattice contribution, we choose to start the
  data-driven tail above $t=2.8\,\fm$. This guarantees that the
  lattice contribution accounts for over $95\%$ of the result.

    \item Beyond reducing the uncertainty on $\amutail$ by an order of
	magnitude, the use of a data-driven tail reduces the finite-volume
	correction that must be applied to the lattice result by a factor of
	$2$ and the associated uncertainty by even more.

    \item As discussed above, for these large times the data-driven
	determinations of $\amutail$ agree very well.

\end{itemize}

Having justified the use of a data-driven approach for determining the
tail contribution, we now explain briefly how it is calculated.  As
mentioned above, the computation is performed following the approach
of \refcite{Davier:2023fpl}.  In that work, the measurements of the
$\pi^+\pi^-$ spectrum by BaBar~\cite{BaBar:2009wpw,BaBar:2012bdw},
KLOE~\cite{KLOE:2008fmq,KLOE:2010qei,KLOE:2012anl,KLOE-2:2017fda},
CMD-3~\cite{CMD-3:2023alj} and via hadronic $\tau$
decays~\cite{Davier:2009ag,Davier:2013sfa} are considered separately.
Outside their centre-of-mass energy ranges and for other hadronic
channels, the data from each experiment are complemented by the
combined experimental and perturbative QCD results compiled in
\refcite{Davier:2019can,Davier:2013sfa}, with a full treatment of
uncertainties and correlations. For BaBar and $\tau$ decays,
centre-of-mass energies range from $0.3\,\gev$ and $2M_{\pi^\pm}$ to
$1.8\,\gev$, and for CMD-3, $0.33\,\gev$ up to $1.2\,\gev$. KLOE
covers the range from $0.32\,\gev$ to $0.97\,\gev$.

The HVPTools
framework~\cite{Davier:2010rnx,Davier:2010nc,Davier:2017zfy,Davier:2019can} is
then used to Laplace transform \cite{Bernecker:2011gh} these four spectra into
the corresponding Euclidean-time correlators. The latter are subsequently
integrated, with the window weights of \refcite{RBC:2018dos}, to give the four
data-driven results for $\amutail$.  As discussed above, the tail contributions
are shown in Figure~\ref{fi:DD_tail} and the comparison of results for the $28-35$
window in Figure~\ref{fi:DD-lat_2835_cmp}.

To obtain our final results for the tail and the $28-35$ window contributions,
we perform weighted averages of the BaBar, KLOE and CMD-3 and $\tau$-decay
determinations, taking into account all correlations.  This procedure yields
our final central values for those quantities, as well as the uncertainties
originating from those in the cross sections.

The agreement of the results using the different data sets
is excellent.  The $\chidof$ for $\amutail$
is less than 1.

The agreement between the results obtained using different data sets fully
extends to those determined via $\tau$ data.  The use of $\tau$ data in this
context requires that one estimates isospin-breaking corrections.  This was
done very carefully in \refcite{Davier:2009ag}.  Nevertheless, we conservatively consider the
absolute value of the full difference between the averages, obtained with and
without the $\tau$-decay data, as an additional uncertainty.  We add it
linearly to the other uncertainties.  In that way, our error bar necessarily
covers the average in which only $e^+e^-$ data are used.  For both quantities
studied in the present section this additional uncertainty is smaller than the
one induced by those in the cross sections.  

Altogether, we obtain
\begin{equation}
    \label{eq:tail_ref}
    a_{\mu,28-\infty}^\mathrm{avg}= \valuesPhenoTail
\end{equation}
as the starting point of our tail-related window result in the data-driven approach. The
first error comes from propagating the uncertainties of the results used in the weighted average. Here no 
PDG-style error rescaling is needed since the two $\chidof$ are less than $1$. The second error is the
additional uncertainty from including or not including the $\tau$ data set. The
third, conservative total error is the first two added linearly.

We consider several sources of systematic uncertainties on the value in
Equation~\eqref{eq:tail_ref}. For this purpose we use the well-tested HVPTools
framework, which combines spectra of different experiments in a local fashion
and integrates the combination. For this error estimation we focus on the
two-pion contribution\footnote{A study of the two-pion contribution to
$\amutail$ using a dispersive approach was recently performed in
\cite{Leplumey:2025kvv}.} in the interval of $\sqrt{s}\in [0.3,\,1.8]$~GeV,
which provides over $97$\% of $a_{\mu,28-\infty}$.  Also we take only the
$e^+e^-\to\pi^+\pi^-$ measurements, but not the $\tau$ data set. Using these
analyses we derive the following systematics:
\begin{enumerate}

    \item In our value in Equation \eqref{eq:tail_ref} we use a weighted
	average of integrals of the two-pion spectra obtained by individual
	experiments. Instead, we could perform the weighted average of the spectra
	first, followed by the integration. The difference between the two
	approaches is $0.12$ in units of $10^{-10}$.

    \item In our value in Equation \eqref{eq:tail_ref} we use four data-sets,
	which have almost full coverage in the relevant energy region. This
	allows for a simple analysis procedure, integrating the spectra first
	and then averaging. These data-sets turn out to agree on the tail
	observable. Agreement is no guarantee of reliability, so we investigate
	adding other experiments with less energy coverage, the principal ones
	being BESIII~\cite{BESIII:2015equ},
	CMD-2~\cite{CMD-2:2003gqi,CMD-2:2005mvb,Aulchenko:2006dxz,CMD-2:2006gxt},
	SND06~\cite{Achasov:2006vp}, SND20~\cite{SND:2020nwa} and
	CLEO~\cite{Xiao:2017dqv}.  They can only be included in an approach
	combining the actual spectra, this was performed in
	\cite{Davier:2023fpl}, which we will refer to as DHLMZ23. The
	difference between the combinations with and without those experiments
	is $0.16$.

    \item In our determination of the tail the recent CMD-3 measurement
	provides a very important contribution, because of its precision in the
	low-mass region. We test the sensitivity of our result to this single
	measurement by removing it from the average. For this purpose we turn
	to the combination performed in \cite{Davier:2019can}, which we will
	refer to as DHMZ19. The difference of the DHLMZ23 and DHMZ19
	combinations, $0.31$, quantifies the effect of removing the CMD-3
	dataset. (Note, that the two combinations also differ in the SND20
	experiment, but it only plays a small role here.)

    \item Even without the CMD-3 experiment there are well-known tensions on
	the total $a_\mu$ (dominated by the $\rho$-meson peak) in the
	data-driven approach between the BaBar and KLOE datasets. As we
	discussed earlier, this is far less important in the tail contribution.
	We quantify its impact by looking at the difference of the DHMZ19
	combination obtained by alternatively removing each of those two
	experiments. This difference is $0.26$. As discussed in WP~’20, the
	BaBar/KLOE tension has a direct impact on the difference between the
	DHMZ19 and an alternative analysis approach KNT19 \cite{Keshavarzi:2019abf}.
	Actually this latter difference is also $0.26$\footnote{We are very
	grateful to A.~Keshavarzi, D.~Nomura and T.~Teubner for sharing their
	KNT19 compilation, which we use to compute this result.}.

\end{enumerate}
We take all these variations and add them up in quadrature to get our
uncertainty estimate
\begin{equation}
    (0.12)_\mathrm{int-avg}(0.16)_\mathrm{other\ exp.}(0.31)_\mathrm{CMD3}(0.26)_\mathrm{BaBar/KLOE} \rightarrow (0.45)
\end{equation}
on the two-pion contribution to the tail observable. Combined with the number in Equation~\eqref{eq:tail_ref}, this gives our final result for the tail:
\begin{equation}
    \label{eq:tail_final}
    \amutail= \valuesPhenoTailnew\ ,
\end{equation}
with a total uncertainty of $1.9$\%. This is the result that we add to our
lattice determination of the complementary window, $a_{\mu,00-28}$, to obtain
our final result for the HVP contribution to the muon. It is important to note
that even if the uncertainty on the tail was multiplied by a factor two, the
final uncertainty on our result for $a_\mu$ would change insignificantly.